\documentclass[10pt, twocolum, journal]{IEEEtran}
\usepackage{mdwmath}
\usepackage{dcolumn}
\usepackage[normalem]{ulem}
\usepackage{multirow}
\usepackage{graphicx,cite,epsfig,amssymb,amsmath,color,epstopdf,epsfig}
\usepackage{xcolor}
\usepackage{setspace}
\graphicspath{{code/}}
\usepackage{extarrows}
\usepackage{amssymb}
\usepackage{amsbsy}
\usepackage{amsmath}
\usepackage{cite}
\usepackage{mathrsfs}
\usepackage{amsfonts}
\usepackage{graphicx}
\usepackage[tight,footnotesize]{subfigure}
\usepackage[10pt]{moresize}
\usepackage{array}
\usepackage{color}
\usepackage{epsfig}
\usepackage{stfloats}
\usepackage{balance}
\usepackage{algorithm}
\usepackage{algorithmic}
\usepackage{booktabs}
\usepackage{multirow}
\usepackage{setspace}
\newtheorem{remark}{Remark}
\usepackage[
            bookmarksnumbered=true,
            bookmarksopen=true,
            colorlinks=false,
            pdfborder={0 0 1},
            citecolor=blue,
            linkcolor=red,
            anchorcolor=green,
            urlcolor=blue,
            breaklinks=true
            ]{hyperref}

\newtheorem{theorem}{\textbf{Theorem}}

\newtheorem{lemma}{\textbf{Lemma}}

\begin{document}
	\title{Optimal Pilots for Anti-Eavesdropping Channel Estimation }
\author
{
	\IEEEauthorblockN
	{
		Qiping Zhu, \emph{Student Member, IEEE,}
		Shuo Wu, \emph{Student Member, IEEE,}
		Yingbo Hua, \emph{Fellow, IEEE}\thanks{The authors are with
			Department of Electrical and Computer Engineering,
			University of California, Riverside, CA 92521, USA. Emails: qzhu005@ucr.edu, swu046@ucr.edu and yhua@ece.ucr.edu. This work was supported in part by the Army Research Office under Grant Number W911NF-17-1-0581. The views and conclusions contained in this
document are those of the author and should not be interpreted as representing the official policies, either expressed or implied, of the Army Research Office or the U.S. Government. The U.S. Government is
authorized to reproduce and distribute reprints for Government purposes notwithstanding any copyright
notation herein.}
	}
	\vspace*{-30pt}
}
	
	\maketitle
	\begin{abstract}
Anti-eavesdropping channel estimation (ANECE) is a method that uses specially designed pilot signals to allow two or more full-duplex radio devices each with one or more antennas to estimate their channel state information (CSI) consistently and at the same time prevent eavesdropper (Eve) with any number of antennas from obtaining its CSI consistently. This paper presents optimal designs of the pilots for ANECE based on two criteria. The first is the mean squared error (MSE) of channel estimation for the users, and the second is the mutual information (MI) between the pilot-driven signals observed by the users. Closed-form optimal pilots are shown under the sum-MSE and sum-MI criteria subject to a symmetric and isotropic condition. Algorithms for computing the optimal pilots are shown for general cases. Fairness issues for three or more users are discussed. The performances of different designs are compared.
	\end{abstract}
\section{Introduction}
Anti-Eavesdropping channel estimation (ANECE) \cite{Hua2019a} is a method that allows two or more legitimate full-duplex radio devices (also called users subsequently) to obtain consistent\footnote{A consistent estimate of a quantity is an estimate which converges to the exact quantity as the signal-to-noise-ratio (SNR) or number of data samples becomes large.} estimates of their receive channel state information (CSI) and at the same time prevents eavesdropper (Eve) from obtaining any consistent estimate of its CSI.
ANECE is useful for the users to maintain a positive secrecy in subsequent transmission of information to each other even if Eve has an unlimited number of antennas. ANECE is unique from many physical layer security approaches as recently surveyed in \cite{Wu2018c} and \cite{Chen2019} where Eve's CSI is assumed to be known not only to Eve but also to users. Only an ``innocent'' Eve would allow users to know its CSI. A ``covert'' Eve would never do that. ANECE can handle not only covert Eve but also ``colluding'' Eves who could form a large antenna array.

At the core of ANECE is the choice of the pilot signals that the full-duplex users transmit to each other simultaneously. As shown in \cite{Hua2019a}, the pilots from all users are such that they
excite all dimensions of the CSI for each user but leave a subspace of Eve's CSI unexcited. In other words, the composite pilot matrix for any user has a full rank that allows consistent estimation of the CSI at this user, but the composite pilot matrix for Eve has a rank deficiency that makes a  subspace of Eve's CSI unobservable by Eve. While sharing a similar goal, ANECE differs from the discriminatory channel estimation (DCE) approach  shown in \cite{Huang2013,Yang2014,Liu2017b} in a number of ways. DCE is designed for user A to: a) assist user B to estimate its CSI, and b) degrade Eve's ability to do the same. DCE requires user A to have more antennas than user B so that artificial noise can be added to the pilot transmitted by user A. In contrast, ANECE does not have the requirement of different numbers of antennas at different users, but ANECE requires the full-duplex capability of users. Also unlike DCE, ANECE is applicable to two or more users simultaneously and allows each and every user to obtain their CSI while keeping Eve blind to its CSI with respect to any user.

When Eve's CSI is unknown to Eve due to use of ANECE, the secrecy capacity of the network against eavesdropping is substantially improved subject to a limited time of information transmission per coherence period as shown in \cite{Hua2019a} and \cite{Sohrabi2019a}.

	%

%

In the literature, there are other works on channel estimation for secret information transmission such as \cite{WangWang2015,YanZhou2018,XiongLiang2016}. But they are not very relevant to this paper as the interest here is to prevent Eve from obtaining its CSI with respect to every transmitter of secret information.

The primary focus of this paper is the optimal design of the pilots for ANECE. We will consider two criteria for optimality: 1) minimizing the mean squared error (MSE) of the estimated channel matrix by each user, and 2) maximizing the mutual information (MI) between the received signals by users.
 The first criterion is useful since the MSE of channel estimation for a user affects the quality of the subsequent operation of information detection by the user. The second criterion is also useful since the MI between two signals observed by two users is the capacity of secret key generation based on the two signals if Eve's knowledge of its CSI is independent of the (reciprocal) CSI between the two users \cite{Maurer1993,bloch2011physical,Lai2012,khisti2016}.

The novelty of this paper includes:
1) the discovery of closed-form optimal pilots  under the sum-MSE and sum-MI criteria and a symmetric and isotropic condition where each user has the same number of antennas, the same noise variance, the same transmit power and the independent and identically distributed (i.i.d.) channel coefficients; and 2) the development of algorithms for computing the (approximately) optimal pilots for any other choices of the above parameters.
The closed-form optimal pilots and the computed optimal pilots are compared with each other and with the previous choice shown in \cite{Hua2019a}. The algorithm for minimum sum-MSE  is  an extension of \cite{Bjornson2010} from two users to more than two users. The algorithm for maximum sum-MI extends \cite{Quist2015} from two users to more than two users.

The rest of the paper is organized as follows. In section \ref{system}, we briefly review ANECE and formulate the pilot design problem. A new insight into the effect of ANECE on Eve's performance are included in Appendix A. In section \ref{sec:MMSE}, the optimal pilots are designed to minimize the sum of MSE for all users, and a discussion for better fairness of MSE among three or more users is also provided.  In section \ref{sec:MI}, the optimal pilots are designed to achieve the maximum sum of the pair-wise MI between the signals observed by all users, and a discussion for better fairness of MI among three or more users is also provided. In section \ref{sec:simulation}, simulation results are shown to compare several types of optimal pilots based on different criteria.

Notations: Vectors and matrices are represented by bold lower case and bold upper case respectively. The $n\times n$ identity matrix is $\mathbf{I}_n$ or simply $\mathbf{I}$ when its dimension is obvious. The trace, expectation, differential, natural logarithm, base-2 logarithm, determinant, transpose, conjugate, conjugated transpose and Kronecker product  are respectively $Tr$, $\mathcal{E}$, $\partial$, $\ln$, $\log_2$, $|\cdot|$, $^T$, $^*$, $^H$ and $\otimes$. The $n\times m$ real field and $n\times m$ complex field are $\mathbb{R}^{n\times m}$ and $\mathbb{C}^{n\times m}$. All other notations are defined in the context.

%
	\section{System Model}\label{system}
		\begin{figure}[t]
			    \vspace{-10pt}
			\includegraphics[width=0.33\textwidth]{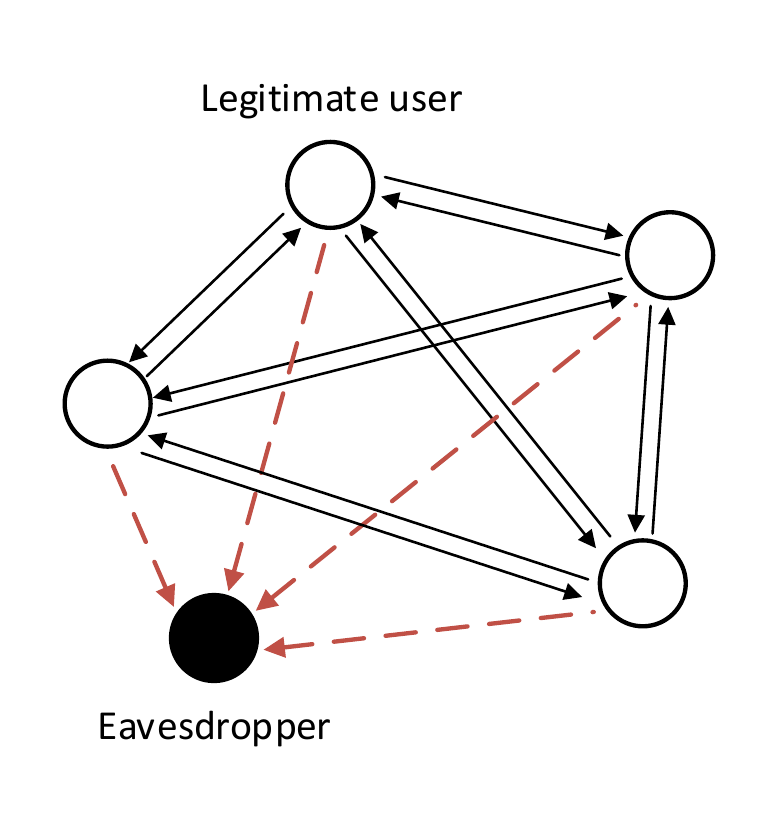}
			\centering
			\caption{Multiple full-duplex multi-antenna users perform ANECE against covert eavesdropper (Eve) with any number of antennas.}
			\label{sys}
		\end{figure}
As illustrated in Fig \ref{sys}, we consider a wireless network of $M$ legitimate full-duplex multi-antenna users and a passive multi-antenna eavesdropper (Eve). Let $N_i$ be the number of antennas on user $i$, and $N_E$ be the number of antennas on Eve. According to ANECE \cite{Hua2019a}, all users concurrently transmit their pilots $\mathbf{p}_i(k)$ over a time window $k=1,\cdots,K$ with $i$ corresponding to user $i$. These pilots are designed in such a way (see below) that all users can reliably estimate their own channel matrices
but Eve cannot.

Specifically, let the signal received by user $i$ over a time window of $K$ sampling intervals be $\mathbf{Y}_i\in \mathbb{C}^{N_i\times K}$, and the signal received by Eve in this window be $\mathbf{Y}_E \in \mathbb{C}^{N_E\times K}$. It follows that
\begin{subequations}\label{mul_sig}
\begin{align}
&\mathbf{Y}_{i} =  \sum_{j\neq i}^{M}\mathbf{R}_{i}^{\frac{1}{2}}\mathbf{H}_{i,j}\mathbf{R}_{j}^{\frac{T}{2}}\mathbf{P}_{j} + \mathbf{N}_{i},\label{mul_sig_user}\\
&\mathbf{Y}_{E} = \sum_{i=1}^{M}\mathbf{H}_{E,i}\mathbf{R}_{i}^{\frac{T}{2}}\mathbf{P}_{i} + \mathbf{N}_{E}\label{mul_sig_eve}
\end{align}
\end{subequations}
where $\mathbf{P}_{i} =[\mathbf{p}_i(1),\cdots,\mathbf{p}_i(K)] \in \mathbb{C}^{N_{i}\times K}$ is the pilot matrix sent by user $i$, $\mathbf{R}_i^{\frac{1}{2}}\mathbf{H}_{i,j}\mathbf{R}_j^{\frac{T}{2}}$ is the overall channel matrix from user $j$ to user $i$, and $\mathbf{H}_{E,i}\mathbf{R}_{i}^{\frac{T}{2}}$ is the overall channel matrix from user $i$ to Eve. Here, we have assumed that all channels between users are reciprocal, the transmit/receive correlation matrix of user $i$ is denoted by $\mathbf{R}_i\in \mathbb{C}^{N_{i}\times N_{i}}$ and the elements in $\mathbf{H}_{i,j}\in \mathbb{C}^{N_{i}\times N_{j}}$ are independent and identical distributed (i.i.d.) with $\mathcal{CN}(0,1)$ entries. We also assume that $\|\mathbf{H}_{E,i}\mathbf{R}_{i}^{\frac{T}{2}}\mathbf{P}_{i}\|$ for any $i$  is  not negligible compared to $\|\mathbf{H}_{E,j}\mathbf{R}_{j}^{\frac{T}{2}}\mathbf{P}_{j}\|$ with $j\neq i$. We will write $\mathbf{R}_i=\mathbf{R}_i^{\frac{1}{2}}\mathbf{R}_i^{\frac{H}{2}}$ which is of full rank and known to all users and Eve. We assume that $\mathbf{H}_{E,j} \in \mathbb{C}^{N_{E}\times N_{j}}$ for any $j$ is independent of $\mathbf{H}_{i,m}$ for any $i$ and $m$.  Finally,
$\mathbf{N}_{i} \in \mathbb{C}^{N_{i}\times K}$ includes all residual self-interference at user $i$ and consists of i.i.d. $\mathcal{CN}(0,\sigma_i^2)$ entries, and $\mathbf{N}_{E} \in \mathbb{C}^{N_{E}\times K}$ consists of i.i.d. $\mathcal{CN}(0,\sigma_E^2)$ entries.

Now define $N_{T} = \sum_{i=1}^{M}N_{i}$,   $\bar{\mathbf{P}} =[\mathbf{P}_1^T,\cdots,\mathbf{P}_M^T]^T \in \mathbb{C}^{N_{T}\times K}$, $\bar{\mathbf{P}}_{(i)} \in \mathbb{C}^{(N_{T}-N_{i})\times K}$ as $\bar{\mathbf{P}}$ without $\mathbf{P}_i$,
$\bar{\mathbf{R}}=diag[\mathbf{R}_1,\cdots,\mathbf{R}_M] \in \mathbb{C}^{N_{T}\times N_{T}}$, $\bar{\mathbf{R}}_{(i)} \in \mathbb{C}^{(N_{T} - N_{i})\times (N_{T} - N_{i})}$ as $\bar{\mathbf{R}}$ without $\mathbf{R}_i$,  $\bar{\mathbf{H}}_{(i)}\in \mathbb{C}^{N_{i}\times (N_{T} - N_{i})}$ as the horizontal stack of $\mathbf{H}_{i,j}$ for all $j \neq i$, and $\bar{\mathbf{H}}_{E}=[\mathbf{H}_{E,1},\cdots,\mathbf{H}_{E,M}]\in \mathbb{C}^{N_{E}\times N_{T}}$.  Also let $P_{i}$ be the transmit power by user $i$ and $P_{T} = \sum_{i=1}^{M}P_{i}$ be the total power by all users. It follows that $Tr(\mathbf{P}_{i}\mathbf{P}_{i}^{H}) \leq KP_{i}$.
 Then \eqref{mul_sig} can be rewritten as
\begin{subequations}\label{mul_sig_2}
	\begin{align}
	&\mathbf{Y}_{i}
=\mathbf{R}_{i}^{\frac{1}{2}}\bar{\mathbf{H}}_{(i)}\bar{\mathbf{R}}_{(i)}^{\frac{T}{2}}
\bar{\mathbf{P}}_{(i)} + \mathbf{N}_{i}\label{mul_sig_user_2},\\
	&\mathbf{Y}_{E} = \bar{\mathbf{H}}_{E}\bar{\mathbf{R}}^{\frac{T}{2}}\bar{\mathbf{P}} + \mathbf{N}_{E}.\label{mul_sig_eve_2}
	\end{align}
\end{subequations}

For ANECE \cite{Hua2019a}, we need to choose the (publicly known) pilots such that $rank(\bar{\mathbf{P}}_{(i)})=N_T-N_i$ (i.e., all rows of $\bar{\mathbf{P}}_{(i)}$ for every $i$ are linearly independent) and $rank(\bar{\mathbf{P}})=r\leq N_T-1$ (i.e., all rows of $\bar{\mathbf{P}}$ are not linearly independent). It is easy to verify from \eqref{mul_sig_2} that
the first rank constraint allows each user to obtain a consistent estimate of its channel matrix while the second rank constraint creates a subspace of Eve's channel matrix for which there is no consistent estimation. Note that since $\bar{\mathbf{P}}_{(i)}$ has a full row rank, user $i$ can estimate $\mathbf{R}_{i}^{\frac{1}{2}}\bar{\mathbf{H}}_{(i)}\bar{\mathbf{R}}_{(i)}^{\frac{T}{2}}$ consistently. And since $\bar{\mathbf{P}}$ has a left null subspace, Eve cannot obtain a consistent estimate of $\bar{\mathbf{H}}_{E}\bar{\mathbf{R}}^{\frac{T}{2}}$. In Appendix \ref{app:EVE_MSE}, the MMSE of Eve's CSI by Eve subject to $rank(\bar{\mathbf{P}})=r\leq N_T-1$ is further discussed.

In the rest of this paper, we will focus on the optimal designs of the pilots subject to the rank conditions required for ANECE. We will consider  two design criteria: one is based on the MSE of users' channel estimation, and the other is based on the MI between users' observations. A discussion of maximum likelihood (ML) channel estimation is included in the end of the next section.
\section{Pilot Designs Based on MSE}\label{sec:MMSE}

Define $\mathbf{S}_{i}$ as the $N_{i}\times N_{T}$ selection matrix such that $\mathbf{S}_{i}\bar{\mathbf{P}} = \mathbf{P}_{i}$, and $\bar{\mathbf{S}}_{(i)}$ as the $(N_{T} - N_{i})\times N_{T}$ matrix which is the vertical stack of $\mathbf{S}_{j}$ for all $j \neq i$. Note that $\bar{\mathbf{R}}_{(i)}^{\frac{T}{2}}
\bar{\mathbf{P}}_{(i)}=\bar{\mathbf{S}}_{(i)}\bar{\mathbf{R}}^{\frac{T}{2}}\bar{\mathbf{P}}$.
Also using $vec(\mathbf{X}\mathbf{Y}\mathbf{Z}) = (\mathbf{Z}^{T}\otimes \mathbf{X})vec(\mathbf{Y})$, \eqref{mul_sig_user_2} becomes
	\begin{align}
	&\mathbf{y}_{i} = \bar{\mathbf{G}}_i^H\bar{\mathbf{h}}_{i} + \mathbf{n}_{i}\label{sig_vec_user}
	\end{align}
where $\mathbf{y}_{i} = vec(\mathbf{Y}_{i})$,  $\bar{\mathbf{h}}_{i} = vec(\bar{\mathbf{H}}_{(i)})$, $\mathbf{n}_{i} = vec(\mathbf{N}_{i})$ and $\bar{\mathbf{G}}_{i} = (\bar{\mathbf{S}}_{(i)}\bar{\mathbf{R}}^{\frac{H}{2}}\bar{\mathbf{P}}^{*}\otimes \mathbf{R}_{i}^{\frac{H}{2}})$.

Let $\mathbf{K}_{\mathbf{x},\mathbf{y}}=\mathcal{E}\{\mathbf{x}\mathbf{y}^H\}$ be the correlation matrix between two random vectors $\mathbf{x}$ and $\mathbf{y}$, and $\mathbf{K}_{\mathbf{x}}=\mathbf{K}_{\mathbf{x},\mathbf{x}}$.
The MMSE estimate of $\bar{\mathbf{h}}_{i}$ by user $i$ is
\begin{align}
&\hat{\bar{\mathbf{h}}}_{i}=\mathbf{K}_{\bar{\mathbf{h}}_{i},\mathbf{y}_{i}}
\mathbf{K}_{\mathbf{y}_{i}}^{-1}\mathbf{y}_{i}
= \bar{\mathbf{G}}_{i}(\bar{\mathbf{G}}_{i}^H\bar{\mathbf{G}}_{i} +  \sigma^{2}_{i}\mathbf{I})^{-1}\mathbf{y}_{i}.\label{MMSE_mul_sig}
\end{align}
Define  $\Delta\bar{\mathbf{h}}_{i} = \bar{\mathbf{h}}_{i} - \hat{\bar{\mathbf{h}}}_{i}$. Then the MSE  of $\hat{\bar{\mathbf{h}}}_{i}$ is
 \begin{align}
 \texttt{MSE}_i&=Tr(\mathcal{E}\{\Delta\bar{\mathbf{h}}_{i}\Delta\bar{\mathbf{h}}_{i}^{H}\})= Tr(\mathbf{K}_{\bar{\mathbf{h}}_{i}} - \mathbf{K}_{\bar{\mathbf{h}}_{i},\mathbf{y}_{i}}\mathbf{K}_{\mathbf{y}_{i}}^{-1}
 \mathbf{K}_{\mathbf{y}_{i},\bar{\mathbf{h}}_{i}})\notag\\
   &= Tr\left(\mathbf{I} - \bar{\mathbf{G}}_{i}(\bar{\mathbf{G}}_{i}^{H}\bar{\mathbf{G}}_{i} +  \sigma_{i}^{2}\mathbf{I})^{-1}\bar{\mathbf{G}}_{i}^{H}\right)\notag\\
 & = Tr\left(\left (\mathbf{I} + \frac{1}{ \sigma^{2}_{i}}\bar{\mathbf{G}}_{i}\bar{\mathbf{G}}_{i}^{H}\right )^{-1}\right)\label{MSE_mul}
 \end{align}
 where the last equality is based on the well known matrix inverse lemma.

Now we consider the following criterion for pilot design:
\begin{align}
&\min_{\bar{\mathbf{P}}}~~ J_{M}=\sum_{i=1}^M \texttt{MSE}_i\label{opt_MSE}\\
s.t. ~&Tr(\mathbf{P}_{i}\mathbf{P}_{i}^{H}) \leq KP_{i},~i = 1,\dots,M,\notag\\
&rank(\bar{\mathbf{P}}) = r\notag
\end{align}
where $N_T-N_{min}\leq r \leq N_T-1$ with $N_{min}=\min_i N_i$.

Since $\bar{\mathbf{R}}$ is known and nonsingular, we can apply the following change of parameters:
\begin{align}
&\bar{\mathbf{R}}^{\frac{H}{2}}\bar{\mathbf{P}}^{*}= \bar{\mathbf{F}}\bar{\mathbf{V}}\label{decom}
\end{align}
where $\bar{\mathbf{V}}\in \mathbb{C}^{r\times K}$ is any  semi-unitary matrix satisfying $\bar{\mathbf{V}}\bar{\mathbf{V}}^{H}  = \mathbf{I}_{r}$, and $\bar{\mathbf{F}} \in \mathbb{C}^{N_{T}\times r}$ is now what we need to design. Namely,
\begin{equation}\label{decom_inv}
\bar{\mathbf{P}} = \bar{\mathbf{R}}^{-\frac{T}{2}}\bar{\mathbf{F}}^{*}\bar{\mathbf{V}}^{*}
\end{equation}
which meets the rank constraint as long as $\bar{\mathbf{F}}$ has a full column rank. To further simplify \eqref{opt_MSE}, we use the eigenvalue decomposition (EVD):
\begin{equation}\label{svd_bar}
\mathbf{R}_{i} = \tilde{\mathbf{U}}_{i}\tilde{\boldsymbol{\Lambda}}_{i}\tilde{\mathbf{U}}^{H}_{i}
\end{equation}
where
 $\tilde{\boldsymbol{\Lambda}}_{i} = diag\{\tilde{\lambda}_{i,1},\dots,\tilde{\lambda}_{i,N_{i}}\}$ with $\sum_{l}\tilde{\lambda}_{i,l} = N_{i}$.  The diagonal elements in $\tilde{\boldsymbol{\Lambda}}_{i}$ are in descending order. From \eqref{svd_bar},  we have $\mathbf{R}_{i}^{\frac{1}{2}} = \tilde{\mathbf{U}}_{i}\tilde{\boldsymbol{\Lambda}}_{i}^{\frac{1}{2}}$.

With \eqref{decom} and \eqref{svd_bar},  the cost function in \eqref{opt_MSE} becomes
\begin{align}
J_{M} = \sum_{i=1}^{M} Tr\left (\left [\mathbf{I} + \frac{1}{ \sigma^{2}_{i}}( \tilde{\boldsymbol{\Lambda}}_{i} \otimes \bar{\mathbf{S}}_{(i)}\bar{\mathbf{F}}\bar{\mathbf{F}}^{H}\bar{\mathbf{S}}_{(i)}^{T} )\right ]^{-1}\right )\label{MSE_mul_2}
\end{align}
where we have used $Tr([\mathbf{I}+\mathbf{X}\otimes\mathbf{Y}]^{-1})=Tr([\mathbf{I}+\mathbf{Y}\otimes\mathbf{X}]^{-1})$,
and hence \eqref{opt_MSE} becomes
\begin{align}
&\min_{\bar{\mathbf{F}}}~~ J_{M}\label{opt_MSE_MulU}\\
s.t. ~&Tr(\mathbf{S}_{i}\bar{\mathbf{R}}^{-\frac{H}{2}}\bar{\mathbf{F}}\bar{\mathbf{F}}^{H}
\bar{\mathbf{R}}^{-\frac{1}{2}}
\mathbf{S}_{i}^{T}) \leq KP_{i},~i = 1,\dots,M\notag
\end{align}
where $\mathbf{S}_{i}\bar{\mathbf{R}}^{-\frac{H}{2}}\bar{\mathbf{F}}\bar{\mathbf{F}}^{H}
\bar{\mathbf{R}}^{-\frac{1}{2}}
\mathbf{S}_{i}^{T}=\mathbf{P}_i^*\mathbf{P}_i^T$.

The problem \eqref{opt_MSE_MulU} is non-convex in general. We will next treat it in three separate situations. We will first present a general algorithm for $M\geq 2$, then a specialized (efficient) algorithm for $M=2$, and finally closed-form solutions of the optimal pilots under the case of $M\geq 2$, $N_i=N$, $P_i=P$, $\sigma_i^2=\sigma^{2}$ and $\mathbf{R}_i=\mathbf{I}_N$. The invariance of the above parameters to $i$ is called a symmetric condition, and $\mathbf{R}_i=\mathbf{I}_N$ is an isotropic condition.
%
\subsection{General algorithm for $M\geq 2$}\label{sec::MSE_m3}

To solve the problem \eqref{opt_MSE_MulU} with $M\geq 2$, we can apply the logarithmic barrier method \cite{Boyd2004a}. With the barrier coefficient $t$, we define
\begin{equation}\label{barrier_ce}
g_{1}(\bar{\mathbf{F}}) = tJ_{M}  + \sum_{i=1}^{M}\mathcal{B}_{P,i}(\bar{\mathbf{F}})
\end{equation}
where
\begin{equation}\label{barrier}
\mathcal{B}_{P,i} (\bar{\mathbf{F}})= - \ln( \psi_{P,i}(\bar{\mathbf{F}}))
\end{equation}
and $\psi_{P,i}(\bar{\mathbf{F}}) = KP_{i}-Tr(\mathbf{S}_{i}\bar{\mathbf{R}}^{-\frac{H}{2}}\bar{\mathbf{F}}\bar{\mathbf{F}}^{H}
\bar{\mathbf{R}}^{-\frac{1}{2}}
	\mathbf{S}_{i}^{T})$. Then, \eqref{opt_MSE_MulU} is approximated by
\begin{align}
\min_{\bar{\mathbf{F}}} &\quad g_{1}(\bar{\mathbf{F}}).\label{opt_MSE_lb}
\end{align}

The gradient of a real-valued function $f(\mathbf{X})$ with respect to a complex matrix $\mathbf{X}$ is denoted and defined as $\nabla f(\mathbf{X})=\frac{\partial f(\mathbf{X})}{\partial \mathbf{X}}=\frac{\partial f(\mathbf{X})}{\partial \Re(\mathbf{X})}+j\frac{\partial f(\mathbf{X})}{\partial \Im(\mathbf{X})}$. One can verify that
$\nabla g_1(\bar{\mathbf{F}}) = t\nabla J_M(\bar{\mathbf{F}})+\sum_{i=1}^M \nabla \mathcal{B}_{P,i} (\bar{\mathbf{F}})$
where
\begin{align}
&\nabla J_M(\bar{\mathbf{F}})=\notag\\
&  -2\sum_{i=1}^{M}\sum_{l = 1}^{N_{i}}\frac{\tilde{\lambda}_{i,l}}{ \sigma^{2}_{i}}\bar{\mathbf{S}}_{(i)}^{T}(\mathbf{I} + \frac{\tilde{\lambda}_{i,l}}{ \sigma^{2}_{i}} \bar{\mathbf{S}}_{(i)}\bar{\mathbf{F}}\bar{\mathbf{F}}^{H}\bar{\mathbf{S}}_{(i)}^{T} )^{-2}\bar{\mathbf{S}}_{(i)}\bar{\mathbf{F}},\label{d1}
\end{align}
\begin{equation}\label{de_b_u}
\nabla \mathcal{B}_{P,i} (\bar{\mathbf{F}}) =  2\left (\frac{\bar{\mathbf{R}}^{-\frac{1}{2}}\mathbf{S}_{i}^{T}\mathbf{S}_{i}\bar{\mathbf{R}}^{-\frac{H}{2}}
\bar{\mathbf{F}}}{\psi_{P,i}(\bar{\mathbf{F}})}\right ).
\end{equation}
Algorithm \ref{Algorithm_g1} shown in the table solves \eqref{opt_MSE_lb}  using gradient descent where  $\bar{\mathbf{F}}$ is initially set to be $\sqrt{\mathbf{D}}\mathbf{Q}_{t}\in \mathbb{C}^{N_T\times r}$, $\mathbf{Q}_{t}$ is the $N_T\times N_T$ discrete Fourier transform (DFT) matrix without the last $(N_T-r)$ columns and $\mathbf{D} = diag\{d_{1}\mathbf{1}_{N_{1}}^{T},\dots,d_{M}\mathbf{1}_{N_{M}}^{T}\}\in \mathbb{R}^{N_{T}\times N_{T}}$ is a positive definite matrix for power control. This initialization is based on the pilots proposed in \cite{Hua2019a}.
 %

\begin{algorithm}[h]
	\caption{Solving \eqref{opt_MSE_lb} with increasing $t$.}
	\label{Algorithm_g1}
	\small
	\begin{algorithmic}[1]
		\renewcommand{\algorithmicrequire}{\textbf{Input:}}
		\REQUIRE ~\\
		$r,\bar{\mathbf{R}},N_{i},\sigma_i,P_{i}$, $T$, for $i=1,\dots,M$;\\
		Accuracy thresholds: $\epsilon_{1}$, $\epsilon_{2}$, $N_{p}$.\\
		Initialization: $t>0$, $\mu>1$, and $\bar{\mathbf{F}}^{(0)} =\sqrt{\mathbf{D}}\mathbf{Q}_{t}$.\\
		\REPEAT
		\STATE  p=0;
		\REPEAT
		\STATE{Compute the derivatives $\frac{\partial g_{1}(\bar{\mathbf{F}}^{(p)})}{\partial \bar{\mathbf{F}}^{(p)}}$}.
		\STATE{Choose step size $\gamma^{(p)}$ via backtracking line search \cite{Boyd2004a}}.
		\STATE{Update $\bar{\mathbf{F}}^{(p+1)}=\bar{\mathbf{F}}^{(p)}-\gamma^{(p)} \nabla g_{1}(\bar{\mathbf{F}}^{(p)})$}.
		\STATE  p = p+1.
		\UNTIL{$\|\nabla g_{1}(\bar{\mathbf{F}}^{(p)}) - \nabla g_{1}(\bar{\mathbf{F}}^{(p-1)})\| \leq \epsilon_{2}$ or $p\geq N_{p}$}
		\STATE{$\bar{\mathbf{F}}^{(0)}=\bar{\mathbf{F}}^{(p)}$, $t = \mu t.$}
		\UNTIL{ $\frac{M}{t} < \epsilon_{1}$}
		\RETURN $\bar{\mathbf{F}}^{(p)}$
	\end{algorithmic}
\end{algorithm}

	\begin{remark}
\label{remark_rank}
If there is a strong channel correlation (i.e., one of $\mathbf{R}_i$ has a high condition number) and $P_T$ is not sufficiently large, Algorithm 1 may converge to a solution where $rank(\bar{\mathbf{P}}_{(i)}) < N_T-N_{i}$ for some $i$ such situation also happens in solving \eqref{opt_MSE_2} and \eqref{opt_SKR_lb} with the proposed methods). This is an undesirable situation which should and can be avoided by either increasing $P_T$ or reducing the ``active'' number $N_i$ of antennas at user $i$. The latter choice would reduce the condition number of $\mathbf{R}_i$.
\end{remark}

\begin{remark}
The problem in \eqref{opt_MSE_MulU} is meaningful as long as the channel conditions for all users are comparable. The result from \eqref{opt_MSE_MulU} is perfectly fair for two users since \eqref{opt_MSE_MulU} with $M=2$ is equivalent to two separate problems for individual users (as shown in next section). But to achieve a better fairness in all situations for three or more users, one may consider the following problem:
\begin{align}
&\min_{\varepsilon,\bar{\mathbf{F}}}~~ \varepsilon, \label{opt_MSE_fair2}\\
s.t.
~&Tr\left (\left [\mathbf{I} + \frac{1}{ \sigma^{2}_{i}}( \tilde{\boldsymbol{\Lambda}}_{i} \otimes \bar{\mathbf{S}}_{(i)}\bar{\mathbf{F}}\bar{\mathbf{F}}^{H}\bar{\mathbf{S}}_{(i)}^{T} )\right ]^{-1}\right ) \leq \varepsilon,\notag\\
&Tr(\mathbf{S}_{i}\bar{\mathbf{R}}^{-\frac{H}{2}}\bar{\mathbf{F}}\bar{\mathbf{F}}^{H}
\bar{\mathbf{R}}^{-\frac{1}{2}}
\mathbf{S}_{i}^{T}) \leq KP_{i},~i = 1,\dots,M.\notag
\end{align}
The constraints in \eqref{opt_MSE_fair2} are non-convex. To solve \eqref{opt_MSE_fair2}, we can define the following logarithm barrier function
\begin{equation}\label{g_f}
g_{1,F}(\varepsilon,\bar{\mathbf{F}}) = t\varepsilon + \sum_{i=1}^{M}\mathcal{B}_{P,i}(\bar{\mathbf{F}}) + \sum_{i=1}^{M}\mathcal{B}_{MSE,i}(\varepsilon,\bar{\mathbf{F}})
\end{equation}
where
\begin{equation}
\mathcal{B}_{MSE,i}(\bar{\mathbf{F}}) = -\ln(\psi_{MSE,i}(\varepsilon,\bar{\mathbf{F}}))
\end{equation}
and $\psi_{MSE,i}(\varepsilon,\bar{\mathbf{F}}) = \varepsilon - Tr\left (\left [\mathbf{I} + \frac{1}{ \sigma^{2}_{i}}( \tilde{\boldsymbol{\Lambda}}_{i} \otimes \bar{\mathbf{S}}_{(i)}\bar{\mathbf{F}}\bar{\mathbf{F}}^{H}\bar{\mathbf{S}}_{(i)}^{T} )\right ]^{-1}\right )$. Then \eqref{opt_MSE_fair2} can be approximated by
\begin{equation}\label{opt_gF}
\min_{\varepsilon,\bar{\mathbf{F}}} g_{1,F}(\varepsilon,\bar{\mathbf{F}}).
\end{equation}
To solve \eqref{opt_gF}, the gradient descent method can be used and all required derivatives can be easily derived based on
\eqref{d1} and \eqref{de_b_u}. However, the gradient search of \eqref{opt_gF} is sensitive to the choices of initial points. In the simulation, we choose $\bar{\mathbf{F}}^{(0)} = \sqrt{\mathbf{D}}\bar{\mathbf{Q}}_{m}$  where $\bar{\mathbf{Q}}_{m}$ is given by Theorem \ref{t1}. We also choose $\varepsilon^{(0)} = max_{i}\{\texttt{MSE}_i^{(0)}\}$ where $\texttt{MSE}_i^{(0)}$ is the corresponding MSE from  $\bar{\mathbf{F}}^{(0)}$. The algorithm to solve \eqref{opt_gF} is similar to Algorithm \ref{Algorithm_g1} and the details of the algorithm are omitted due to space limitation.
\end{remark}

\subsection{Special algorithm for $M=2$}\label{sec:MMSE_2}
When $M=2$, we can develop an efficient algorithm with guaranteed global optimality. This algorithm has a simple connection with that in \cite{Bjornson2010} as shown next.

Denote the two users by the indices $i = 1$ and $i = 2$. Now the cost function is $J_2$ given by \eqref{MSE_mul_2} with $M=2$.
Notice that $\bar{\mathbf{S}}_{(1)}\bar{\mathbf{F}} = \mathbf{S}_{2}\bar{\mathbf{F}} \in \mathbb{C}^{N_{2}\times r}$ and $\bar{\mathbf{S}}_{(2)}\bar{\mathbf{F}} = \mathbf{S}_{1}\bar{\mathbf{F}} \in \mathbb{C}^{N_{1}\times r}$, which do not have any shared entry.
Let us now use the following singular value decompositions (SVDs) to reparameterize $\bar {\mathbf{F}}$:
\begin{equation}\label{svd1}
\left \{
\begin{aligned}
&\bar{\mathbf{S}}_{(2)}\bar{\mathbf{F}} = \mathbf{U}_{1}\boldsymbol{\Lambda}_{1}\mathbf{V}_{1}^{H},\\
&\bar{\mathbf{S}}_{(1)}\bar{\mathbf{F}} = \mathbf{U}_{2}\boldsymbol{\Lambda}_{2}\mathbf{V}_{2}^{H}		
\end{aligned}
\right .
\end{equation}
where $\mathbf{U}_{1} \in \mathbb{C}^{N_{1}\times N_{1}}$, $\boldsymbol{\Lambda}_{1}\in \mathbb{R}^{N_{1} \times r}$, $\mathbf{V}_{1} \in \mathbb{C}^{r \times r}$, $\mathbf{U}_{2} \in \mathbb{C}^{N_{2}\times N_{2}}$, $\boldsymbol{\Lambda}_{2}\in \mathbb{R}^{N_{2} \times r}$ and $\mathbf{V}_{2} \in \mathbb{C}^{r \times r}$. All of these matrices need to be optimized as they all affect the pilots. With $r \geq \max\{N_1, N_2\}$, we denote the singular value matrices in \eqref{svd1} as $\boldsymbol{\Lambda}_{1} = [diag\{\lambda_{1,1},\dots,\lambda_{1,N_{1}}\},\mathbf{0}_{N_{1}\times(r - N_{1})}]$ and $\boldsymbol{\Lambda}_{2} = [diag\{\lambda_{2,1},\dots,\lambda_{2,N_{2}}\},\mathbf{0}_{N_{2}\times(r - N_{2})}]$ where the diagonal elements in each matrix are in descending order. Using \eqref{decom_inv} and \eqref{svd1}, we have
\begin{equation}\label{svdr}
\begin{aligned}
& \bar{\mathbf{P}} = \bar{\mathbf{R}}^{-\frac{T}{2}}[(\mathbf{U}_{1}\boldsymbol{\Lambda}_{1}\mathbf{V}_{1}^{H})^{T}, (\mathbf{U}_{2}\boldsymbol{\Lambda}_{2}\mathbf{V}_{2}^{H})^{T}]^{H}\bar{\mathbf{V}}^{*}.
\end{aligned}
\end{equation}
Let  $\boldsymbol{\Lambda}_{1}^{2} = diag\{\lambda_{1,1}^{2},\dots,\lambda_{1,N_{1}}^{2}\}$ and  $\boldsymbol{\Lambda}_{2}^{2} = diag\{\lambda_{2,1}^{2},\dots,\lambda_{2,N_{2}}^{2}\}$.
Also let $\mathbf{C}_{1} = \tilde{\boldsymbol{\Lambda}}_{1}^{-1}\boldsymbol{\Lambda}_{1}^{2}$ and $\mathbf{C}_{2} = \tilde{\boldsymbol{\Lambda}}_{2}^{-1}\boldsymbol{\Lambda}_{2}^{2}$. Then one can verify that $J_2$ becomes
\begin{align}
J_{2} &=  Tr((\mathbf{I} + \frac{1}{ \sigma^{2}_{1}}( \tilde{\boldsymbol{\Lambda}}_{1}\otimes \mathbf{C}_{2}\tilde{\boldsymbol{\Lambda}}_{2})^{-1})\notag\\
&~  +Tr((\mathbf{I} + \frac{1}{ \sigma^{2}_{2}}( \tilde{\boldsymbol{\Lambda}}_{2}\otimes \mathbf{C}_{1}\tilde{\boldsymbol{\Lambda}}_{1}))^{-1})\label{MSE_2_svd}
\end{align}
which is invariant to  $\mathbf{U}_{1}$, $\mathbf{V}_{1}$, $\mathbf{U}_{2}$ and $\mathbf{V}_{2}$. Only  $\mathbf{C}_{1}$ and $\mathbf{C}_{2}$ remain to be optimized as far as the cost function is concerned.

For the power constraints in \eqref{opt_MSE_MulU}, we see that for $i=1,2$,
\begin{equation}\label{po1}
{\small
\begin{aligned}
Tr(\mathbf{P}_{i}\mathbf{P}_{i}^{H}) &= Tr(\tilde{\boldsymbol{\Lambda}}_{i}^{-1}\mathbf{U}_{i}\boldsymbol{\Lambda}_{i}^{2}\mathbf{U}_{i}^{H}) \geq Tr(\tilde{\boldsymbol{\Lambda}}_{i}^{-1}\boldsymbol{\Lambda}_{i}^{2}) = Tr(\mathbf{C}_{i})
\end{aligned}}
\end{equation}
where the equality in ``$\geq$'' holds when $\mathbf{U}_{i} = \mathbf{I}_{N_{i}}$ \cite[H.1.h]{Marshalla}.

Therefore, both the cost and the power constraints are optimized by choosing $\mathbf{U}_{i}$ and $\mathbf{V}_{i}$ with $i=1,2$ to be the identity matrices. So, \eqref{opt_MSE_MulU} becomes
\begin{align}
&\min_{\mathbf{C}_{1},\mathbf{C}_{2}}~J_{2}\label{opt_MSE_2}\\
&s.t. ~Tr(\mathbf{C}_{1}) \leq KP_{1},~ Tr(\mathbf{C}_{2}) \leq KP_{2}\notag
\end{align}
where $J_2$ is shown in \eqref{MSE_2_svd}
Here  $\mathbf{C}_{1}$ and $\mathbf{C}_{2}$ are completely decoupled from each other. Each of the two decoupled problems can be solved by following \cite{Bjornson2010,Zhu2019}.
It is obvious that if $\tilde{\boldsymbol{\Lambda}}_{i}$ is proportional to the identity matrix, so is the optimal $\mathbf{C}_i$ with $i=1,2$.
%

\subsection{Closed-form solution}
For $M\geq 2$, we now consider the (previously mentioned) symmetric and isotropic case, i.e.,
$N_{i} = N$, $P_{i} = P$, $\sigma_i^2=\sigma^2$ and  $\mathbf{R}_{i} =  \mathbf{I}_{N}$. Furthermore, we consider $r = (M-1)N$ which yields the maximal dimensional of the subspace of Eve's CSI that is not identifiable by Eve. Then from \eqref{MSE_mul_2},  $J_M= N\sum_{i=1}^{M}Tr\big((\mathbf{I} + \frac{1}{\sigma^{2}}  \bar{\mathbf{S}}_{(i)}\bar{\mathbf{F}}\bar{\mathbf{F}}^{H}\bar{\mathbf{S}}_{(i)}^{T} )^{-1}\big)$. Also the power constraints become $Tr(\mathbf{S}_{i}\bar{\mathbf{F}}\bar{\mathbf{F}}^{H}\mathbf{S}_{i}^{T}) \leq KP,~i = 1,\dots,M$. The corresponding Lagrangian function is
\begin{equation}\label{L_MCEE_spCase}
\begin{aligned}
\mathcal{L} &= J_{M} + \sum_{i=1}^{M}\mu_{i}( Tr(\mathbf{S}_{i}\bar{\mathbf{F}}\bar{\mathbf{F}}^{H}\mathbf{S}_{i}^{T}) -KP)
\end{aligned}
\end{equation}
and the KKT conditions \cite{Boyd2004a} are
\begin{equation}\label{kkt_mul_sc}
{
	\left\{
	\begin{aligned}
	&\frac{\partial \mathcal{L}}{\partial \bar{\mathbf{F}}} = \frac{\partial J_{M}}{\partial \bar{\mathbf{F}}} + 2\sum_{i=1}^{M}\mu_{i}\mathbf{S}_{i}^{T}\mathbf{S}_{i}\bar{\mathbf{F}} = 0,\\
	&Tr(\mathbf{S}_{i}\bar{\mathbf{F}}\bar{\mathbf{F}}^{H}\mathbf{S}_{i}^{T}) \leq KP,~ i = 1,\dots,M,\\
	&\mu_{i}( Tr(\mathbf{S}_{i}\bar{\mathbf{F}}\bar{\mathbf{F}}^{H}\mathbf{S}_{i}^{T}) - KP) = 0,~ \mu_{i} \geq 0,~ i = 1,\dots,M.
	\end{aligned}\right.}
\end{equation}
It is shown below that a set of (equally optimal) solutions to \eqref{kkt_mul_sc} are given by the $NM\times NM$ discrete Fourier transform (DFT) matrix $\mathbf{Q}$ with any $N$ equally spaced columns removed.
\begin{theorem}\label{t1}
Let $\mathbf{Q}$ be such that its $(l+1,k+1)$th element is  $(\mathbf{Q})_{l+1,k+1} =  w_{NM}^{lk}$ with $w_{NM}=e^{-j2\pi\frac{1}{MN}}$, $0\leq l\leq NM-1$ and $0\leq k \leq NM-1$. Let $\mathbf{Q}_{m}$ consist of $N$ equally spaced columns of $\mathbf{Q}$ as follows:
	\begin{equation}\label{Qc}
	\begin{aligned}
	&\mathbf{Q}_{m} = \\
	&{\small\begin{bmatrix}
	1&1&\cdots&1\\
	w_{MN}^{m} & w_{MN}^{m+M} &\cdots&w_{MN}^{m+(N-1)M}\\
	\vdots&&&\vdots\\
	w_{MN}^{m(NM-1)} & w_{MN}^{(m+M)(NM-1)} &\cdots&w_{MN}^{(m+(N-1)M)(NM-1)}
	\end{bmatrix}}
	\end{aligned}.
	\end{equation}
Also let $\mathbf{\bar Q}_m$ be $\mathbf{Q}$ without the columns in $\mathbf{Q}_{m}$. Then, a solution to \eqref{kkt_mul_sc} is
$\bar{\mathbf{F}} = \sqrt{\frac{KP}{ N^2(M-1)}}\mathbf{\bar Q}_m$ where $m$ can be any integer in $[0,M-1]$.
\end{theorem}
\begin{IEEEproof}
	See Appendix \ref{app_mse}
\end{IEEEproof}
For $M=2$, the theorem yields $\mathbf{P}_i=\mathbf{S}_i\bar{\mathbf{F}}^*\bar{\mathbf{V}}^*$  that satisfies $\mathbf{P}_i\mathbf{P}_i^H=\frac{KP}{N}\mathbf{I}_N$ where $i=1,2$ (easy to verify). These pilots are known to be globally optimal. For $M\geq 3$, our numerical simulations using the previously developed algorithm did not yield any result better than that from Theorem \ref{t1} subject to the conditions in the theorem.

\subsubsection{For optimal ML channel estimation}
The ML estimate of $\bar{\mathbf{h}}_i$ is $\hat{\bar{\mathbf{h}}}_{i,ML}=(\bar{\mathbf{G}}_i\bar{\mathbf{G}}_i^H)^{-1}\bar{\mathbf{G}}_i\mathbf{y}_i$ and its covariance matrix is $\mathbf{C}_{i,ML}=\sigma_i^2(\bar{\mathbf{G}}_i\bar{\mathbf{G}}_i^H)^{-1}=\sigma_i^2(\bar{\mathbf{S}}_{(i)}\bar{\mathbf{F}}
\bar{\mathbf{F}}^{H}\bar{\mathbf{S}}_{(i)}^{T}\otimes \mathbf{R}_i^{\frac{H}{2}}\mathbf{R}_i^{\frac{1}{2}})^{-1}$. We can design the optimal pilots by minimizing $J_{M,ML}=\sum_{i=1}^M Tr(\mathbf{C}_{i,ML})$ subject to the same power constraints as before.

If $N_{i} = N$, $P_{i} = P$, $\sigma_i^2=\sigma^2$,  $\mathbf{R}_{i} =  \mathbf{I}_{N}$ and  $r = (M-1)N$,  one can verify that $J_{M,ML}$ equals $J_M$ as $\sigma^2$ becomes small or equivalently $KP$ becomes large. Hence, the optimal pilots from Theorem \ref{t1} also apply here (which can also be proved directly by following a similar procedure used for Theorem \ref{t1}).
%

\section{Pilot Designs Based on MI}\label{sec:MI}
 Given $\mathbf{Y}_i$ at user $i$ for all $i$ as shown in \eqref{mul_sig_user_2}, every pair of users can follow a secret key generation protocol \cite{Maurer1993,bloch2011physical,Lai2012,khisti2016} to produce a (shared) secret key. This secret key can be a useful by-product of ANECE which was originally designed to protect the information directly transmitted between users \cite{Hua2019a}. If $\mathbf{Y}_E$ received by Eve as shown in \eqref{mul_sig_eve_2} or equivalently the Eve's channel matrix $\bar{\mathbf{H}}_E$ is independent of all channel matrices between users, the capacity of the secret key (in bits per channel coherence period) achievable between user $i$ and user $j$ is known \cite[Th. 4.1]{bloch2011physical} to be $I(\mathbf{Y}_i;\mathbf{Y}_j)$ which is the mutual information between $\mathbf{Y}_i$ and $\mathbf{Y}_j$. So, it is also meaningful to design the optimal pilots as follows:
\begin{align}
&\max_{\bar{\mathbf{P}}}~~ I_{M}=\sum_{i=1}^{M-1}\sum_{j=i+1}^{M}I(\mathbf{Y}_{i};\mathbf{Y}_{j})\label{opt_SKR}\\
s.t. ~&Tr(\mathbf{P}_{i}\mathbf{P}_{i}^{H}) \leq KP_{i},~i = 1,\dots,M, \notag\\
&rank(\bar{\mathbf{P}}) = r,~ \notag 
\end{align}
with $N_T-N_{min}\leq r\leq N_T-1$. Like \eqref{opt_MSE}, the above problem is also non-convex. We will treat it next in three separate situations as before.

\subsection{General algorithm for $M\geq 2$}\label{sec::MI_m3}
From \eqref{mul_sig_user}, we can write
\begin{equation}\label{mul_ij_vec}
\left \{
	\begin{aligned}
	&\mathbf{y}_{i} = \sum_{j\neq i}^{M}(\bar{\mathbf{P}}^{T}\bar{\mathbf{R}}^{\frac{1}{2}}\mathbf{S}_{j}^{T}\otimes
\mathbf{R}_{i}^{\frac{1}{2}})\mathbf{h}_{i,j} + \mathbf{n}_{i},\\
	&\mathbf{y}_{T,j} = \sum_{i\neq j}^{M}(\mathbf{R}_{j}^{\frac{1}{2}}\otimes \bar{\mathbf{P}}^{T}\bar{\mathbf{R}}^{\frac{1}{2}}\mathbf{S}_{i}^{T})\mathbf{h}_{i,j} + \mathbf{n}_{T,j}
	\end{aligned}\right .
\end{equation}
where $\mathbf{y}_{i} = vec(\mathbf{Y}_{i})$, $\mathbf{y}_{T,j} = vec(\mathbf{Y}_{j}^{T})$, $\mathbf{H}_{i,j} = \mathbf{H}_{j,i}^{T}$, $\mathbf{h}_{i,j} = vec(\mathbf{H}_{i,j})$, $\mathbf{n}_{i} = vec(\mathbf{N}_{i})$ and $\mathbf{n}_{T,j} = vec(\mathbf{N}_{j}^{T})$. Clearly we have  $I(\mathbf{Y}_{i};\mathbf{Y}_{j}) = I(\mathbf{y}_{i};\mathbf{y}_{T,j})$.

Recall $\bar{\mathbf{G}}_{i} = (\bar{\mathbf{S}}_{(i)}\bar{\mathbf{R}}^{\frac{H}{2}}\bar{\mathbf{P}}^{*}\otimes \mathbf{R}_{i}^{\frac{H}{2}})$. Also define $\bar{\mathbf{G}}_{T,j} = ( \mathbf{R}_{j}^{\frac{H}{2}} \otimes \bar{\mathbf{S}}_{(j)}\bar{\mathbf{R}}^{\frac{H}{2}}\bar{\mathbf{P}}^{*})$, $\mathbf{G}_{i,j} = (\mathbf{S}_{j}\bar{\mathbf{R}}^{\frac{H}{2}}\bar{\mathbf{P}}^{*}\otimes \mathbf{R}_{i}^{\frac{H}{2}})$ and $\mathbf{G}_{T,j,i} = ( \mathbf{R}_{j}^{\frac{H}{2}} \otimes \mathbf{S}_{i}\bar{\mathbf{R}}^{\frac{H}{2}}\bar{\mathbf{P}}^{*})$.
From \eqref{mul_ij_vec} , one can verify that
\begin{equation}\label{}
  \mathbf{K}_{\mathbf{y}_{i}} = \sigma^{2}_{i}\mathbf{I} + \bar{\mathbf{G}}_{i}^{H}\bar{\mathbf{G}}_{i},
\end{equation}
\begin{equation}\label{}
  \mathbf{K}_{\mathbf{y}_{T,j}} = \sigma^{2}_{j}\mathbf{I} + \bar{\mathbf{G}}_{T,j}^{H}\bar{\mathbf{G}}_{T,j},
\end{equation}
\begin{equation}\label{}
  \mathbf{K}_{\mathbf{y}_{i},\mathbf{y}_{T,j}}  = \mathbf{G}_{i,j}^{H}\mathbf{G}_{T,j,i},
\end{equation}
\begin{equation}\label{}
  \mathbf{K}_{\mathbf{y}_{T,j},\mathbf{y}_{i}} = \mathbf{G}_{T,j,i}^{H}\mathbf{G}_{i,j}.
\end{equation}
Also note
\begin{subequations}
	\begin{align}
	I(\mathbf{y}_{i};\mathbf{y}_{T,j})& = h(\mathbf{y}_{i}) + h(\mathbf{y}_{T,j}) - h(\mathbf{y}_{i},\mathbf{y}_{T,j})\notag\\
	& = \log_{2}|\mathbf{K}_{\mathbf{y}_{i}}| + \log_{2}|\mathbf{K}_{\mathbf{y}_{T,j}}| - \log_{2}|\mathbf{K}_{\{\mathbf{y}_{i},\mathbf{y}_{T,j}\}}|\notag\\
	& = -\log_{2}|\mathbf{I} - \mathbf{K}_{\mathbf{y}_{T,j}}^{-1}\mathbf{K}_{\mathbf{y}_{T,j}, \mathbf{y}_{i}}	\mathbf{K}_{\mathbf{y}_{i}}^{-1}\mathbf{K}_{\mathbf{y}_{i}, \mathbf{y}_{T,j}}|\label{Mu_ij}
\\
	& = -\log_{2}|\mathbf{I} -(\sigma^{2}_{j}\mathbf{I} + \bar{\mathbf{G}}_{T,j}^{H}\bar{\mathbf{G}}_{T,j})^{-1} \mathbf{G}_{T,j,i}^{H}\mathbf{G}_{i,j}\notag
\\
	&\quad \cdot (\sigma^{2}_{i}\mathbf{I} + \bar{\mathbf{G}}_{i}^{H}\bar{\mathbf{G}}_{i})^{-1}\mathbf{G}_{i,j}^{H}\mathbf{G}_{T,j,i}|
\label{Mu_ij.2}
	\end{align}
\end{subequations}
where
\begin{equation}
\mathbf{K}_{\{\mathbf{y}_{i},\mathbf{y}_{T,j}\}} =
\begin{bmatrix}
\mathbf{K}_{\mathbf{y}_{i}} & \mathbf{K}_{\mathbf{y}_{i}, \mathbf{y}_{T,j}}\\
\mathbf{K}_{\mathbf{y}_{T,j}, \mathbf{y}_{i}} & \mathbf{K}_{\mathbf{y}_{T,j}}
\end{bmatrix}
\end{equation}
and the last equality in \eqref{Mu_ij} is based on the fact that $\left |\begin{bmatrix}
\mathbf{X}&\mathbf{Y}\\\mathbf{Y}^{H}&\mathbf{Z}
\end{bmatrix}\right | = |\mathbf{X}||\mathbf{Z} - \mathbf{Y}^{H}\mathbf{X}^{-1}\mathbf{Y}| = |\mathbf{Z}||\mathbf{X} - \mathbf{Y}\mathbf{Z}^{-1}\mathbf{Y}^{H}|$ with invertible $\mathbf{X}$ and $\mathbf{Z}$.

From \eqref{mul_ij_vec}, we can express the MMSE estimates of $\mathbf{h}_{i,j}$ by users $i$ and $j$, respectively, as
\begin{equation}\label{MMSE_Hij}
\left \{
\begin{aligned}
&\hat{\mathbf{h}}_{ij,i} = \mathbf{K}_{\mathbf{h}_{i,j},\mathbf{y}_{i}}\mathbf{K}_{\mathbf{y}_{i}}^{-1}\mathbf{y}_{i} = \mathbf{G}_{i,j}(\sigma^{2}_{i}\mathbf{I} + \bar{\mathbf{G}}_{i}^{H}\bar{\mathbf{G}}_{i})^{-1}\mathbf{y}_{i},\\
&\hat{\mathbf{h}}_{ij,j} = \mathbf{K}_{\mathbf{h}_{i,j},\mathbf{y}_{T,j}}\mathbf{K}_{\mathbf{y}_{T,j}}^{-1}\mathbf{y}_{T,j} \\
&\quad\quad = \mathbf{G}_{T,j,i}(\sigma^{2}_{j}\mathbf{I} + \bar{\mathbf{G}}_{T,j}^{H}\bar{\mathbf{G}}_{T,j})^{-1}\mathbf{y}_{T,j}.
\end{aligned}
\right .
\end{equation}
The following lemma is a generalization of a SISO result shown in \cite{Chou2012}. It also complements the fact that $I(\mathbf{y}_{i};\mathbf{y}_{T,j})$ equals to the mutual information between the ML estimates of $\mathbf{h}_{i,j}$ by users $i$ and $j$ \cite{Lai2012}.
\begin{lemma}\label{lemma1}
For each pair of $i$ and $j$, if $\mathbf{S}_{j}\bar{\mathbf{R}}^{\frac{H}{2}}\bar{\mathbf{P}}^{*}$, $\mathbf{S}_{i}\bar{\mathbf{R}}^{\frac{H}{2}}\bar{\mathbf{P}}^{*}$, $\mathbf{R}_{i}$, $\mathbf{R}_{j}$ have all full row ranks (which requires $K \geq \max\{N_{i}, N_{j}\}$), then  we have $I(\mathbf{y}_{i};\mathbf{y}_{T,j}) = I(\hat{\mathbf{h}}_{ij,i};\hat{\mathbf{h}}_{ij,j})$.
\end{lemma}
\begin{IEEEproof}
	With the stated conditions, we have  $\mathbf{K}_{\hat{\mathbf{h}}_{ij,i}} = \mathbf{G}_{i,j}(\sigma^{2}_{i}\mathbf{I} + \bar{\mathbf{G}}_{i}^{H}\bar{\mathbf{G}}_{i})^{-1}\mathbf{G}_{i,j}^{H}$, $\mathbf{K}_{\hat{\mathbf{h}}_{ij,j}} = \mathbf{G}_{T,j,i}(\sigma^{2}_{j}\mathbf{I} + \bar{\mathbf{G}}_{T,j}^{H}\bar{\mathbf{G}}_{T,j})^{-1}\mathbf{G}_{T,j,i}^{H}$, and
	$\mathbf{K}_{\hat{\mathbf{h}}_{ij,i}, \hat{\mathbf{h}}_{ij,j}} =\mathbf{G}_{i,j}(\sigma^{2}_{i}\mathbf{I} + \bar{\mathbf{G}}_{i}^{H}\bar{\mathbf{G}}_{i})^{-1}\mathbf{G}_{i,j}^{H}\mathbf{G}_{T,j,i}
(\sigma^{2}_{j}\mathbf{I} + \bar{\mathbf{G}}_{T,j}^{H}\bar{\mathbf{G}}_{T,j})^{-1}\mathbf{G}_{T,j,i}^{H}$. Also, $ \mathbf{K}_{\hat{\mathbf{h}}_{ij,i}, \hat{\mathbf{h}}_{ij,j}}=  \mathbf{K}_{\hat{\mathbf{h}}_{ij,i}}\mathbf{K}_{\hat{\mathbf{h}}_{ij,j}}$. Then,
	\begin{align}
	&I(\hat{\mathbf{h}}_{ij,i};\hat{\mathbf{h}}_{ij,j})\notag\\
	& = -\log_{2}|\mathbf{I} - \mathbf{K}_{\hat{\mathbf{h}}_{ij,j}}^{-1}\mathbf{K}_{\hat{\mathbf{h}}_{ij,j}, \hat{\mathbf{h}}_{ij,i}}	\mathbf{K}_{\hat{\mathbf{h}}_{ij,i}}^{-1}\mathbf{K}_{\hat{\mathbf{h}}_{ij,i}, \hat{\mathbf{h}}_{ij,j}}|\notag\\
	& = -\log_{2}|\mathbf{I} - \mathbf{K}_{\hat{\mathbf{h}}_{ij,i}}\mathbf{K}_{\hat{\mathbf{h}}_{ij,j}}|\notag\\
& = -\log_2|\mathbf{I}-\mathbf{G}_{i,j}(\sigma^{2}_{i}\mathbf{I} + \bar{\mathbf{G}}_{i}^{H}\bar{\mathbf{G}}_{i})^{-1}\mathbf{G}_{i,j}^{H}\mathbf{G}_{T,j,i}\notag\\
& \cdot (\sigma^{2}_{j}\mathbf{I} + \bar{\mathbf{G}}_{T,j}^{H}\bar{\mathbf{G}}_{T,j})^{-1}\mathbf{G}_{T,j,i}^{H}|=
I(\mathbf{y}_i;\mathbf{y}_{T,j})
	\end{align}
where the last equation follows from \eqref{Mu_ij.2} using $\log_2|\mathbf{I}-\mathbf{X}\mathbf{Y}|=\log_2|\mathbf{I}-\mathbf{Y}\mathbf{X}|$.
\end{IEEEproof}

Define $\boldsymbol{\Gamma}_{i,j} = \mathbf{G}_{i,j}(\sigma^{2}_{i}\mathbf{I} + \bar{\mathbf{G}}_{i}^{H}\bar{\mathbf{G}}_{i})^{-1}\mathbf{G}_{i,j}^{H}$ and $\boldsymbol{\Gamma}_{T,j,i} = \mathbf{G}_{T,j,i}(\sigma^{2}_{j}\mathbf{I} + \bar{\mathbf{G}}_{T,j}^{H}\bar{\mathbf{G}}_{T,j})^{-1}\mathbf{G}_{T,j,i}^{H}$. Also using \eqref{decom} and \eqref{svd_bar}, one can verify that
\begin{align}
\boldsymbol{\Gamma}_{i,j}
 =& (\mathbf{S}_{j}\bar{\mathbf{F}}\otimes\tilde{\boldsymbol{\Lambda}}_{i}^{\frac{1}{2}})
(\sigma^{2}_{i}\mathbf{I} +  \bar{\mathbf{F}}^{H}\bar{\mathbf{S}}_{(i)}^{T}\bar{\mathbf{S}}_{(i)}\bar{\mathbf{F}}\otimes \tilde{\boldsymbol{\Lambda}}_{i})^{-1}\notag\\
&(\bar{\mathbf{F}}^{H}\mathbf{S}_{j}^{T}\otimes
\tilde{\boldsymbol{\Lambda}}_{i}^{\frac{1}{2}}),\label{gammaij.1}
\end{align}
\begin{align}
\boldsymbol{\Gamma}_{T,j,i}
 = &(\tilde{\boldsymbol{\Lambda}}_{j}^{\frac{1}{2}}\otimes \mathbf{S}_{i}\bar{\mathbf{F}})(\sigma^{2}_{j}\mathbf{I} +  \tilde{\boldsymbol{\Lambda}}_{j}\otimes\bar{\mathbf{F}}^{H}\bar{\mathbf{S}}_{(j)}^{T}
\bar{\mathbf{S}}_{(j)}\bar{\mathbf{F}})^{-1}\notag\\
&(\tilde{\boldsymbol{\Lambda}}_{j}^{\frac{1}{2}}
\otimes\bar{\mathbf{F}}^{H}\mathbf{S}_{i}^{T}).\label{gammaji.1}
\end{align}

The rank constraint on $\bar{\mathbf{P}}$ is satisfied by using $\bar{\mathbf{F}}$ defined in \eqref{decom}. With \eqref{gammaij.1} and \eqref{gammaji.1}, we have
\begin{equation}\label{IM_F}
I_{M} = -\sum_{i=1}^{M-1}\sum_{j = i+1}^{M} \log_{2}|\mathbf{I} - \boldsymbol{\Gamma}_{i,j}\boldsymbol{\Gamma}_{T,j,i}|
\end{equation}
and \eqref{opt_SKR} becomes
\begin{align}
&\max_{\bar{\mathbf{F}}}~~ I_{M}\label{opt_mskr_MulU}\\
s.t. ~&Tr(\mathbf{S}_{i}\bar{\mathbf{R}}^{-\frac{H}{2}}\bar{\mathbf{F}}\bar{\mathbf{F}}^{H}\bar{\mathbf{R}}^{-\frac{1}{2}}
	\mathbf{S}_{i}^{T})\leq KP_{i},~i = 1,\dots,M.\notag
\end{align}

To solve \eqref{opt_mskr_MulU} by using the logarithmic barrier method, we let
\begin{equation}\label{barrier_key}
g_{2}(\bar{\mathbf{F}}) = -tI_{M} + \sum_{i=1}^{M}\mathcal{B}_{P,i}(\bar{\mathbf{F}})
\end{equation}
where $t$ is the barrier coefficient and $\mathcal{B}_{P,i}(\bar{\mathbf{F}})$ is shown in \eqref{barrier}. Then we can solve \eqref{opt_mskr_MulU} by solving the following (with an increasing $t$):
\begin{equation}\label{opt_SKR_lb}
\begin{aligned}
\min_{\bar{\mathbf{F}}} &\quad  g_{2}(\bar{\mathbf{F}}).
\end{aligned}
\end{equation}
The algorithm to solve \eqref{opt_SKR_lb} is similar to Algorithm \ref{Algorithm_g1} and hence omitted here. The way to find the gradient of $g_{2}(\bar{\mathbf{F}})$ is shown in Appendix \ref{sec:ap:skr}.
\begin{remark} For $M=2$, the previous method is perfectly fair.
For a better fairness of MI for all pairs among three or more users, we can consider the following problem
\begin{align}
&\min_{\varepsilon,\bar{\mathbf{F}}}~~ \varepsilon,\label{opt_mskr_fair}\\
s.t. ~&Tr(\mathbf{S}_{i}\bar{\mathbf{R}}^{-\frac{H}{2}}\bar{\mathbf{F}}\bar{\mathbf{F}}^{H}\bar{\mathbf{R}}^{-\frac{1}{2}}
\mathbf{S}_{i}^{T}) \leq KP_{i},~i= 1,\dots,M,\notag\\
&  \log_{2}|\mathbf{I} - \boldsymbol{\Gamma}_{i,j}\boldsymbol{\Gamma}_{T,j,i}| \leq \varepsilon, \forall \{i,j\}\notag
\end{align}
where $-\log_{2}|\mathbf{I} - \boldsymbol{\Gamma}_{i,j}\boldsymbol{\Gamma}_{T,j,i}|$ is the mutual information for the user pair $\{i,j\}$.

The constraints in \eqref{opt_mskr_fair} are non-convex. To solve this problem using the logarithm barrier method, we define
\begin{equation}
g_{2,F}(\varepsilon,\bar{\mathbf{F}}) = t\varepsilon + \sum_{i=1}^{M}\mathcal{B}_{P,i}(\bar{\mathbf{F}}) + \sum_{i=1}^{M-1}\sum_{j = i+1}^{M}\mathcal{B}_{MI,i}(\varepsilon,\bar{\mathbf{F}})
\end{equation}
where $\mathcal{B}_{MI,i}(\varepsilon,\bar{\mathbf{F}}) = -\ln(\varepsilon - \log_{2}|\mathbf{I} - \boldsymbol{\Gamma}_{i,j}\boldsymbol{\Gamma}_{T,j,i}|)$. Then \eqref{opt_mskr_fair} can be approximated by
\begin{equation}\label{g2f}
\min_{\varepsilon,\bar{\mathbf{F}}} ~g_{2,F}(\varepsilon,\bar{\mathbf{F}})
\end{equation}
which can be solved by gradient descent. This algorithm is similar to Algorithm \ref{Algorithm_g1}. But for initialization, we will use  $\bar{\mathbf{F}}^{(0)} = \sqrt{\mathbf{D}}\bar{\mathbf{Q}}_{m}$ and $\varepsilon^{(0)} = max_{\{i,j\}}\{\log_{2}|\mathbf{I} - \boldsymbol{\Gamma}_{i,j}^{(0)}\boldsymbol{\Gamma}_{T,j,i}^{(0)}|\}$. All required derivatives can be easily obtained using results in Appendix \ref{sec:ap:skr}. The details are omitted.

\end{remark}
\subsection{Special algorithm for $M=2$}\label{sec:skr_M2}
For $M = 2$, the problem is similar to one addressed in \cite{Quist2015} where an algorithm was developed and its local optimality is stated there. In this following, we effectively readdress the same problem but show some new insights. One of them is the establishment of optimality of two matrices heuristically chosen in \cite{Quist2015}. Furthermore, we will present an asymptotical analysis to show the globally optimal solution in high or low power region.

For $M = 2$, we know $\bar{\mathbf{S}}_{(1)} = \mathbf{S}_{2}$ and $\bar{\mathbf{S}}_{(2)} = \mathbf{S}_{1}$.
Using \eqref{svd1}, \eqref{gammaij.1} and \eqref{gammaji.1}, we have
	\begin{align}
	&\boldsymbol{\Gamma}_{1,2}\notag\\ &=(\mathbf{S}_{2}\bar{\mathbf{F}}\otimes\tilde{\boldsymbol{\Lambda}}_{1}^{\frac{1}{2}})(\sigma^{2}_{1}
\mathbf{I} +  \bar{\mathbf{F}}^{H}\bar{\mathbf{S}}_{(1)}^{T}\bar{\mathbf{S}}_{(1)}\bar{\mathbf{F}}\otimes \tilde{\boldsymbol{\Lambda}}_{1})^{-1}(\bar{\mathbf{F}}^{H}\mathbf{S}_{2}^{T}\otimes
\tilde{\boldsymbol{\Lambda}}_{1}^{\frac{1}{2}})\notag\\
	& = (\mathbf{U}_{2}\otimes \mathbf{I})( \boldsymbol{\Lambda}_{2} \otimes\tilde{\boldsymbol{\Lambda}}_{1}^{\frac{1}{2}})(\sigma_{1}^{2}\mathbf{I} + \boldsymbol{\Lambda}_{2}^{2}\otimes \tilde{\boldsymbol{\Lambda}}_{1} )^{-1}\notag\\
	&\qquad\cdot ( \boldsymbol{\Lambda}_{2}^{T}\otimes\tilde{\boldsymbol{\Lambda}}_{1}^{\frac{1}{2}})(\mathbf{U}_{2}^{H}\otimes \mathbf{I}),\label{psi2}
	\end{align}
	\begin{align}
	&\boldsymbol{\Gamma}_{T,2,1}\notag\\
	&=(\tilde{\boldsymbol{\Lambda}}_{2}^{\frac{1}{2}}\otimes \mathbf{S}_{1}\bar{\mathbf{F}})(\sigma^{2}_{2}\mathbf{I} +  \tilde{\boldsymbol{\Lambda}}_{2}\otimes\bar{\mathbf{F}}^{H}\bar{\mathbf{S}}_{(2)}^{T}\bar{\mathbf{S}}_{(2)}
\bar{\mathbf{F}})^{-1}(\tilde{\boldsymbol{\Lambda}}_{2}^{\frac{1}{2}}\otimes\bar{\mathbf{F}}^{H}
\mathbf{S}_{1}^{T})\notag\\
	& =(\mathbf{I}\otimes \mathbf{U}_{1})(\tilde{\boldsymbol{\Lambda}}_{2}^{\frac{1}{2}}\otimes \boldsymbol{\Lambda}_{1})(\sigma_{2}^{2}\mathbf{I} + \tilde{\boldsymbol{\Lambda}}_{2}\otimes \boldsymbol{\Lambda}_{1}^{T}\boldsymbol{\Lambda}_{1})^{-1}\notag\\
	&\qquad\cdot(\tilde{\boldsymbol{\Lambda}}_{2}^{\frac{1}{2}}\otimes \boldsymbol{\Lambda}_{1}^{T})(\mathbf{I}\otimes \mathbf{U}_{1}^{H}).\label{psi1}	
	\end{align}
It is obvious that both $I_2= I(\mathbf{y}_{1};\mathbf{y}_{T,2})
=-\log_2|\mathbf{I}-\boldsymbol{\Gamma}_{1,2}\boldsymbol{\Gamma}_{T,2,1}|$ and $Tr(\mathbf{P}_i\mathbf{P}_i^H)$ are invariant to $\mathbf{V}_i$ in \eqref{svd1} where $i=1,2$. We can set $\mathbf{V}_i=\mathbf{I}_r$.
Now  we reformulate \eqref{opt_mskr_MulU}  to
\begin{align}
&\max_{\mathbf{U}_{1},\mathbf{U}_{2},\boldsymbol{\Lambda}_{1},\boldsymbol{\Lambda}_{2}}~~I_{2}\label{opt_mu_2}\\
s.t.~  &Tr(\tilde{\boldsymbol{\Lambda}}_{1}^{-1}\mathbf{U}_{1}\boldsymbol{\Lambda}_{1}^{2}\mathbf{U}_{1}^{H})\leq KP_{1},~ Tr(\tilde{\boldsymbol{\Lambda}}_{2}^{-1}\mathbf{U}_{2}\boldsymbol{\Lambda}_{2}^{2}\mathbf{U}_{2}^{H}) \leq KP_{2},\notag\\
&\quad \boldsymbol{\Lambda}_{1} \succ \mathbf{0}, ~\boldsymbol{\Lambda}_{2} \succ \mathbf{0}.\notag
\end{align}
In \eqref{opt_mu_2}, we have introduced the positive definite constraints on $\boldsymbol{\Lambda}_{1}$ and $\boldsymbol{\Lambda}_{2}$. The reasons are: 1) the optimal $\mathbf{U}_{1}$ and $\mathbf{U}_{2}$ subject to those positive definite constraints are the identity matrices  (which is shown next); 2) those constraints barely change the solution from \eqref{opt_mskr_MulU} in terms of the objective function and the power constraints;  and 3) with those constraints each user is able to have consistent estimate of its channel.

With $\boldsymbol{\Lambda}_{1} \succ \mathbf{0}$ and $\boldsymbol{\Lambda}_{2} \succ \mathbf{0}$, \eqref{psi1} and \eqref{psi2} become $\boldsymbol{\Gamma}_{1,2} = (\mathbf{I} + \sigma^{2}_{1}(\mathbf{U}_{2}\boldsymbol{\Lambda}_{2}^{2}\mathbf{U}_{2}^{H} \otimes\tilde{\boldsymbol{\Lambda}}_{1} )^{-1})^{-1}$ and $\boldsymbol{\Gamma}_{T,2,1} = (\mathbf{I} + \sigma^{2}_{2}(\tilde{\boldsymbol{\Lambda}}_{2}\otimes\mathbf{U}_{1}\boldsymbol{\Lambda}_{1}^{2}
\mathbf{U}_{1}^{H} )^{-1})^{-1}$, and then the cost function in \eqref{opt_mu_2} becomes
\begin{subequations}\label{det}
		\begin{align}
		I_{2}
		& = \log_{2}\big|\mathbf{I} + \sigma^{2}_{2}(\tilde{\boldsymbol{\Lambda}}_{2} \otimes \tilde{\mathbf{U}}_{1}\boldsymbol{\Lambda}_{1}^{2}\tilde{\mathbf{U}}_{1}^{H})^{-1}\big|\notag\\
		&~+ \log_{2}\big|\mathbf{I}  + \sigma^{2}_{1}( \mathbf{U}_{2}\boldsymbol{\Lambda}_{2}^{2}\mathbf{U}_{2}^{H}\otimes \tilde{\boldsymbol{\Lambda}}_{1})^{-1}\big|\notag\\
		&~- \log_{2}\big|(\mathbf{I}  + \sigma^{2}_{2}(\tilde{\boldsymbol{\Lambda}}_{2} \otimes \tilde{\mathbf{U}}_{1}\boldsymbol{\Lambda}_{1}^{2}\tilde{\mathbf{U}}_{1}^{H})^{-1})\notag\\
		&\qquad\cdot(\mathbf{I}  + \sigma^{2}_{1}( \mathbf{U}_{2}\boldsymbol{\Lambda}_{2}^{2}\mathbf{U}_{2}^{H}\otimes \tilde{\boldsymbol{\Lambda}}_{1})^{-1})- \mathbf{I}\big|\label{det1}\\
		& = \log_{2}|\sigma^{2}_{2}\mathbf{I} + \tilde{\boldsymbol{\Lambda}}_{2} \otimes \boldsymbol{\Lambda}_{1}^{2}|+ \log_{2}|\sigma^{2}_{1}\mathbf{I} +  \boldsymbol{\Lambda}_{2}^{2} \otimes \tilde{\boldsymbol{\Lambda}}_{1}| \notag\\
		&{\fontsize{9.5}{9.5} \selectfont~-\log_{2}|\sigma^{2}_{1}\sigma^{2}_{2}\mathbf{I} + \sigma^{2}_{1}\tilde{\boldsymbol{\Lambda}}_{2} \otimes \tilde{\mathbf{U}}_{1}\boldsymbol{\Lambda}_{1}^{2}\tilde{\mathbf{U}}_{1}^{H}+ \sigma^{2}_{2}\mathbf{U}_{2}\boldsymbol{\Lambda}_{2}^{2}\mathbf{U}_{2}^{H}\otimes \tilde{\boldsymbol{\Lambda}}_{1}|}\label{det2}\\
		& = \log_{2}|\sigma^{2}_{2}\mathbf{I} + \tilde{\boldsymbol{\Lambda}}_{2} \otimes \boldsymbol{\Lambda}_{1}^{2}| + \log_{2}|\sigma^{2}_{1}\mathbf{I} +  \boldsymbol{\Lambda}_{2}^{2} \otimes \tilde{\boldsymbol{\Lambda}}_{1}|\notag\\
		&~-\log_{2}|\sigma^{2}_{1}\sigma^{2}_{2}\mathbf{I} + \sigma^{2}_{1}\tilde{\boldsymbol{\Lambda}}_{2} \otimes \boldsymbol{\Lambda}_{1}^{2} + \sigma^{2}_{2}\mathbf{U}(\boldsymbol{\Lambda}_{2}^{2}\otimes\tilde{\boldsymbol{\Lambda}}_{1})
\mathbf{U}^{H}|\label{det3}
		\end{align}
\end{subequations}
where $\mathbf{U} \triangleq \mathbf{U}_{2} \otimes\tilde{\mathbf{U}}_{1}^{H}$. Here, \eqref{det1} is due to $-\log_{2}|\mathbf{I} - \mathbf{A}^{-1}\mathbf{B}^{-1}| = \log_{2}|\mathbf{A}| + \log_{2}|\mathbf{B}| - \log_{2}|\mathbf{A}\mathbf{B} - \mathbf{I}|$,  and \eqref{det2} is due to $\log_{2}|\mathbf{I} + \mathbf{A}^{-1}| = \log_{2}|\mathbf{I} + \mathbf{A}| - \log_{2}|\mathbf{A}|$. Then the optimal  $\mathbf{U}_{1}$ and $\mathbf{U}_{2}$ that maximize \eqref{det} are given by
\begin{equation}\label{u}
\begin{aligned}
&\{\mathbf{U}_{1,opt},\mathbf{U}_{2,opt}\}\\
& = {\fontsize{9.5}{9.5} \selectfont arg\min_{\mathbf{U}_{1},\mathbf{U}_{2}}\log_{2}|\sigma^{2}_{1}\sigma^{2}_{2}\mathbf{I} + \sigma^{2}_{1}\tilde{\boldsymbol{\Lambda}}_{2} \otimes \boldsymbol{\Lambda}_{1}^{2}+\sigma^{2}_{2}\mathbf{U}(\boldsymbol{\Lambda}_{2}^{2}\otimes
\tilde{\boldsymbol{\Lambda}}_{1})\mathbf{U}^{H}|}.
\end{aligned}
\end{equation}
According to \cite{Fiedler1971}, we have:
\begin{lemma}\label{lemma:lm1}
	Given Hermitian matrices $\mathbf{A}, \mathbf{C} \in \mathbb{C}^{n\times n}$ and $\mathbf{B}, \mathbf{D}\in \mathbb{C}^{m\times m}$ with the corresponding diagonal eigenvalue matrices $\boldsymbol{\Lambda}_{a}$, $\boldsymbol{\Lambda}_{c}$, $\boldsymbol{\Lambda}_{b}$, $\boldsymbol{\Lambda}_{d}$ where the diagonal elements in each diagonal matrix are in descending order. Then
	\begin{subequations}
		\begin{align}
		&|\mathbf{A}\otimes \mathbf{B} + \mathbf{C}\otimes\mathbf{D}| \geq
		\min_{P_{1},P_{2}}|\boldsymbol{\Lambda}_{a}\otimes \boldsymbol{\Lambda}_{b} + \boldsymbol{\Lambda}_{c,P_{1}}\otimes\boldsymbol{\Lambda}_{d,P_{2}}|,\label{lm1.l}\\
		&|\mathbf{A}\otimes \mathbf{B} + \mathbf{C}\otimes\mathbf{D}| \leq
		\max_{P_{1},P_{2}}|\boldsymbol{\Lambda}_{a}\otimes \boldsymbol{\Lambda}_{b} + \boldsymbol{\Lambda}_{c,P_{1}}\otimes\boldsymbol{\Lambda}_{d,P_{2}}|\label{lm1.u}
		\end{align}
	\end{subequations}	
	where the minimum or maximum are taken over all possible (diagonal-wise) permutations $\{P_{1},P_{2}\}$.
\end{lemma}
From Lemma \ref{lemma:lm1}, we have:
\begin{lemma}\label{lemma3}
	Let $\mathbf{A}, \mathbf{B}, \mathbf{C}, \mathbf{D}$ be positive semi-definite Hermitian matrices with the corresponding eigenvalue matrices $\boldsymbol{\Lambda}_{a}$, $\boldsymbol{\Lambda}_{b}$, $\boldsymbol{\Lambda}_{c}$, $\boldsymbol{\Lambda}_{d}$ each of descending diagonal elements. Then
	\begin{subequations}
		\begin{align}
		&|\mathbf{A}\otimes \mathbf{B} + \mathbf{C}\otimes\mathbf{D}| \geq
		|\boldsymbol{\Lambda}_{a}\otimes \boldsymbol{\Lambda}_{b} + \boldsymbol{\Lambda}_{c}\otimes\boldsymbol{\Lambda}_{d}|,\label{lm1.l1}\\
		&|\mathbf{A}\otimes \mathbf{B} + \mathbf{C}\otimes\mathbf{D}| \leq
		|\boldsymbol{\Lambda}_{a}\otimes \boldsymbol{\Lambda}_{b} + \bar{\boldsymbol{\Lambda}}_{c}\otimes\bar{\boldsymbol{\Lambda}}_{d}|\label{lm1.u1}
		\end{align}
	\end{subequations}
	where $\bar{\boldsymbol{\Lambda}}_{c}$ and $\bar{\boldsymbol{\Lambda}}_{d}$ are respectively $\boldsymbol{\Lambda}_{c}$ and $\boldsymbol{\Lambda}_{d}$ but with reversed order of diagonal elements.
\end{lemma}
\begin{IEEEproof}
See Appendix \ref{app:lemma3}
\end{IEEEproof}

Applying \eqref{lm1.l1} to \eqref{u} and from \eqref{po1}, we have:

\begin{theorem}\label{t3}
  For $M=2$, $\mathbf{U}_{1,opt} = \mathbf{I}$ and $\mathbf{U}_{2,opt} = \mathbf{I}$ are respectively the globally optimal solutions of $\mathbf{U}_{1}$ and $\mathbf{U}_{2}$ (defined in \eqref{svd1}) to the MI based problem \eqref{opt_mu_2}.
\end{theorem}

The above choices of $\mathbf{U}_{1}$ and $\mathbf{U}_{2}$ were also used in \cite{Quist2015} but they could not establish their optimality. Also note that the optimality of the above choice of $\mathbf{U}_{1}$ and $\mathbf{U}_{2}$ was rather obvious (see the discussions of \eqref{MSE_2_svd} and \eqref{po1}) for the MSE based problem \eqref{opt_MSE}.

Let $\mathbf{C}_{1} = \tilde{\boldsymbol{\Lambda}}_{1}^{-1}\boldsymbol{\Lambda}_{1}^{2}$ and $\mathbf{C}_{2} = \tilde{\boldsymbol{\Lambda}}_{2}^{-1}\boldsymbol{\Lambda}_{2}^{2}$ with their diagonal elements denoted by $c_{1,l} = \lambda_{1,l}^{2}/\tilde{\lambda}_{1,l}$ and $c_{2,k} = \lambda_{2,k}^{2}/\tilde{\lambda}_{2,k}$. Then  \eqref{det3} becomes
\begin{equation}\label{Ikey}
	\begin{aligned}
	&I_{2}\\
	& = \log_{2}|\sigma^{2}_{2}\mathbf{I} + \tilde{\boldsymbol{\Lambda}}_{2} \otimes \mathbf{C}_{1}\tilde{\boldsymbol{\Lambda}}_{1}| + \log_{2}|\sigma^{2}_{1}\mathbf{I} +  \mathbf{C}_{2} \tilde{\boldsymbol{\Lambda}}_{2} \otimes \tilde{\boldsymbol{\Lambda}}_{1}| \\
	&\quad-\log_{2}|\sigma^{2}_{1}\sigma^{2}_{2}\mathbf{I} + \sigma^{2}_{1}\tilde{\boldsymbol{\Lambda}}_{2} \otimes \mathbf{C}_{1}\tilde{\boldsymbol{\Lambda}}_{1} + \sigma^{2}_{2}\mathbf{C}_{2} \tilde{\boldsymbol{\Lambda}}_{2}\otimes\tilde{\boldsymbol{\Lambda}}_{1}|\\
	&= \sum_{k=1}^{N_{2}}\sum_{l=1}^{N_{1}}\log_{2}\bigg(\frac{(\sigma_{2}^{2} + \tilde{\lambda}_{1,l}\tilde{\lambda}_{2,k}c_{1,l})(\sigma_{1}^{2} +  \tilde{\lambda}_{1,l}\tilde{\lambda}_{2,k}c_{2,k})}{\sigma_{1}^{2}\sigma_{2}^{2} + \sigma_{1}^{2}\tilde{\lambda}_{1,l}\tilde{\lambda}_{2,k}c_{1,l} + \sigma_{2}^{2}\tilde{\lambda}_{1,l}\tilde{\lambda}_{2,k}c_{2,k}}\bigg)\\
	& \triangleq \sum_{k=1}^{N_{2}}\sum_{l=1}^{N_{1}}f_{l,k}(c_{1,l},c_{2,k}).
	\end{aligned}
\end{equation}
Let $\mathbf{c}_{1}$ and $\mathbf{c}_{2}$ be the vectors of the diagonal elements from $\mathbf{C}_{1}$ and $\mathbf{C}_{2}$ respectively. Then \eqref{opt_mu_2} is transformed to
	\begin{align}
	\max_{\mathbf{c}_{1}>\mathbf{0},\mathbf{c}_{2}>\mathbf{0}} & \quad \sum_{k=1}^{N_{2}}\sum_{l=1}^{N_{1}}f_{l,k}(c_{1,l},c_{2,k})\label{opt2}\\
	s.t. &~~  \sum_{l = 1}^{N_{1}}c_{1,l} \leq  KP_{1},~\sum_{k = 1}^{N_{2}}c_{2,k}\leq KP_{2}.\notag
	\end{align}
It is easy to verify that $f(c_{1,l},c_{2,k})$ is a monotonically increasing function of $c_{1,l}$ and $c_{2,k}$ respectively. So, the optimal solutions must satisfy $\sum_{l= 1}^{N_{1}}c_{1,l} =  KP_{1}$ and $\sum_{k= 1}^{N_{2}}c_{2,k}= KP_{2}$.

However, $-f_{l,k}(c_{1,l},c_{2,k})$ is not always convex of $c_{1,l}$ and $c_{2,k}$. The Hessian matrix of $-f_{l,k}(c_{1,l},c_{2,k})$ is
\begin{equation}\label{Herimi}
\begin{bmatrix}
\frac{\tilde{\lambda}_{1,l}^{2}\tilde{\lambda}_{2,k}^{2}(\vartheta_{l,k} - \sigma^{4}_{1}\theta_{1,l,k})}{\theta_{1,l,k}\vartheta_{l,k}} & -\frac{\sigma^{2}_{1}\sigma^{2}_{2}\tilde{\lambda}_{1,l}^{2}\tilde{\lambda}_{2,k}^{2}}{\vartheta_{l,k}}\\
-\frac{\sigma^{2}_{1}\sigma^{2}_{2}\tilde{\lambda}_{1,l}^{2}\tilde{\lambda}_{2,k}^{2}}{\vartheta_{l,k}} &
\frac{\tilde{\lambda}_{1,l}^{2}\tilde{\lambda}_{2,k}^{2}(\vartheta_{l,k} - \sigma^{4}_{2}\theta_{2,l,k})}{\theta_{2,l,k}\vartheta_{l,k}}
\end{bmatrix}
\end{equation}
where $\theta_{1,l,k} = (\sigma^{2}_{2}+\tilde{\lambda}_{1,l}\tilde{\lambda}_{2,k}c_{1,l})^{2}$, $\theta_{2,l,k} = (\sigma^{2}_{1}+\tilde{\lambda}_{1,l}\tilde{\lambda}_{2,k}c_{2,k})^{2}$ and $\vartheta_{l,k} =(\sigma^{2}_{1}\sigma^{2}_{2} + \sigma^{2}_{1}\tilde{\lambda}_{1,l}\tilde{\lambda}_{2,k}c_{1,l} + \sigma^{2}_{2}\tilde{\lambda}_{1,l}\tilde{\lambda}_{2,k}c_{2,k})^{2}$.
This matrix is positive semidefinite if and only if  $c_{1,l}c_{2,k} \geq \frac{\sigma^{2}_{1}\sigma^{2}_{2}}{2\tilde{\lambda}_{1,l}^{2}\tilde{\lambda}_{2,k}^{2}}$. This means that when $KP_1$ and $KP_2$ are large, the Hessian matrix of $-f_{l,k}(c_{1,l},c_{2,k})$ is typically positive definite and hence $-f_{l,k}(c_{1,l},c_{2,k})$ is typically convex. In this high power case, the problem \eqref{opt2} is convex and the globally optimal solution is available.
In general,  $-f_{l,k}(c_{1,l},c_{2,k})$ is a convex function with respect to $c_{1,l}$ and $c_{2,k}$ individually. To obtain locally optimal solution to \eqref{opt2}, we can apply a two-phase iteration method, i.e., optimizing $\mathbf{c}_{1}$ and $\mathbf{c}_{2}$ alternately until convergence. The discussion of the following two-phase algorithm is similar to that in \cite{Quist2015}.

In  phase one, the Lagrangian function with respect to $c_{1,l}$ is
\begin{equation}\label{L1}
\begin{aligned}
\mathcal{L} = \sum_{k=1}^{N_{2}}\sum_{l=1}^{N_{1}}f_{l,k}(c_{1,l},c_{2,k}) - \mu\bigg( \sum_{l= 1}^{N_{1}}c_{1,l}  - KP_{1}\bigg) + \boldsymbol{\alpha}^{T}\mathbf{c}_{1}.
\end{aligned}
\end{equation}
And the corresponding KKT conditions are
\begin{equation}\label{kkt1}
\left \{ \begin{aligned}
&\frac{\partial\mathcal{L}}{\partial c_{1,l}} = \frac{1}{\ln 2}\sum_{k= 1}^{N_{2}} f_{l,k}'(c_{1,l},c_{2,k}) - \mu = 0 ,\\
& \sum_{l= 1}^{N_{1}}c_{1,l} \leq KP_{1},~\mu(\sum_{l= 1}^{N_{1}}c_{1,l} - KP_{1}) = 0,~\mu \geq 0,\\
&\mathbf{c}_{1}>\mathbf{0},~\boldsymbol{\alpha}^{T}\mathbf{c}_{1}=0,~\boldsymbol{\alpha}\geq \mathbf{0}
\end{aligned}
\right .
\end{equation}
where
\begin{equation}\label{g}
\begin{aligned}
&f_{l,k}'(x,y)\\
& = \frac{\sigma^{2}_{2}\tilde{\lambda}_{1,l}^{2}\tilde{\lambda}_{2,k}^{2}y}{(\sigma^{2}_{2} + \tilde{\lambda}_{1,l}\tilde{\lambda}_{2,k}x)(\sigma^{2}_{1}\sigma^{2}_{2} + \sigma^{2}_{1}\tilde{\lambda}_{1,l}\tilde{\lambda}_{2,k}x + \sigma^{2}_{2}\tilde{\lambda}_{1,l}\tilde{\lambda}_{2,k}y)}.
\end{aligned}
\end{equation}
In phase two, similar KKT conditions can be found.
From \eqref{kkt1}, we see that $\mu$ is a monotonically decreasing function of $c_{1,l}$. Therefore, we can use a bisection search to solve \eqref{kkt1}.
An efficient algorithm to solve \eqref{opt2} is shown in
Algorithm \ref{Algorithm1}.
\begin{algorithm}[t]
	\caption{ Bisection section search to solve \eqref{kkt1}}
	\label{Algorithm1}
	\small
	\begin{algorithmic}[1]
		\renewcommand{\algorithmicrequire}{\textbf{Input:}}
		\REQUIRE ~\\
		$\tilde{\boldsymbol{\Lambda}}_{1}, \tilde{\boldsymbol{\Lambda}}_{2}, P_{1}, P_{2}$, K;\\
		Accuracy threshold $\epsilon_{1}$, $\epsilon_{2}$.\\
		Initialization $p = 0$, $\mathbf{c}_{1}^{(p)} = \frac{KP_{1}}{N_{1}}\mathbf{1}_{N_{1}}, \mathbf{c}_{2}^{(p)} =\frac{KP_{2}}{N_{2}}\mathbf{1}_{N_{2}}$.\\
		\REPEAT
		\STATE {Given $\mathbf{c}_{2}^{(p)}$, do bisection search of $\mu$ and obtain solution $\mathbf{c}_{1}^{(p+1)}$ to meet the power constraint $|\sum_{l = 1}^{N_{1}}c_{1,l} - KP_{1}| \leq \epsilon_{1}$; Given $\mathbf{c}_{1}^{(p+1)}$, do bisection search of $\nu$ and obtain solution $\mathbf{c}_{2}^{(p+1)}$ to meet the power constraint $|\sum_{k= 1}^{N_{2}}c_{2,k} - KP_{2}| \leq \epsilon_{1}$}.
		\STATE{$p = p + 1.$}
		\UNTIL{ $\|[\mathbf{c}_{1}^{(p)},\mathbf{c}_{2}^{(p)}] - [\mathbf{c}_{1}^{(p-1)},\mathbf{c}_{2}^{(p-1)}]\| \leq \epsilon_{2}$}
		\RETURN $\{\mathbf{c}_{1}^{(p)},\mathbf{c}_{2}^{(p)}\}$
	\end{algorithmic}
\end{algorithm}
From \eqref{g}, we know that $f_{l,k}'(c_{1,l},c_{2,k})$ is an increasing function of $\tilde{\lambda}_{1,l}$ and a decreasing function of $c_{1,l}$. Given any $\mathbf{c}_{2}$,  the solution from \eqref{kkt1} is $\mathbf{c}_{1}^{*}$, which must satisfy $\sum_{k= 1}^{N_{2}} f_{l,k}'(c^{*}_{1,l},c_{2,k}) = \mu\ln 2$. Hence, one can verify that $c_{1,l}^{*} \geq c_{1,l+1}^{*}$. (If $c_{1,l}^{*} < c_{1,l+1}^{*}$ then $\mu\ln 2 =
\sum_{k= 1}^{N_{2}} f_{l,k}'(c^{*}_{1,l},c_{2,k}) > \sum_{k= 1}^{N_{2}} f_{l,k}'(c^{*}_{1,l+1},c_{2,k})\geq \sum_{k= 1}^{N_{2}} f'_{l+1,k}(c^{*}_{1,l+1},c_{2,k}) = \mu\ln 2
$, which is not possible.) Similarly, $c_{2,k}^{*} \geq c_{2,k+1}^{*}$. Therefore, the diagonal elements of the optimal solutions of $\boldsymbol{\Lambda}_{1}^{2}$ and $\boldsymbol{\Lambda}_{2}^{2}$ are also in descending order respectively.

\subsubsection{Asymptotic Analysis}

The following theorem shows the globally optimal solution to \eqref{opt_SKR} in high or low power region. These solutions are also given by Algorithm \ref{Algorithm1}.

\begin{theorem}\label{t4}
  Let $P_{1} = P_{2}=P$. If $P$ is arbitrarily large, the globally optimal $c_{1,l}$ and $c_{2,k}$ (defined before \eqref{Ikey}) are invariant to $l$ and $k$ (which will be called ``uniform power'' allocation), and a less correlated channel yields a higher secret key rate. If $P$ is arbitrarily small, the globally optimal $c_{1,l}$ and $c_{2,k}$ are all arbitrarily small except for $l=k=1$, and a higher correlated channel yields a higher secret key rate.
\end{theorem}
\begin{IEEEproof}
See Appendix \ref{proof_t4}.
\end{IEEEproof}
%
\subsection{Closed-form solution}\label{sec:key_sc}
For $M\geq 2$, we now consider the same symmetric and isotropic case considered before. Without loss of generality, also let $\sigma=1$. Then applying the matrix inverse lemma to \eqref{gammaij.1} and \eqref{gammaji.1}, we have
\begin{equation}\label{gammaij_sc}
	\begin{aligned}
	&\boldsymbol{\Gamma}_{i,j}
	 = (\mathbf{S}_{j}\bar{\mathbf{F}}\bar{\mathbf{F}}^{H}\mathbf{S}_{j}^{T}\otimes\mathbf{I} )\\
	&- \big((\mathbf{S}_{j}\bar{\mathbf{F}}\bar{\mathbf{F}}^{H}\bar{\mathbf{S}}_{(i)}^{T})(\mathbf{I} + \bar{\mathbf{S}}_{(i)}\bar{\mathbf{F}}\bar{\mathbf{F}}^{H}\bar{\mathbf{S}}_{(i)}^{T})^{-1}(\bar{\mathbf{S}}_{(i)}\bar{\mathbf{F}}\bar{\mathbf{F}}^{H}\mathbf{S}_{j}^{T})\big)\otimes\mathbf{I},
	\end{aligned}
\end{equation}
\begin{equation}\label{gammaji_sc}
\begin{aligned}
&\boldsymbol{\Gamma}_{T,j,i}
 = (\mathbf{I} \otimes\mathbf{S}_{i}\bar{\mathbf{F}}\bar{\mathbf{F}}^{H}\mathbf{S}_{i}^{T})\\
& - \mathbf{I} \otimes\big((\mathbf{S}_{i}\bar{\mathbf{F}}\bar{\mathbf{F}}^{H}\bar{\mathbf{S}}_{(j)}^{T})(\mathbf{I} + \bar{\mathbf{S}}_{(j)}\bar{\mathbf{F}}\bar{\mathbf{F}}^{H}\bar{\mathbf{S}}_{(j)}^{T} )^{-1}(\bar{\mathbf{S}}_{(j)}\bar{\mathbf{F}}\bar{\mathbf{F}}^{H}\mathbf{S}_{i}^{T})\big).
\end{aligned}
\end{equation}
Note that $I(\mathbf{y}_{i};\mathbf{y}_{T,j}) = -\log_{2}|\mathbf{I} - \boldsymbol{\Gamma}_{i,j}\boldsymbol{\Gamma}_{T,j,i}|$, $I_{M} = \sum_{i=1}^{M-1}\sum_{j = i+1}^{M} I(\mathbf{y}_{i};\mathbf{y}_{T,j})$ and the power and rank constraints in \eqref{opt_SKR} become $Tr(\mathbf{S}_{i}\bar{\mathbf{F}}\bar{\mathbf{F}}^{H}\mathbf{S}_{i}^{T}) \leq KP,~i = 1,\dots,M$. Then the Lagrangian function  is now
\begin{equation}\label{L_MSKR_spCase}
\mathcal{L} = I_{M} - \sum_{i = 1}^{M}\mu_{i} (Tr(\mathbf{S}_{i}\bar{\mathbf{F}}\bar{\mathbf{F}}^{H}\mathbf{S}_{i}^{T}) -KP)
\end{equation}
and the KKT conditions are
\begin{equation}\label{kkt_mul_sk_sc}
	\left\{
	\begin{aligned}
	&\frac{\partial \mathcal{L}}{\partial \bar{\mathbf{F}}} =  \frac{\partial I_{M}}{\partial \bar{\mathbf{F}}}- \sum_{i=1}^{M}2\mu_{i}\mathbf{S}_{i}^{T}\mathbf{S}_{i}\bar{\mathbf{F}} = 0,\\
	&Tr(\mathbf{S}_{i}\bar{\mathbf{F}}\bar{\mathbf{F}}^{H}\mathbf{S}_{i}^{T}) \leq KP,~ i = 1,\dots,M,\\
	&\mu_{i}( Tr(\mathbf{S}_{i}\bar{\mathbf{F}}\bar{\mathbf{F}}^{H}\mathbf{S}_{i}^{T}) - KP) = 0,~ \mu_{i} \geq 0,~ i = 1,\dots,M.
	\end{aligned}\right.
\end{equation}
\begin{theorem}\label{t2}
	The solutions to \eqref{kkt_mul_sc} as shown in Theorem \ref{t1} are also solutions to  \eqref{kkt_mul_sk_sc}.
\end{theorem}
\begin{IEEEproof}
	See Appendix \ref{app_skr}.
\end{IEEEproof}
For $M=2$, the pilots from this theorem satisfy $\mathbf{P}_i\mathbf{P}_i^H=\frac{KP}{N}\mathbf{I}_N$ where $i=1,2$, and these pilots are known to be globally optimal for maximal MI \cite{Jorswieck2013} under the symmetric and isotropic condition. Also note that for $M\geq 3$, our numerical simulations did not yield any result better than that from Theorem \ref{t2} subject to the symmetric and isotropic condition.

 \section{Simulation results}\label{sec:simulation}

To show some simulation results, we let $P_{i} = P$, $\sigma^{2}_{i} = 1$, $N_{i} = 4$, $\mathbf{R}_i=\mathbf{R}$,  $r = (M-1)N$ and $K \geq r$. We choose the channel correlation matrix to be such that $(\mathbf{R})_{l,k} = R^{|l-k|}$ where $R\in[0,1]$ is the correlation coefficient.
\subsection{Comparison of user's channel MSE}
We first use the normalized MSE (per element of each channel matrix):
\begin{equation}\label{}
\mathcal{J}_{M}=\frac{J_{M}}{M(M-1)N^2}
\end{equation}
to compare three different choices of pilots. Since $\mathcal{J}_{M}$ depends on $R$, we will also write $\mathcal{J}_{M}=\mathcal{J}_{M}(R)$. More specifically, we use $\mathcal{J}_{M,MSE-opt}(R)$ for the optimal pilots computed from algorithm 1, $\mathcal{J}_{M,c-opt}(R)$  for the conditionally optimal pilots from Theorem 1, $\mathcal{J}_{M,first}(R)$  for the pilots proposed in \cite{Hua2019a} (which coincides with that from Theorem 1 if $N_i=N=1$) and $\mathcal{J}_{M,MI-opt}(R)$ for the pilots that maximizes MI from \text{(29)}.
\begin{figure}[h]
	\includegraphics[width=0.4\textwidth]{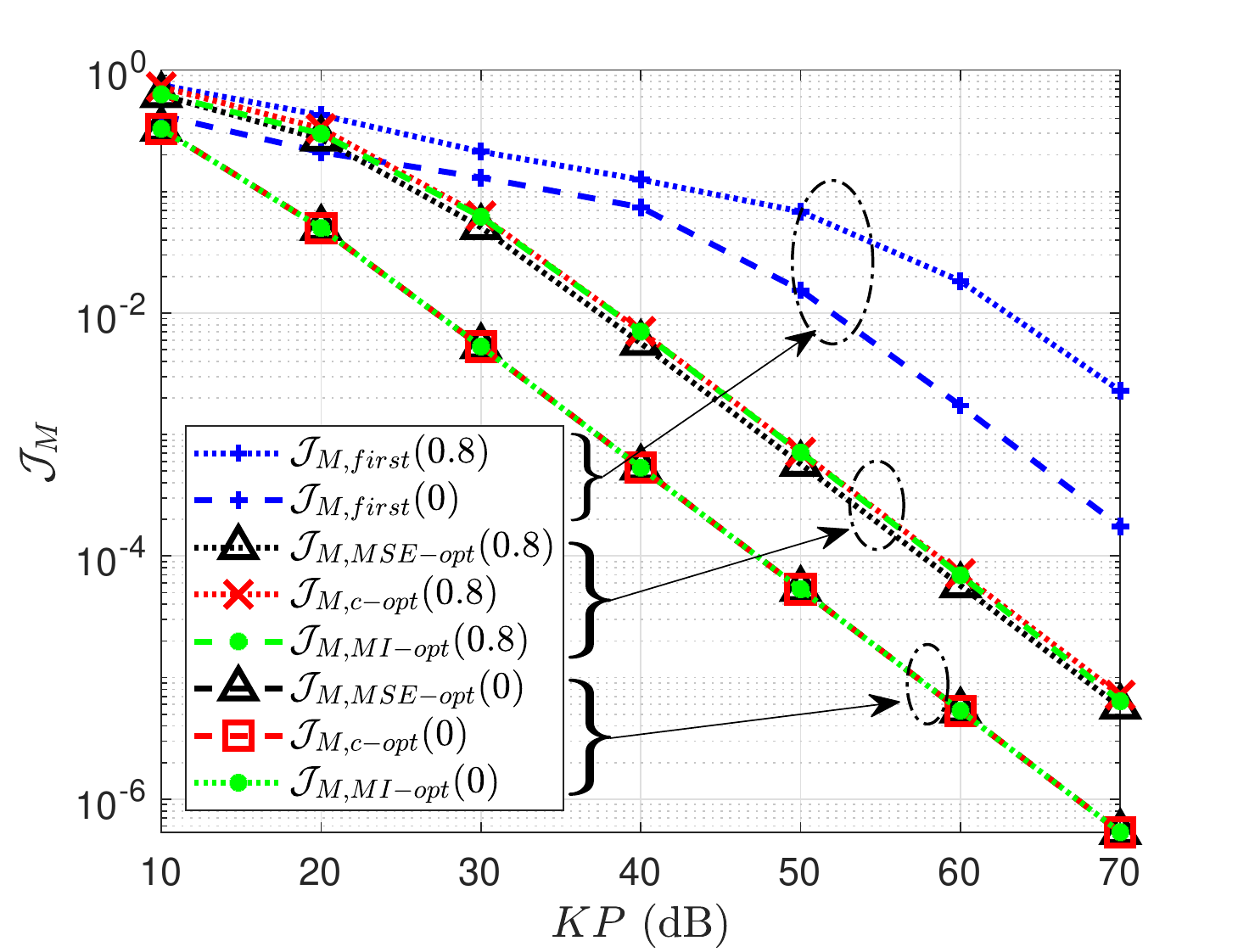}
	\centering
	\caption{Normalized MSE vs $10\texttt{dB}\leq KP \leq 70\texttt{dB}$ where $M = 3$.}
	\label{f_m_mse_0_70}
\end{figure}

\begin{figure}[h]
	\includegraphics[width=0.4\textwidth]{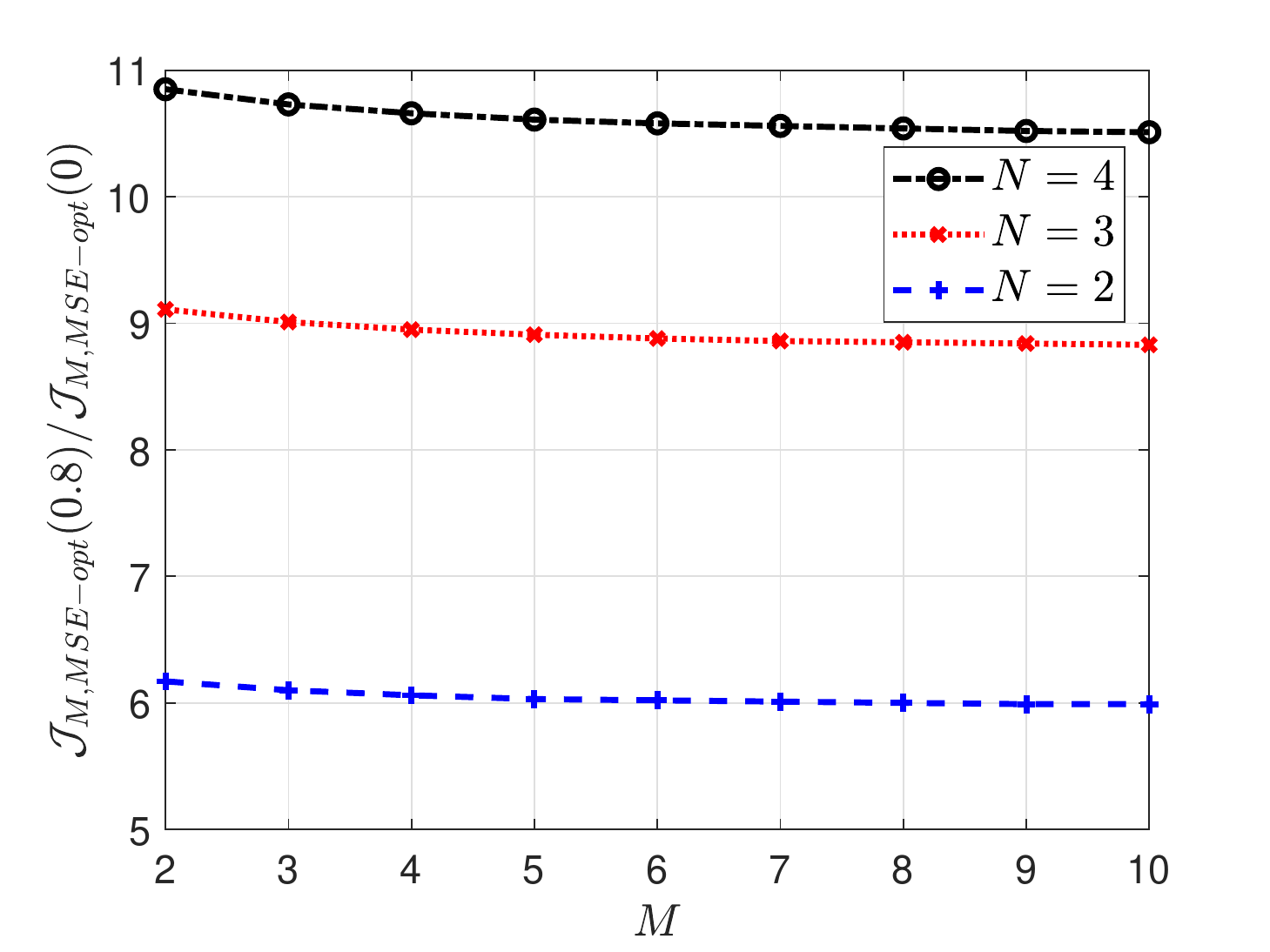}
	\centering
	\caption{$\frac{\mathcal{J}_{M,MSE-opt}(0.8)}{\mathcal{J}_{M,MSE-opt}(0)}$ vs $M$ and $N$ with $KP = 60\mathtt{dB}$.}
	\label{xi_mse}
\end{figure}

For $M=3$, Fig. \ref{f_m_mse_0_70} shows the normalized MSE vs $10\texttt{dB}\leq KP \leq 70\texttt{dB}$. We see that for high $KP$ all curves of the normalized MSE in log-scale vs $KP$ in dB become parallel straight lines. This is expected since for large enough $KP$ the MSE is proportional to  $\frac{1}{KP}$.
It is also expected that $\mathcal{J}_{M,MSE-opt}(0)=\mathcal{J}_{M,c-opt}(0) = \mathcal{J}_{M,MI-opt}(0)$. But we see that  $\mathcal{J}_{M,MSE-opt}(R)$,  $\mathcal{J}_{M,c-opt}(R)$ and $\mathcal{J}_{M,MI-opt}(R)$ are still rather close to each other even for $R=0.8$ and they all are substantially better than $\mathcal{J}_{M,first}(R)$ especially at high $KP$. The above results suggest that the pilots from maximizing MI is a good sub-optimal solution for minimizing MSE.

Using the pilots from Theorem 1, we know that  $J_{M,MSE-opt}(0)= N\sum_{i=1}^{M}Tr\big((\mathbf{I} + \frac{KP}{N^2(M-1)} \bar{\mathbf{S}}_{(i)}\bar{\mathbf{Q}}_{m}\bar{\mathbf{Q}}_{m}^{H}\bar{\mathbf{S}}_{(i)}^{T} )^{-1}\big)$, and hence one can verify that
\begin{equation}\label{}
\lim_{KP \rightarrow \infty} \mathcal{J}_{M,MSE-opt}(0) = 2N(1-\frac{1}{M})\frac{1}{KP}
\end{equation}
which is invariant to large $M$. But this limit increases linearly as $N$ increases (because the per-antenna power is $\frac{P}{N}$).

Fig. \ref{xi_mse} shows $\frac{\mathcal{J}_{M,MSE-opt}(0.8)}{\mathcal{J}_{M,MSE-opt}(0)}$ vs $M$ and $N$ where $KP=60$dB. Note that $\frac{\mathcal{J}_{M,MSE-opt}(0.8)}{\mathcal{J}_{M,MSE-opt}(0)}$ is invariant to large $KP$. From this and other similar plots that we have obtained but not shown here, we have observed that $\mathcal{J}_{M,MSE-opt}(R)$ is also invariant to large $M$ but increases as $N$ increases. Furthermore, $\mathcal{J}_{M,MSE-opt}(R)$ increases as $R$ increases within $[0,1)$ in the high power region.

\subsection{Comparison of user's channel MI}
We also use the normalized MI (per pair and per degree-of-freedom):
\begin{equation}
\mathcal{I}_M = \frac{I_M}{\frac{M(M-1)N^2}{2}}
\end{equation}
to compare four different choices of pilots. Let $\mathcal{I}_M =\mathcal{I}_M (R)$. We use $\mathcal{I}_{M,MI-opt}(R)$ for the pilots that maximizes the MI from  \eqref{opt_SKR}, $\mathcal{I}_{M,c-opt}(R)$ for the pilots from Theorem \ref{t2}, $\mathcal{I}_{M,first}(R)$ for the pilots initially suggested in \cite{Hua2019a} and $\mathcal{I}_{M,MSE-opt}(R)$ for the pilots that minimized MSE from \eqref{opt_MSE}.
%
\begin{figure}[t]
	\includegraphics[width=0.4\textwidth]{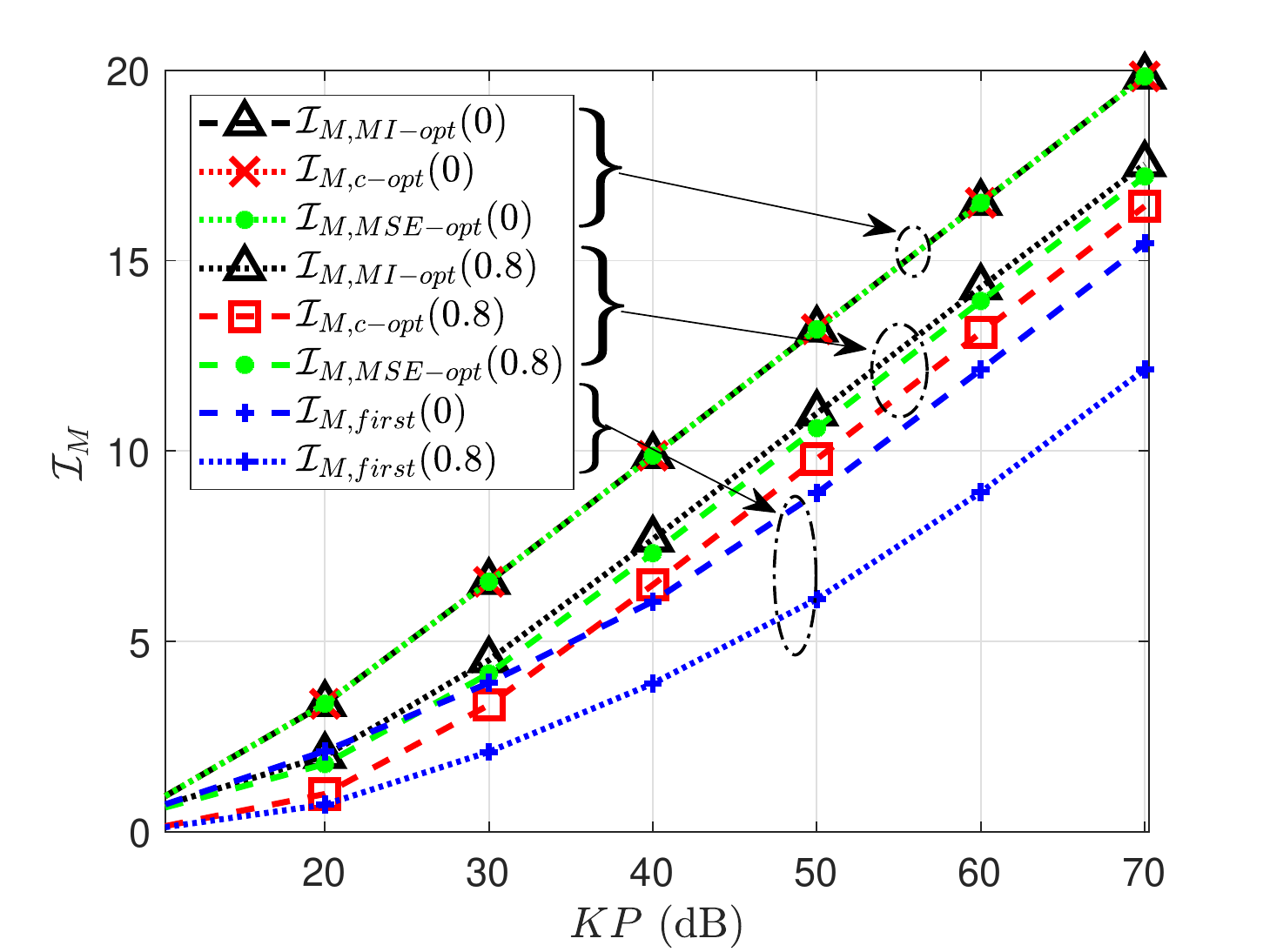}
	\centering
	\caption{Normalized MI $10\texttt{dB}\leq KP \leq 70\texttt{dB}$ with $M=3$.}
	\label{f_m_skr_0_70}
\end{figure}
\begin{figure}[t]
	\includegraphics[width=0.4\textwidth]{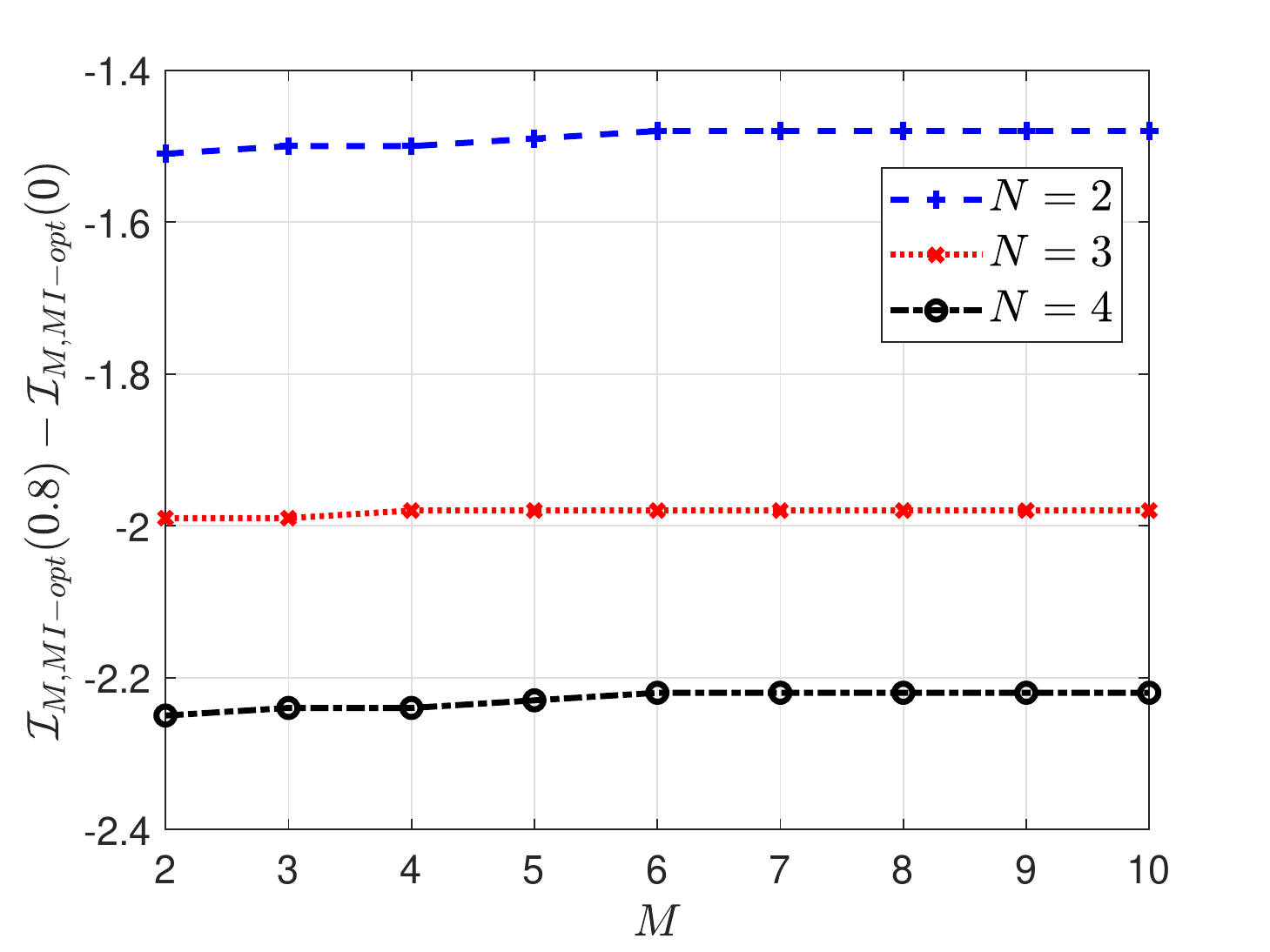}
	\centering
	\caption{$\mathcal{I}_{M,MI-opt}(0.8)-\mathcal{I}_{M,MI-opt}(0)$ vs $M$ and $N$ with $KP = 60\mathtt{dB}$.}
	\label{xi_mi}
\end{figure}

For $M=3$, Fig. \ref{f_m_skr_0_70} shows $\mathcal{I}_M(R)$ vs $10\texttt{dB}\leq KP \leq 70\texttt{dB}$. Since $\mathcal{I}_M(R)$ is a constant plus $\log_{2} (KP)$ at high $KP$, we see that all curves here become parallel straight lines when $KP$ is large. As expected, we see that $\mathcal{I}_{M,MI-opt}(0) = \mathcal{I}_{M,c-opt}(0) = \mathcal{I}_{M,MSE-opt}(0)$. But $\mathcal{I}_{M,MI-opt}(R)$, $\mathcal{I}_{M,c-opt}(R)$, $\mathcal{I}_{M,MSE-opt}(R)$ are still rather close to each other even for $R = 0.8$ and they are all significantly better than $\mathcal{I}_{M,first}(R)$. Such results suggest that the pilots from minimizing MSE is a good sub-optimal solution for maximizing MI.

One can verify by using  \eqref{pair_ij_3} and $I_{M,MI-opt}(0) = -N^2\log_{2}(1 - \Gamma^{2})$ that
\begin{equation}\label{}
\lim_{KP \rightarrow \infty}\mathcal{I}_{M,MI-opt}(0) = \log_{2}(\frac{1}{4N}(1 + \frac{1}{M-1}))+\log_2(KP)
\end{equation}
which is invariant to large $M$ but decreases as $N$ increases.

Fig. \ref{xi_mi} shows $\mathcal{I}_{M,MI-opt}(0.8)-\mathcal{I}_{M,MI-opt}(0)$ vs $M$ and $N$ where $KP=60$dB. Note that $\mathcal{I}_{M,MI-opt}(0.8)-\mathcal{I}_{M,MI-opt}(0)$ is invariant to large $KP$. From this and other similar plots not shown here, we have observed that $\mathcal{I}_{M,MI-opt}(R)$ is also invariant to large $M$ but decreases as $N$ increases. And $\mathcal{I}_{M,MI-opt}(R)$ decreases as $R$ increases within $[0,1)$ in the high power region.
\begin{figure}[t]
	\includegraphics[width=0.4\textwidth]{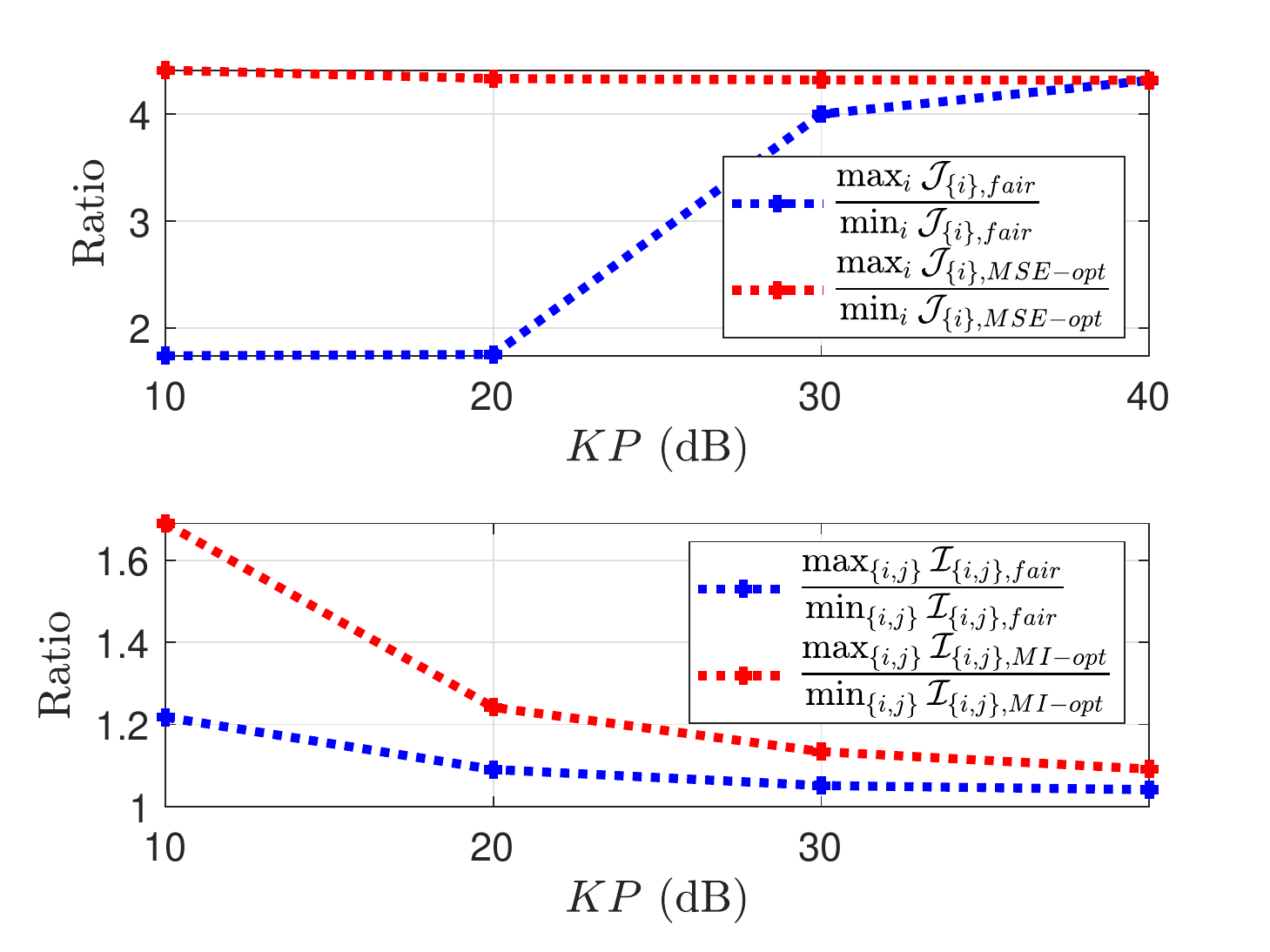}
	\centering
	\caption{Fairness ratios of $\mathcal{J}_{\{i\},fair}$, $\mathcal{J}_{\{i\},MSE-opt}$, $\mathcal{I}_{\{i,j\},fair}$,  $\mathcal{I}_{\{i,j\},MI-opt}$ for the case $\{\sigma_{1}^2 = 1,~\sigma_{2}^2 =0.6,~\sigma_{3}^2 = 0.1\}$ vs $10\texttt{dB}\leq KP \leq 40\texttt{dB}$.}
	\label{fair_s}
\end{figure}

\begin{figure}[t]
	\includegraphics[width=0.4\textwidth]{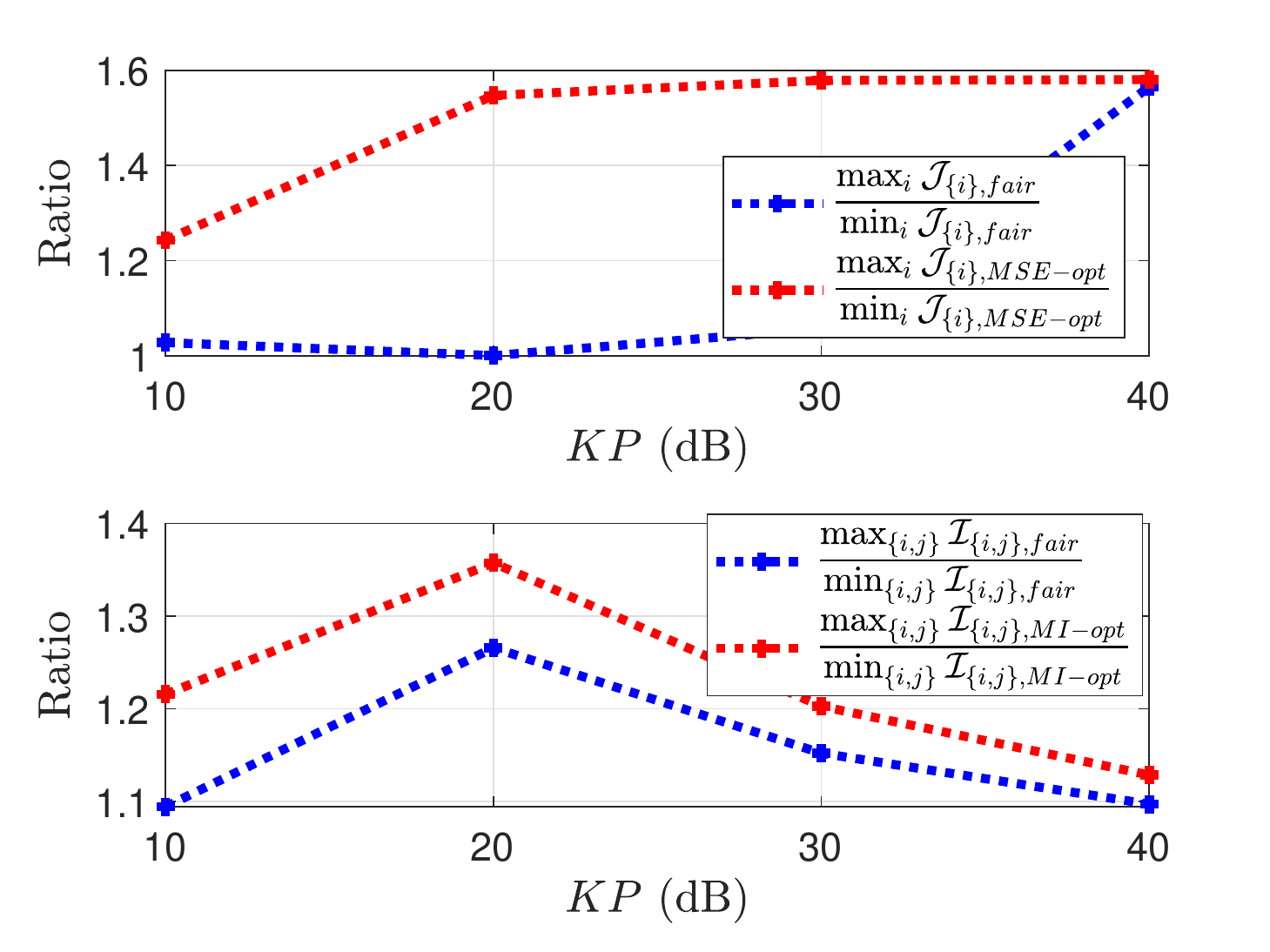}
	\centering
	\caption{Fairness ratios of $\mathcal{J}_{\{i\},fair}$, $\mathcal{J}_{\{i\},MSE-opt}$, $\mathcal{I}_{\{i,j\},fair}$,  $\mathcal{I}_{\{i,j\},MI-opt}$ for the case $\{R_{1} = 0.8,~R_{2} =0.4,~R_{3} = 0\}$ vs $10\texttt{dB}\leq KP \leq 40\texttt{dB}$.}
	\label{fair_r}
\end{figure}
\subsection{Comparison of user's channel fairness}
We now compare the results from \eqref{opt_MSE_fair2} and  \eqref{opt_mskr_fair} with those based on the sum of MSE and the sum of MI.   We consider two situations with three users: 1) different noise variances $\sigma_{1}^2 = 1$, $\sigma_{2}^2 =0.6$, $\sigma_{3}^2 = 0.1$ with the same channel correlation $R_{i} = 0,\forall i$, and 2) different channel correlations $R_{1} = 0.8$, $R_{2} = 0.4$, $R_{3} = 0$ with the same noise variance $\sigma_{i}^2 = 1,\forall i$.
We use $\mathcal{J}_{\{i\},fair}$ and $\mathcal{J}_{\{i\},MSE-opt}$ to denote the normalized MSE for the $i$th user based on  \eqref{opt_MSE_fair2} and \eqref{opt_MSE} respectively, and use $\mathcal{I}_{\{i,j\},fair}$ and $\mathcal{I}_{\{i,j\},MI-opt}$ to denote the normalized MI for the distinct pair of users $\{i,j\}$ based on \eqref{opt_mskr_fair} and \eqref{opt_SKR} respectively.

In Fig. \ref{fair_s} and Fig. \ref{fair_r}, we shows the  ``fairness ratios'' $ \frac{\max_{i}\mathcal{J}_{\{i\},MSE-opt}}{\min_{i}\mathcal{J}_{\{i\},MSE-opt}}$, $\frac{\max_{i}\mathcal{J}_{\{i\},fair}}{\min_{i}\mathcal{J}_{\{i\},fair}}$,  $\frac{\max_{\{i,j\}}\mathcal{I}_{\{i,j\},MI-opt}}{\min_{\{i,j\}}\mathcal{I}_{\{i,j\},MI-opt}}$ and $\frac{\max_{\{i,j\}}\mathcal{I}_{\{i,j\},fair}}{\min_{\{i,j\}}\mathcal{I}_{\{i,j\},fair}}$ vs $10\texttt{dB}\leq KP \leq 40\texttt{dB}$ for the situation of different noise variances and the situation of different channel correlations respectively.
As expected, results based on criteria aimed for better fairness have smaller fairness ratios. But we also see that as the power or $KP$ increases, the ``worst case'' based algorithms (i.e., \eqref{opt_MSE_fair2} and \eqref{opt_mskr_fair}) and the ``equally weighted'' algorithms (i.e., \eqref{opt_MSE} and \eqref{opt_SKR}) yield the same fairness ratios.

\begin{figure}[t]
	\includegraphics[width=0.4\textwidth]{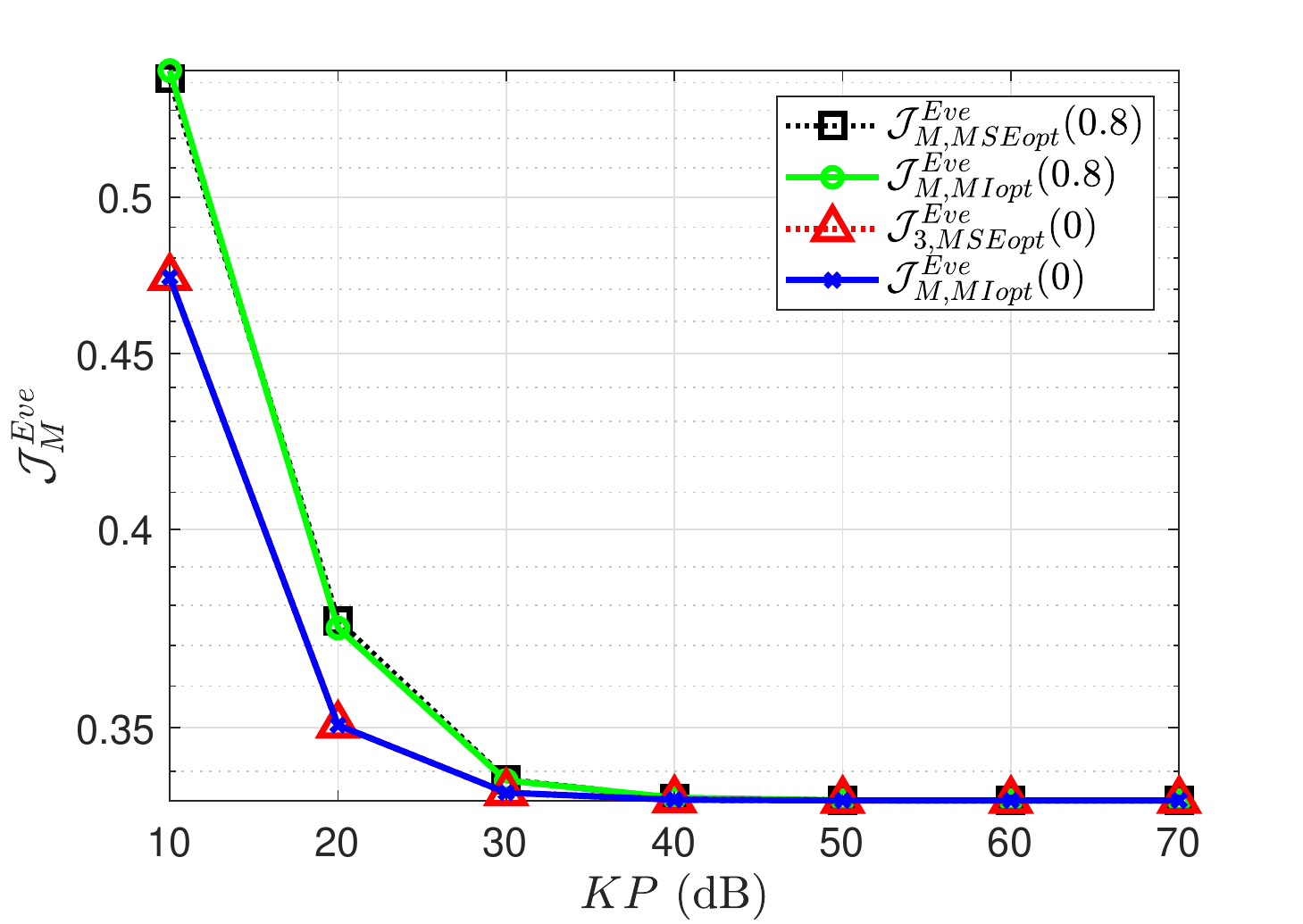}
	\centering
	\caption{Average normalized MSE for Eve vs $10\texttt{dB}\leq KP \leq 70\texttt{dB}$ with $M=3$.}
	\label{J_eve}
\end{figure}

\subsection{Comparison of Eve's channel MSE}
To illustrate the performance of the channel estimation by Eve, we define the following normalized MSE
\begin{equation}\label{eve_mse}
\mathcal{J}^{Eve}_{M} = \frac{1}{M}\sum_{i=1}^{M}\frac{Tr(\mathbf{K}_{\Delta\mathbf{h}_{E,i}})}{N_{E}N_{i}}
\end{equation}
where $Tr(\mathbf{K}_{\Delta\mathbf{h}_{E,i}})$ is from \eqref{MSE2} and we assume $\sigma_{Eve,i} = 1, \forall i$. Also note that we can write $\mathcal{J}^{Eve}_{M} = \mathcal{J}^{Eve}_{M}(R)$ where $R$ is the users' channel correlation. We compare two different pilots: 1) $\mathcal{J}^{Eve}_{M,MSE-opt}(R)$ for the MSE based pilots from \eqref{opt_MSE}, and 2) $\mathcal{J}^{Eve}_{M,MI-opt}(R)$ for the MI based pilots from \eqref{opt_SKR}.

In Fig. \ref{J_eve}, we can see that both $\mathcal{J}^{Eve}_{M,MSE-opt}(R)$ and $\mathcal{J}^{Eve}_{M,MI-opt}(R)$ become saturated as $KP$ increases, and both are lower bounded by a significant constant. We also see that each of $\mathcal{J}_{Eve,MI-opt}(R)$ and $\mathcal{J}_{Eve,MSE-opt}(R)$ is almost invariant to $R$. These results indicate that both MSE and MI based designs have a similar detrimental impact on Eve's channel estimation. The key reason for this is because of the reduced-rank constraint on the pilots.

\subsection{Two-user case}
For the two-user case,  we use $\mathcal{J}_{2,MSE}(R)$ and $\mathcal{I}_{2,MSE}(R)$ for the MSE based pilots from \cite{Bjornson2010},  $\mathcal{J}_{2,MI}(R)$ and  $\mathcal{I}_{2,MI}(R)$ for the  MI based pilots from \text{(50)}, and $\mathcal{J}_{2,u}(R)$ and $\mathcal{I}_{2,u}(R)$ for the pilots based on the ``uniform power'' allocation, i.e. $\mathbf{c}_{1} = \mathbf{c}_{2} = \frac{KP}{N}\mathbf{1}$.

From \cite{Jorswieck2013,Zhu2019}, we know that $\mathcal{J}_{2,MSE}(0) = \mathcal{J}_{2,MI}(0) = \mathcal{J}_{2,u}(0)$ and $\mathcal{I}_{2,MSE}(0) = \mathcal{I}_{2,MI}(0) = \mathcal{I}_{2,u}(0)$.

\begin{figure}[t]
	\includegraphics[width=0.4\textwidth]{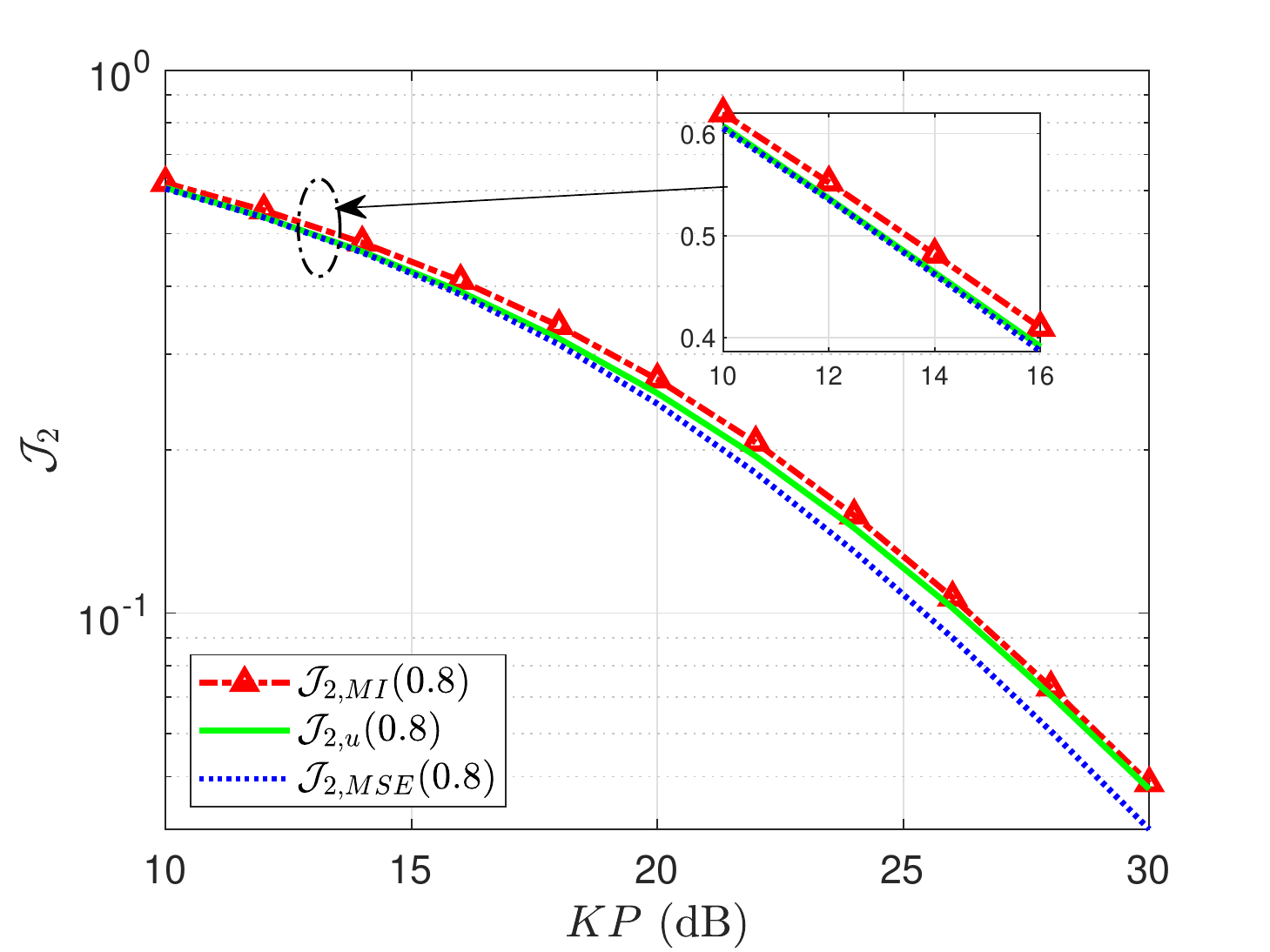}
	\centering
	\caption{Normalized MSE for $10\texttt{dB}\leq KP \leq 30\texttt{dB}$ with $M=2$.}
	\label{2mse}
\end{figure}

\begin{figure}[t]
	\includegraphics[width=0.4\textwidth]{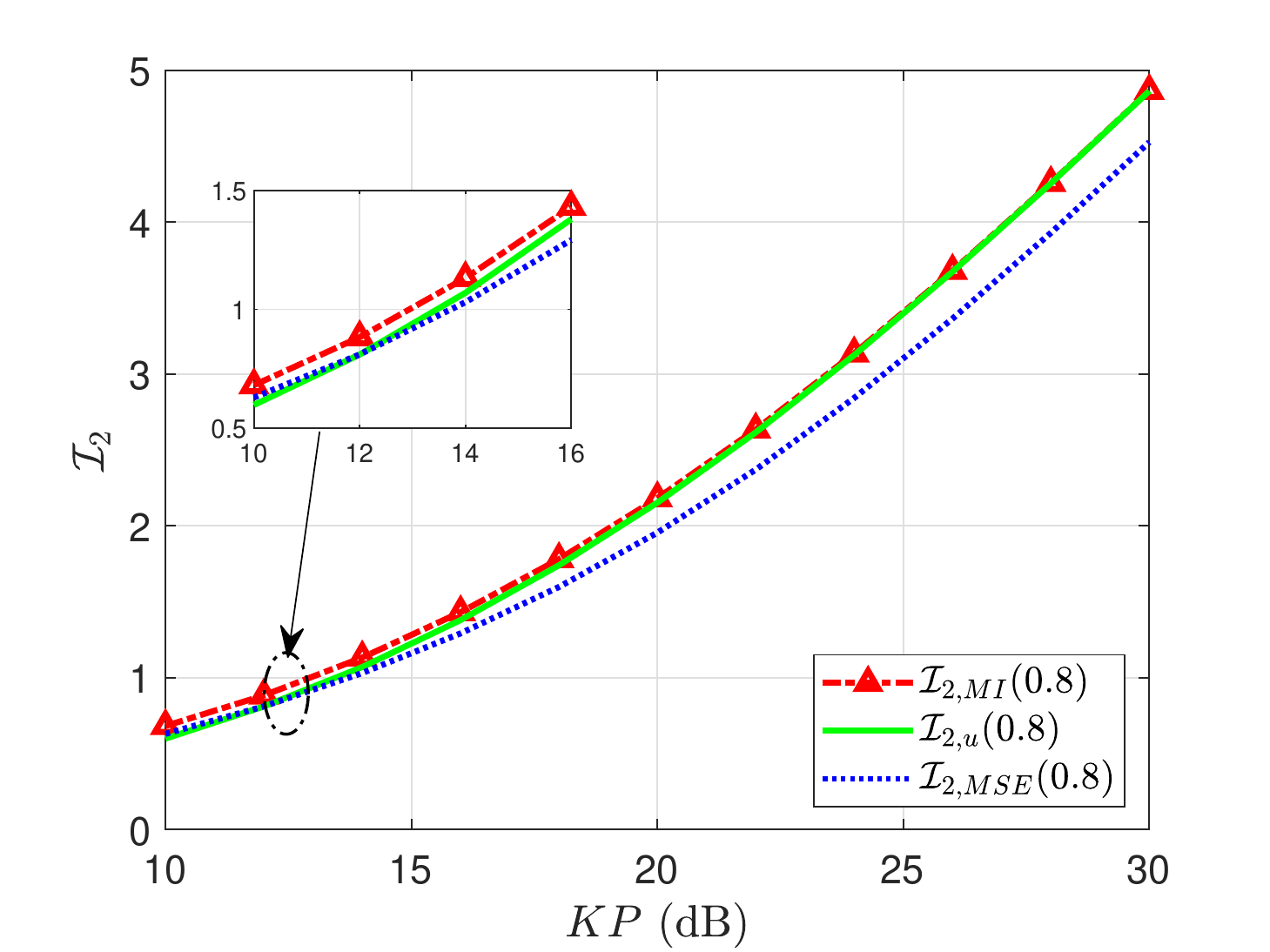}
	\centering
	\caption{Normalized MI for $10\texttt{dB}\leq KP \leq 30\texttt{dB}$ with $M=2$.}
	\label{2mi}
\end{figure}

But for the correlated channels, the normalized MSE is shown in Fig. \ref{2mse},   and the normalized MI is shown in Fig. \ref{2mi}. We see that $\mathcal{J}_{2,MI}(R)$ and $\mathcal{I}_{2,MI}(R)$ are rather close to $\mathcal{J}_{2,MSE}(R)$ and $\mathcal{I}_{2,MSE}(R)$ respectively. Also $\mathcal{J}_{2,MI}(R)$ and  $\mathcal{I}_{2,MI}(R)$ overlap with $\mathcal{J}_{2,u}(R)$ and $\mathcal{I}_{2,u}(R)$  respectively in the high power region.

\begin{figure}[t]
	\includegraphics[width=0.4\textwidth]{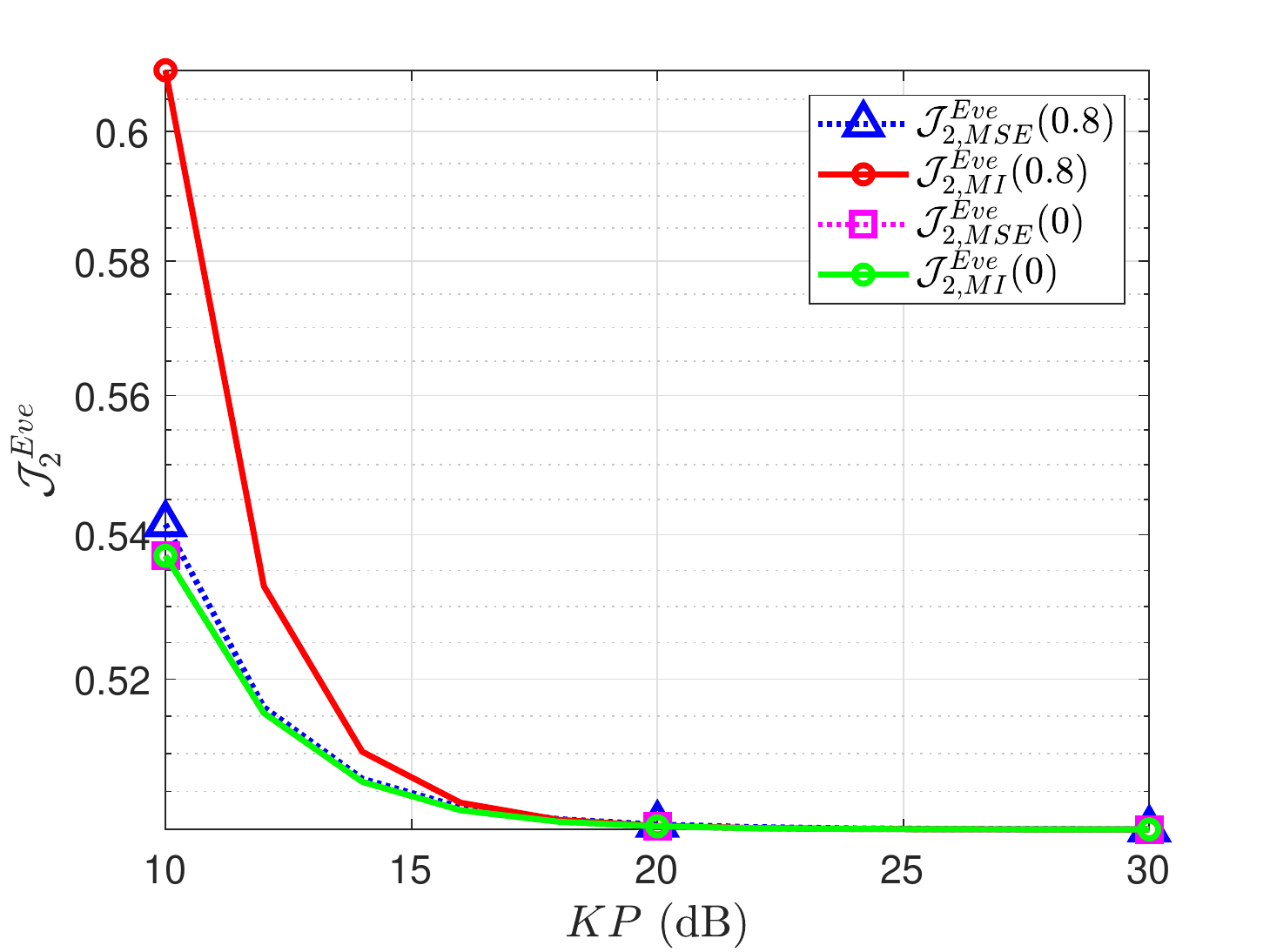}
	\centering
	\caption{Average normalized MSE for Eve vs $10\texttt{dB}\leq KP \leq 30\texttt{dB}$ with $M=2$.}
	\label{eve2}
\end{figure}

Finally, to show the corresponding normalized MSE at Eve for the two-user case, we use $\mathcal{J}^{Eve}_{2,MI}(R)$ for the pilots from \eqref{opt_mu_2} and $\mathcal{J}^{Eve}_{2,MSE}(R)$ for the pilots given by \cite{Bjornson2010}.
In Fig. \ref{eve2}, we show $\mathcal{J}^{Eve}_{2}(R)$ vs $10\texttt{dB}\leq KP \leq 30\texttt{dB}$. As expected, both $\mathcal{J}^{Eve}_{2,MI}(R)$ and $\mathcal{J}^{Eve}_{2,MSE}(R)$ get saturated to a significant constant as $KP$ increases.

\section{Conclusion}
We have developed algorithms for computing the optimal pilots for ANECE under MSE and MI criteria. Each channel matrix is modelled by a known correlation matrix and a matrix of i.i.d. complex Gaussian entries. While the logarithmic-barrier based gradient method was used to develop algorithms for more than two users, more efficient algorithms were developed for two users. Under a symmetric and isotropic condition, a closed-form expression of the optimal pilots was shown (in Theorems \ref{t1} and \ref{t2}) for both sum-MSE and sum-MI criteria. While this closed-form expression coincides with that proposed in \cite{Hua2019a} for three or more single-antenna users, this is a significant discovery for three or more multi-antenna users.  The general algorithms developed for three or more multi-antenna users are also significant contributions beyond the prior works shown in \cite{Bjornson2010} and \cite{Quist2015}.

We have shown that although the sum-MSE and sum-MI criteria yield the same optimal pilots under the symmetric and isotropic condition or under a lower transmit power condition, they do not yield the same optimal pilots in general  but each criterion yields a good sub-optimal solution for the other. In terms of computational complexity, the algorithms based on both criteria are nearly the same.

We should note however that although the optimal pilots developed in this paper meet the KKT conditions of non-convex problems and there is no other known design that performs better, the global optimality of the optimal pilots from this work is not yet established for most situations of three or more users. One strategy to prove the global optimality (if true) of the solutions in Theorems \ref{t1} and \ref{t2} is to find all solutions to the KKT conditions of the non-convex problems and rule out the possibility of better solutions. This is a challenge not yet met.
\appendix
%
\subsection{MMSE of Eve's CSI by Eve}\label{app:EVE_MSE}
In this section, we show that Eve cannot obtain a consistent estimate of its CSI by MMSE when users apply ANECE.
To simplify the analysis, we assume that the receive correlation matrix at Eve is the identity matrix  and $\mathbf{H}_{E,i}$ consists of i.i.d. $\mathcal{CN}(0,\sigma_{E,i}^{2})$ entries. Corresponding to the pilots sent by all users, the signal received by Eve as shown in \eqref{mul_sig_eve_2} can be rewritten as
\begin{equation}\label{eve_vec}
\mathbf{y}_{E} =\sum_{i=1}^{M}(\bar{\mathbf{P}}^{T}\bar{\mathbf{R}}^{\frac{1}{2}}\mathbf{S}_{i}^{T}\otimes \mathbf{I})\mathbf{h}_{E,i} + \mathbf{n}_{E}
\end{equation}
where $\mathbf{y}_{E} = vec(\mathbf{Y}_{E})$,  $\mathbf{h}_{E,i} = vec(\mathbf{H}_{E,i})$, $\mathbf{n}_{E} = vec(\mathbf{N}_{E})$ and $\mathbf{S}_{i} \in \mathbb{R}^{N_{i}\times N_{T}}$ is the selection matrix defined in section \ref{sec:MMSE}.

Since $\mathbf{h}_{E,i}$ for all $i$ are independent of each other and $\mathbf{h}_{E,i}$ has the covariance matrix $\sigma_{E,i}^2\mathbf{I}$,  Eve's MMSE of $\mathbf{h}_{E,i}$ is
\begin{align}
\hat{\mathbf{h}}_{E,i}&=\mathbf{K}_{\mathbf{h}_{E,i},\mathbf{y}_{E}}\mathbf{K}_{\mathbf{y}_{E}}^{-1}
\mathbf{y}_{E}\notag\\
&= \sigma^{2}_{E,i}(\mathbf{S}_{i}\bar{\mathbf{R}}^{\frac{H}{2}}\bar{\mathbf{P}}^{*}\otimes \mathbf{I})\big(\bar{\mathbf{P}}^{T}\bar{\mathbf{R}}^{\frac{1}{2}}\boldsymbol{\Sigma}_{E}
\bar{\mathbf{R}}^{\frac{H}{2}}\bar{\mathbf{P}}^{*}\otimes \mathbf{I} + \mathbf{I}\big)^{-1}\mathbf{y}_{E}\label{eve_esi}
\end{align}
where $\boldsymbol{\Sigma}_{E} = diag\{\sigma_{E,1}^{2}\mathbf{I}_{N_1},\dots,\sigma_{E,M}^{2}\mathbf{I}_{N_M}\}$. Then we  know that the covariance matrix of $\hat{\mathbf{h}}_{E,i}$ is
\begin{align}
\mathbf{K}_{\hat{\mathbf{h}}_{E,i}} &= \mathbf{K}_{\mathbf{h}_{E,i},\mathbf{y}_{E}}\mathbf{K}_{\mathbf{y}_{E}}^{-1}
\mathbf{K}_{\mathbf{h}_{E,i},\mathbf{y}_{E}}^H\notag\\ &=\sigma^{4}_{E,i}(\mathbf{S}_{i}\bar{\mathbf{R}}^{\frac{H}{2}}\bar{\mathbf{P}}^{*}\otimes \mathbf{I})\notag\\
&\quad\cdot\big(\bar{\mathbf{P}}^{T}\bar{\mathbf{R}}^{\frac{1}{2}}\boldsymbol{\Sigma}_{E}
\bar{\mathbf{R}}^{\frac{H}{2}}\bar{\mathbf{P}}^{*}\otimes \mathbf{I} + \mathbf{I}\big)^{-1}(\bar{\mathbf{P}}^{T}\bar{\mathbf{R}}^{\frac{1}{2}}\mathbf{S}_{i}^{T}\otimes \mathbf{I})\notag\\
& = \sigma^{4}_{E,i}(\mathbf{S}_{i}\boldsymbol{\Sigma}_{E}^{-\frac{1}{2}}\boldsymbol{\Phi}\boldsymbol{\Sigma}_{E}^{-\frac{1}{2}}\mathbf{S}_{i}^{T}\otimes \mathbf{I})\notag\\
&=\sigma^{2}_{E,i}(\mathbf{S}_{i}\boldsymbol{\Phi}\mathbf{S}_{i}^{T}\otimes \mathbf{I}) \label{Kh}
\end{align}
where
\begin{equation}\label{}
  \boldsymbol{\Phi} = \boldsymbol{\Sigma}_{E}^{\frac{1}{2}}\bar{\mathbf{R}}^{\frac{H}{2}}\bar{\mathbf{P}}^{*}
\left(\bar{\mathbf{P}}^{T}\bar{\mathbf{R}}^{\frac{1}{2}}\boldsymbol{\Sigma}_{E}
\bar{\mathbf{R}}^{\frac{H}{2}}\bar{\mathbf{P}}^{*}+\mathbf{I}\right)^{-1}\bar{\mathbf{P}}^{T}
\bar{\mathbf{R}}^{\frac{1}{2}}\boldsymbol{\Sigma}_{E}^{\frac{1}{2}}.
\end{equation}
Let $\hat{\mathbf{H}}_{E,i} = ivec(\hat{\mathbf{h}}_{E,i})$. It can be verified from \eqref{Kh}  that the $k$th and $l$th columns in $\hat{\mathbf{H}}_{E,i}$ are correlated and the elements in each column of $\hat{\mathbf{H}}_{E,i}$ are i.i.d. complex Gaussian. Because $rank(\boldsymbol{\Sigma}_{E}^{\frac{1}{2}}\bar{\mathbf{R}}^{\frac{H}{2}}\bar{\mathbf{P}}^{*}) = r < N_{T}$, the (thin) SVD of $\boldsymbol{\Sigma}_{E}^{\frac{1}{2}}\bar{\mathbf{R}}^{\frac{H}{2}}\bar{\mathbf{P}}^{*}$ can be expressed as $\boldsymbol{\Sigma}_{E}^{\frac{1}{2}}\bar{\mathbf{R}}^{\frac{H}{2}}\bar{\mathbf{P}}^{*} = \check{\mathbf{U}}[\check{\boldsymbol{\Lambda}}~\mathbf{0}_{r\times (K-r)}]\check{\mathbf{V}}^{H}$ where $\check{\mathbf{U}} \in \mathbb{C}^{N_{T}\times r}$, $\check{\boldsymbol{\Lambda}} \in \mathbb{R}^{r\times r}$ and $\check{\mathbf{V}} \in \mathbb{C}^{K \times K}$. It follows that
\begin{align}\label{}
  &\boldsymbol{\Phi} = \check{\mathbf{U}}[\check{\boldsymbol{\Lambda}}~\mathbf{0}_{r\times (K-r)}](diag(\check{\boldsymbol{\Lambda}}^{2},~ \mathbf{0}_{K-r}) + \mathbf{I})^{-1}\notag\\
  &\qquad \cdot [\check{\boldsymbol{\Lambda}}~\mathbf{0}_{r\times (K-r)}]^{T}\check{\mathbf{U}}^{H} \notag\\
  &= \check{\mathbf{U}}\check{\boldsymbol{\Lambda}}^{2}(\check{\boldsymbol{\Lambda}}^2 + \mathbf{I})^{-1}\check{\mathbf{U}}^{H}.
\end{align}
 It is known that $\Delta\mathbf{h}_{E,i} = \mathbf{h}_{E,i} - \hat{\mathbf{h}}_{E,i}$ has the covariance matrix
$\mathbf{K}_{\Delta\mathbf{h}_{E,i}} = \mathbf{K}_{\mathbf{h}_{E,i}} - \mathbf{K}_{\mathbf{h}_{E,i},\mathbf{y}_{E}}\mathbf{K}_{\mathbf{y}_{E}}^{-1}
\mathbf{K}_{\mathbf{y}_{E},\mathbf{h}_{E,i}}=\mathbf{K}_{\mathbf{h}_{E,i}} -\mathbf{K}_{\hat{\mathbf{h}}_{E,i}}$. Define the semi-unitary matrix  $\check{\mathbf{U}}_{n} \in \mathbb{C}^{N_{T}\times (N_{T} - r)}$ such that $\check{\mathbf{U}}_{n}^{H} \check{\mathbf{U}} = \mathbf{0}$. It follows that
\begin{align}
Tr(\mathbf{K}_{\Delta\mathbf{h}_{E,i}})
& = \sigma^{2}_{E,i}Tr\left((\mathbf{I} - (\mathbf{S}_{i}\boldsymbol{\Phi}\mathbf{S}_{i}^{T}\otimes \mathbf{I}))\right)\notag\\
& = \sigma^{2}_{E,i}Tr\left((\mathbf{S}_{i}(\mathbf{I} - \boldsymbol{\Phi})\mathbf{S}_{i}^{T}\otimes \mathbf{I}))\right)\notag\\
& =  \sigma^{2}_{E,i}N_{E}Tr\left(\mathbf{S}_{i}\check{\mathbf{U}}(\mathbf{I} - \check{\boldsymbol{\Lambda}}^{2}(\check{\boldsymbol{\Lambda}}^2 + \mathbf{I})^{-1})\check{\mathbf{U}}^{H}\mathbf{S}_{i}^{T}\right)\notag\\
&\quad + \sigma^{2}_{E,i}N_{E}Tr\left((\mathbf{S}_{i}\check{\mathbf{U}}_{n}\check{\mathbf{U}}_{n}^{H}\mathbf{S}_{i}^{T}\right).\label{MSE2}
\end{align}
%
%
From the definition of $\check{\boldsymbol{\Lambda}}$ shown above, we know that each element in $\check{\boldsymbol{\Lambda}}$ is propositional to the total transmit power $P_{T}$. Therefore, the first term in \eqref{MSE2} reduces to zero as $P_{T}$ increases. But the second term in \eqref{MSE2} is independent of $P_T$. In general, $\mathbf{S}_{i}\check{\mathbf{U}}_{n} \neq \mathbf{0}$ given $r<N_T$, and hence Eve is unable to obtain a consistent estimate of $\mathbf{h}_{E,i}$ for any $i$.
%

\subsection{Proof of Theorem \ref{t1}}\label{app_mse}
From \eqref{Qc}, the $(l+1,k+1)$th element of $\mathbf{Q}_m\mathbf{Q}_m^{H}$ is
\begin{align}
(\mathbf{Q}_m\mathbf{Q}_m^{H})_{l+1,k+1} &= \sum_{n=0}^{N-1}e^{-j2\pi\frac{(l-k)(m+ nM)}{NM}}\notag\\
& = e^{-j2\pi\frac{(l-k)m}{NM}}\sum_{n=0}^{N-1}e^{-j2\pi\frac{(l-k)n}{N}}\notag\\
& = \left\{
\begin{aligned}
&0,&&|l-k| \neq vN\\
&Ne^{-j2\pi\frac{(l-k)m}{NM}},&&|l-k| = vN
\end{aligned}
\right.\label{Fc_e}
\end{align}
where $v$ is an integer satisfying $0\leq v\leq M-1$. From \eqref{Fc_e}, we know that there are only $M$ non-zero elements on each column or row of $\mathbf{Q}_m\mathbf{Q}_m^{H}$. More specifically, using $w_M=e^{-j2\pi\frac{1}{M}}$, we have
\begin{align}
&\mathbf{Q}_m\mathbf{Q}_m^{H}\notag\\
&=N\begin{bmatrix}
1 & w_{M}^{-m}&\cdots & w_{M}^{-(M-1)m}\\
w_{M}^{m} &1 &\cdots&w_{M}^{-(M-2)m}\\
\vdots &\vdots&\ddots&\vdots\\
w_{M}^{(M-1)m}& w_{M}^{(M-2)m}&\cdots &1
\end{bmatrix}\otimes \mathbf{I}_{N}\\
& = N\mathbf{q}_m\mathbf{q}_m^{H}\otimes \mathbf{I}_{N}\label{q}
\end{align}
where $\mathbf{q}_m = [1, w_{M}^{m}, \dots, w_{M}^{(M-1)m}]^{T}$. Since $\mathbf{Q}_m^{H}\mathbf{\bar Q}_m = 0$, we have $(\mathbf{q}_m\mathbf{q}_m^H\otimes \mathbf{I}_N)\mathbf{\bar Q}_m = 0$.

For $N_i=N$, we have $\bar{\mathbf{S}}_{(i)} = \mathbf{I}_{M,i}\otimes \mathbf{I}_{N}$ where $\mathbf{I}_{M,i}$ $\mathbf{I}_{M}$ without its $i$th row, and $\mathbf{S}_{i} = \mathbf{e}_{i}^{T}\otimes \mathbf{I}_{N},i = 1,\dots,M$ where $\mathbf{e}_{i}$ is the $M\times 1$ vector with its $i$th element equal to one.
Now assume $\bar{\mathbf{F}} = \sqrt{\alpha_{d}}\mathbf{\bar Q}_m$. Then $\bar{\mathbf{F}}\bar{\mathbf{F}}^{H} =\alpha_{d}\mathbf{\bar Q}_m\mathbf{\bar Q}_m^{H} =  \alpha_{d}(MN\mathbf{I}_{MN} -\mathbf{Q}_m\mathbf{Q}_m^H)= \alpha_{d}(MN\mathbf{I}_{MN} - N\mathbf{q}_m\mathbf{q}_m^{H}\otimes\mathbf{I}_{N}) = \alpha_{d}(MN\mathbf{I}_{M} - N\mathbf{q}_m\mathbf{q}_m^{H})\otimes\mathbf{I}_{N}$, and
{\small
\begin{equation}\label{b_inv}
\begin{aligned}
&(\mathbf{I}_{(M-1)N} +  \bar{\mathbf{S}}_{(i)}\bar{\mathbf{F}}\bar{\mathbf{F}}^{H}\bar{\mathbf{S}}_{(i)}^{T})^{-1}\\
& = [\mathbf{I}_{(M-1)N} +  \alpha_{d}  (\mathbf{I}_{M,i}\otimes\mathbf{I}_N) (NM\mathbf{I} - N\mathbf{q}_m\mathbf{q}_m^{H}\otimes \mathbf{I}_{N})\\
&\quad\cdot (\mathbf{I}_{M,i}^{T}\otimes \mathbf{I}_N)]^{-1}\\
& = ((1 + NM\alpha_{d})\mathbf{I}_{(M-1)N} - N\alpha_{d}  \big(\mathbf{I}_{M,i}\mathbf{q}_m\mathbf{q}_m^{H}\mathbf{I}_{M,i}^{T}\big)\otimes \mathbf{I}_{N})^{-1}\\
& = \frac{\big(\mathbf{I}_{M-1} - \frac{N\alpha_{d}}{1 + NM\alpha_{d}}  \mathbf{I}_{M,i}\mathbf{q}_m\mathbf{q}_m^{H}\mathbf{I}_{M,i}^{T}\big)^{-1}\otimes\mathbf{I}_{N}}{1 + NM\alpha_{d}}\\
& = \frac{\big(\mathbf{I}_{M-1} + \frac{N\alpha_{d}}{1 + N\alpha_{d}}  \mathbf{I}_{M,i}\mathbf{q}_m\mathbf{q}_m^{H}\mathbf{I}_{M,i}^{T}\big)\otimes\mathbf{I}_{N}}{1 + NM\alpha_{d}}
\end{aligned}
\end{equation}
}
where the last equality in \eqref{b_inv} is based on $(\mathbf{I} + \bar{\mathbf{x}}\mathbf{y}^{H})^{-1} = \mathbf{I} - \frac{1}{1 + \mathbf{y}^{H}\bar{\mathbf{x}}}\bar{\mathbf{x}}\mathbf{y}$ and $\mathbf{q}_m^{H}\mathbf{I}_{M,i}^{T}\mathbf{I}_{M,i}\mathbf{q}_m = M-1$.

Without loss of generality, we now set $\sigma^2=1$ since $P$ can be any positive number. Then from \eqref{d1} and the conditions of the theorem, we have
\begin{equation}\label{de_J}
\frac{\partial J_{M}}{\partial \bar{\mathbf{F}}} =  -2N\sum_{i=1}^{M}\bar{\mathbf{S}}_{(i)}^{T}\big(\mathbf{I} +   \bar{\mathbf{S}}_{(i)}\bar{\mathbf{F}}\bar{\mathbf{F}}^{H}\bar{\mathbf{S}}_{(i)}^{T}\big)^{-2}
\bar{\mathbf{S}}_{(i)}\bar{\mathbf{F}}
\end{equation}
where, using \eqref{b_inv}, we have
\begin{equation}\label{sum_inv}
\begin{aligned}
&\sum_{i=1}^{M}\bar{\mathbf{S}}_{(i)}^{T}\big(\mathbf{I}_{(M-1)N} +   \bar{\mathbf{S}}_{(i)}\bar{\mathbf{F}}\bar{\mathbf{F}}^{H}\bar{\mathbf{S}}_{(i)}^{T}\big)^{-2}\bar{\mathbf{S}}_{(i)}\\
& = \frac{\sum_{i=1}^{M}\bar{\mathbf{S}}_{(i)}^{T}\big((\mathbf{I}_{M-1} +  \frac{N\alpha_{d}}{1  + N\alpha_{d}}\mathbf{I}_{M,i}\mathbf{q}_m\mathbf{q}_m^{H}\mathbf{I}_{M,i}^{T})^{2}\otimes \mathbf{I}_{N}\big)\bar{\mathbf{S}}_{(i)}}{(1 + NM\alpha_{d}  )^{2}}\\
& = \frac{\sum_{i=1}^{M}\bar{\mathbf{S}}_{(i)}^{T}\big((\mathbf{I}_{M-1} + \beta\mathbf{I}_{M,i}\mathbf{q}_m\mathbf{q}_m^{H}\mathbf{I}_{M,i}^{T})\otimes \mathbf{I}_{N}\big)\bar{\mathbf{S}}_{(i)}}{(1 + NM\alpha_{d}  )^{2}}\\
& =  \frac{\sum_{i=1}^{M}\big(\mathbf{I}_{M,i}^{T}\mathbf{I}_{M,i} + \beta\mathbf{I}_{M,i}^{T}\mathbf{I}_{M,i}\mathbf{q}_m\mathbf{q}_m^{H}\mathbf{I}_{M,i}^{T}\mathbf{I}_{M,i}\big)\otimes \mathbf{I}_{N}}{(1 + NM\alpha_{d}  )^{2}}\\
& =  \frac{\big((M-1 + \beta)\mathbf{I}_{M} + \beta(M-2)\mathbf{q}_m\mathbf{q}_m^{H}  \big)\otimes \mathbf{I}_{N}}{(1 + NM\alpha_{d}  )^{2}}
\end{aligned}
\end{equation}
where $\beta = \frac{2N\alpha_{d}(1+N\alpha_{d}) + N^2\alpha_{d}^2(M-1)}{(1 + N\alpha_{d})^2} >0$. The last equality in \eqref{sum_inv} has used $\sum_{i=1}^{M}\mathbf{I}_{M,i}^{T}\mathbf{I}_{M,i} = (M-1)\mathbf{I}_{M}$ and
\begin{equation}\label{qq_sum}
\sum_{i=1}^{M}\mathbf{I}_{M,i}^{T}\mathbf{I}_{M,i}\mathbf{q}_m\mathbf{q}_m^{H}\mathbf{I}_{M,i}^{T}\mathbf{I}_{M,i}\ = \mathbf{I}_{M} + (M-2)\mathbf{q}_m\mathbf{q}_m^{H}.
\end{equation}
Using $(\mathbf{q}_m\mathbf{q}_m^{H}\otimes \mathbf{I}_{N})\bar{\mathbf{F}} = \mathbf{Q}_m^{H}\mathbf{\bar Q}_m = 0$, \eqref{de_J} and \eqref{sum_inv} yield
\begin{equation}\label{dft_}
\nabla J_M = -2N\frac{(M-1 + \beta)}{(1 + NM\alpha_{d}  )^{2}}\bar{\mathbf{F}}.
\end{equation}
Also note that $\sum_{i=1}^M\mathbf{S}_i^T\mathbf{S}_i = (\sum_{i=1}^M \mathbf{e}_i\mathbf{e}_i^T)\otimes \mathbf{I}_N = \mathbf{I}_M\otimes \mathbf{I}_N=\mathbf{I}_{MN}$. Therefore, the first KKT condition in \eqref{kkt_mul_sc} is satisfied by $\mu_{i} = \frac{N(M-1 + \beta)}{(1 + NM\alpha_{d}  )^{2}} >0$, and all the other KKT conditions are satisfied by
 $\alpha_{d} = \frac{KP}{ N^2(M-1)}$. Therefore, $\bar{\mathbf{F}} =\sqrt{\frac{KP}{ N^2(M-1)}}\mathbf{\bar Q}_m$ is a solution to \eqref{kkt_mul_sc}.

\subsection{The gradient of $g_2(\bar{\mathbf{F}})$ in \eqref{barrier_key}}\label{sec:ap:skr}
It follows from \eqref{barrier_key} that $\nabla g_2(\bar{\mathbf{F}})=t\sum_{i=1}^{M-1}\sum_{j=i+1}^M \nabla \log_2|\mathbf{I}-\boldsymbol{\Gamma}_{i,j}\boldsymbol{\Gamma}_{T,j,i}|+\sum_{i=1}^M \nabla \mathcal{B}_i(\bar{\mathbf{F}})$. Here, $\nabla \mathcal{B}_i(\bar{\mathbf{F}})$ is given by \eqref{de_b_u}. To show $\nabla \log_2|\mathbf{I}-\boldsymbol{\Gamma}_{i,j}\boldsymbol{\Gamma}_{T,j,i}|$, we first consider
\begin{align}
&\nabla \log_{2}|\mathbf{I} - \boldsymbol{\Gamma}_{i,j}\boldsymbol{\Gamma}_{T,j,i}|\notag\\
& = -\frac{1}{\ln 2\partial \bar{\mathbf{F}}}Tr\left(\boldsymbol{\Gamma}_{T,j,i}(\mathbf{I} - \boldsymbol{\Gamma}_{i,j}\boldsymbol{\Gamma}_{T,j,i})^{-1}\partial \boldsymbol{\Gamma}_{i,j}\right)\notag\\
&~  - \frac{1}{\ln 2 \partial \bar{\mathbf{F}}}Tr\left((\mathbf{I} - \boldsymbol{\Gamma}_{i,j}\boldsymbol{\Gamma}_{T,j,i})^{-1}\boldsymbol{\Gamma}_{i,j}\partial \boldsymbol{\Gamma}_{T,j,i}\right)\label{mul_deri}
\end{align}
where we have applied $\partial \ln |\mathbf{X}| = Tr(\mathbf{X}^{-1}\partial \mathbf{X})$, $\partial (\mathbf{X}\mathbf{Y})=\partial\mathbf{X}\cdot \mathbf{Y}+ \mathbf{X}\cdot \partial \mathbf{Y}$ and $Tr(\mathbf{X}\mathbf{Y})=Tr(\mathbf{Y}\mathbf{X})$.

Using the matrix inverse lemma, \eqref{gammaij.1} can be rewritten as
\begin{align}
&\boldsymbol{\Gamma}_{i,j}\notag\\
& = \frac{1}{\sigma^{2}_{i}}(\mathbf{S}_{j}\bar{\mathbf{F}}\bar{\mathbf{F}}^{H}\mathbf{S}_{j}^{T})\otimes \tilde{\boldsymbol{\Lambda}}_{i} - \frac{1}{\sigma^{4}_{i}}((\mathbf{S}_{j}\bar{\mathbf{F}}\bar{\mathbf{F}}^{H}
\bar{\mathbf{S}}_{(i)}^{T})\otimes \tilde{\boldsymbol{\Lambda}}_{i})\notag\\
&\quad \cdot (\mathbf{I} + \frac{1}{\sigma^{2}_{i}}\bar{\mathbf{S}}_{(i)}\bar{\mathbf{F}}
\bar{\mathbf{F}}^{H}\bar{\mathbf{S}}_{(i)}^{T}\otimes \tilde{\boldsymbol{\Lambda}}_{i})^{-1}((\bar{\mathbf{S}}_{(i)}
\bar{\mathbf{F}}\bar{\mathbf{F}}^{H}\mathbf{S}_{j}^{T})\otimes \tilde{\boldsymbol{\Lambda}}_{i})
\label{gammaij.2}
\end{align}
where each factor or term is a function of $\bar{\mathbf{F}}\bar{\mathbf{F}}^{H}$, which is useful to simplify the gradient expressions. For example, with respect to the complex matrix $\mathbf{X}$, $\nabla Tr(\mathbf{A}\mathbf{X}\mathbf{X}^H\mathbf{B})=2\mathbf{B}\mathbf{A}\mathbf{X}$.
Let $\mathbf{T}_{i,j}$ be such a permutation matrix that $\mathbf{T}_{i,j}^T[(\mathbf{S}_{j}\bar{\mathbf{F}}\bar{\mathbf{F}}^{H}\mathbf{S}_{j}^{T})\otimes \tilde{\boldsymbol{\Lambda}}_{i}]\mathbf{T}_{i,j}=\tilde{\boldsymbol{\Lambda}}_{i}\otimes
(\mathbf{S}_{j}\bar{\mathbf{F}}\bar{\mathbf{F}}^{H}\mathbf{S}_{j}^{T})$.
Also define $\tilde{\boldsymbol{\Gamma}}_{i,j} =\mathbf{T}_{i,j}^{T}\boldsymbol{\Gamma}_{T,j,i}(\mathbf{I} - \boldsymbol{\Gamma}_{i,j}\boldsymbol{\Gamma}_{T,j,i})^{-1}\mathbf{T}_{i,j}$. Then, one can verify (after a slightly tedious process) that the first term in \eqref{mul_deri} can be written as (without the coefficient $1/\ln 2$):
\begin{equation}\label{mul_deri_2}
\begin{aligned}
&\frac{1}{\partial \bar{\mathbf{F}}}Tr\left(\mathbf{T}_{i,j}\tilde{\boldsymbol{\Gamma}}_{i,j}\mathbf{T}_{i,j}^{T}\partial \boldsymbol{\Gamma}_{i,j}\right) = 2\big(\boldsymbol{\Gamma}_{i,j}^{(0)} - \boldsymbol{\Gamma}_{i,j}^{(1)}+\boldsymbol{\Gamma}_{i,j}^{(2)}- \boldsymbol{\Gamma}_{i,j}^{(3)}\big)\bar{\mathbf{F}}
\end{aligned}
\end{equation}
where
\begin{equation}\label{mul_d1}
\boldsymbol{\Gamma}_{i,j}^{(0)} = \sum_{l=1}^{N_{i}}\frac{\tilde{\lambda}_{i,l}}{\sigma_{i}^{2}}\mathbf{S}_{j}^{T}(\tilde{\boldsymbol{\Gamma}}_{i,j})_{l}\mathbf{S}_{j},
\end{equation}
\begin{align}
\boldsymbol{\Gamma}_{i,j}^{(1)}&= \sum_{l=1}^{N_{i}}\frac{\tilde{\lambda}_{i,l}^2}{\sigma_{i}^{4}}
\bar{\mathbf{S}}_{(i)}^{T}(\mathbf{I} + \frac{\tilde{\lambda}_{i,l}}{\sigma^{2}_{i}}\bar{\mathbf{S}}_{(i)}\bar{\mathbf{F}}\bar{\mathbf{F}}^{H}
\bar{\mathbf{S}}_{(i)}^{T} )^{-1}\notag\\
&\quad\cdot  \bar{\mathbf{S}}_{(i)}\bar{\mathbf{F}}\bar{\mathbf{F}}^{H}\mathbf{S}_{j}^{T}
(\tilde{\boldsymbol{\Gamma}}_{i,j})_{l}\mathbf{S}_{j},\label{mul_d2}
\end{align}

\begin{align}
&\boldsymbol{\Gamma}_{i,j}^{(2)} =\sum_{l=1}^{N_{i}}\frac{\tilde{\lambda}_{i,l}^3}{\sigma_{i}^{6}}\bar{\mathbf{S}}_{(i)}^{T}
(\mathbf{I} + \frac{\tilde{\lambda}_{i,l}}{\sigma^{2}_{i}}\bar{\mathbf{S}}_{(i)}\bar{\mathbf{F}}
\bar{\mathbf{F}}^{H}\bar{\mathbf{S}}_{(i)}^{T} )^{-1} \bar{\mathbf{S}}_{(i)}\bar{\mathbf{F}}\bar{\mathbf{F}}^{H}\mathbf{S}_{j}^{T}\notag\\
&\cdot(\tilde{\boldsymbol{\Gamma}}_{i,j})_{l}\mathbf{S}_{j}
\bar{\mathbf{F}}\bar{\mathbf{F}}^{H}\bar{\mathbf{S}}_{(i)}^{T}(\mathbf{I} + \frac{\tilde{\lambda}_{i,l}}{\sigma^{2}_{i}}\bar{\mathbf{S}}_{(i)}\bar{\mathbf{F}}\bar{\mathbf{F}}^{H}
\bar{\mathbf{S}}_{(i)}^{T})^{-1}\bar{\mathbf{S}}_{(i)},\label{mul_d3}
\end{align}
\begin{align}
\boldsymbol{\Gamma}_{i,j}^{(3)}&= \sum_{l=1}^{N_{i}}\frac{\tilde{\lambda}_{i,l}^2}{\sigma_{i}^{4}}
\mathbf{S}_{j}^{T}(\tilde{\boldsymbol{\Gamma}}_{i,j})_{l}
\mathbf{S}_{j}\bar{\mathbf{F}}\bar{\mathbf{F}}^{H}\bar{\mathbf{S}}_{(i)}^{T}\notag \\
& \quad\cdot (\mathbf{I} + \frac{\tilde{\lambda}_{i,l}}{\sigma^{2}_{i}}\bar{\mathbf{S}}_{(i)}\bar{\mathbf{F}}
\bar{\mathbf{F}}^{H}\bar{\mathbf{S}}_{(i)}^{T} )^{-1}\bar{\mathbf{S}}_{(i)}\label{mul_d4}
\end{align}
and $(\tilde{\boldsymbol{\Gamma}}_{i,j})_{l}$ is the $l$th $N_{j}\times N_{j}$ diagonal block of $\tilde{\boldsymbol{\Gamma}}_{i,j}$.

 A similar procedure can be applied to obtain the corresponding (explicit) expression of the second term in \eqref{mul_deri}. The details are omitted here.

\subsection{Proof of Lemma \ref{lemma3}}\label{app:lemma3}
To prove \eqref{lm1.l1}, we start with \eqref{lm1.l} which can rewritten as
	\begin{equation}\label{deKProof}
	\begin{aligned}
	&|\mathbf{A}\otimes \mathbf{B} + \mathbf{C}\otimes\mathbf{D}|\geq  \min_{P_{1},P_{2}}\prod_{k=1}^{m}\prod_{l=1}^{n}(\lambda_{a,l}\lambda_{b,k} + \lambda_{c,P_{1},l}\lambda_{d,P_{2},k})
	\end{aligned}
	\end{equation}
where $\lambda_{a,l}$ is the $l$th diagonal element of $\boldsymbol{\Lambda}_{a}$, and $\lambda_{b,k}$, $\lambda_{c,P_{1},l}$ and $\lambda_{d,P_{2},k}$ are defined similarly. Every permutation of the diagonal elements of a diagonal matrix can be represented by a sequence of pair-wise permutations (each involving two diagonal elements). To prove \eqref{lm1.l1}, we only need to prove that (1) for every pair of diagonal elements of $\boldsymbol{\Lambda}_{a}$ (which are descending) the corresponding pair of diagonal elements of $\boldsymbol{\Lambda}_{c,P_1}$ must be descending to minimize the right side of \eqref{deKProof}, and (2) for every pair of $\boldsymbol{\Lambda}_{b}$ (which are descending) the corresponding pair of diagonal elements of $\boldsymbol{\Lambda}_{d,P_2}$ must be descending to minimize the right side of \eqref{deKProof}. The proofs of the above two statements are virtually the same. So, we only need to prove the first.

	Let $\lambda_{c,P_1,s}$ and $\lambda_{c,P_1,l}$ be two diagonal elements in $\boldsymbol{\Lambda}_{c,P_{1}}$ where $s < l$ and $\lambda_{c,P_1,s}\geq \lambda_{c,P_1,l}$ (descending). Let $P'_1$ be another permutation that differs from $P_1$ only for these two elements, i.e., $\lambda_{c,P'_1,s}\leq \lambda_{c,P'_1,l}$ (ascending), $\lambda_{c,P_1,s}= \lambda_{c,P'_1,l}$ and $\lambda_{c,P_1,l}= \lambda_{c,P'_1,s}$. To compare the two permutations $P_1$ and $P'_1$, we only need to compare the two factors in \eqref{deKProof} that are affected from $P_1$ to $P'_1$. The difference between the products of the two factors  is
	\begin{equation}
	\begin{aligned}
	&(\lambda_{a,s}\lambda_{b,k} + \lambda_{c,P_1,s}\lambda_{d,P_{2},k})(\lambda_{a,l}\lambda_{b,k} + \lambda_{c,P_1,l}\lambda_{d,P_{2},k})\\
	&\quad - 	(\lambda_{a,s}\lambda_{b,k} + \lambda_{c,P'_1,s}\lambda_{d,P_{2},k})(\lambda_{a,l}\lambda_{b,k} + \lambda_{c,P'_1,l}\lambda_{d,P_{2},k})\\
	& = \lambda_{a,s}\lambda_{b,k}\lambda_{c,P_1,l}\lambda_{d,P_{2},k} + \lambda_{c,P_1,s}\lambda_{d,P_{2},k}\lambda_{a,l}\lambda_{b,k}\\
	&\quad - \lambda_{a,s}\lambda_{b,k}\lambda_{c,P'_1,l}\lambda_{d,P_{2},k} - \lambda_{c,P'_1,s}\lambda_{d,P_{2},k}\lambda_{a,l}\lambda_{b,k}\\
	& = \lambda_{d,P_{2},k}\lambda_{b,k}(\lambda_{a,s} - \lambda_{a,l})(\lambda_{c,P_1,l} - \lambda_{c,P_1,s}) \leq 0.
	\end{aligned}
	\end{equation}
	This proves the first statement. The second statement can be proved similarly.  Hence \eqref{lm1.l1} is proven.

The proof of \eqref{lm1.u1} can be done in a similar manner.

\subsection{Proof of Theorem \ref{t4}}\label{proof_t4}
Define $\check{c}_{1,l}=\frac{c_{1,l}}{KP}$ and $\check{c}_{2,k}=\frac{c_{2,k}}{KP}$. Then, the power constraints become
$\sum_{l= 1}^{N_{1}}\check{c}_{1,l} =  1$ and $\sum_{k= 1}^{N_{2}}\check{c}_{2,k}= 1$. And \eqref{Ikey} now becomes
\begin{equation}\label{Ip}
{\small
\begin{aligned}
&I_{2}=\\
&\sum_{k=1}^{N_{2}}\sum_{l=1}^{N_{1}}\log_{2}\bigg(\frac{(\sigma^{2}_{2} + KP\tilde{\lambda}_{1,l}\tilde{\lambda}_{2,k}\check{c}_{1,l})(\sigma^{2}_{1} +  KP\tilde{\lambda}_{1,l}\tilde{\lambda}_{2,k}\check{c}_{2,k})}{\sigma^{2}_{1}\sigma^{2}_{2} + KP\sigma^{2}_{1}\tilde{\lambda}_{1,l}\tilde{\lambda}_{2,k}\check{c}_{1,l} + KP\sigma^{2}_{2}\tilde{\lambda}_{1,l}\tilde{\lambda}_{2,k}\check{c}_{2,k}}\bigg).
\end{aligned}}
\end{equation}
\paragraph{High Power Case}
For large $P$, \eqref{Ip} can be approximated as
\begin{equation}\label{Ihp}
\begin{aligned}
&I_{2}\\
&\approx \sum_{k=1}^{N_{2}}\sum_{l=1}^{N_{1}}\log_{2}(\frac{   KP\tilde{\lambda}_{1,l}\tilde{\lambda}_{2,k}\check{c}_{1,l}\check{c}_{2,k}}{\sigma_{1}^{2}
\check{c}_{1,l} + \sigma_{2}^{2}\check{c}_{2,k}})\\
& = \sum_{k=1}^{N_{2}}\sum_{l=1}^{N_{1}}\log_{2}(\frac{\check{c}_{1,l}\check{c}_{2,k}}{\sigma_{1}^{2}
\check{c}_{1,l} + \sigma_{2}^{2}\check{c}_{2,k}}) + \sum_{k=1}^{N_{2}}\sum_{l=1}^{N_{1}}\log_{2}(KP\tilde{\lambda}_{1,l}\tilde{\lambda}_{2,k})\\
& \triangleq \phi_{1}(\check{\mathbf{c}}_{1},\check{\mathbf{c}}_{2},\boldsymbol{\tilde{\lambda}}_{1},
\boldsymbol{\tilde{\lambda}}_{2}).
\end{aligned}
\end{equation}

From \eqref{Ihp}, we know that the degrees of freedom per channel realization is  $\lim_{P\rightarrow\infty}\frac{\phi_{1}(\check{\mathbf{c}}_{1},\check{\mathbf{c}}_{2},
\boldsymbol{\tilde{\lambda}}_{1},\boldsymbol{\tilde{\lambda}}_{2})}{\log_{2}P} = N_{1}N_{2}$.

Also, $-\frac{\partial^{2} \phi_{1}}{\partial \check{c}_{1,l}^{2}} = -\sum_{j}(\frac{\sigma_{1}^{4}}{(\sigma_{1}^{2}\check{c}_{1,l} + \sigma_{2}^{2}\check{c}_{2,k})^{2}} - \frac{1}{\check{c}_{1,l}^{2}}) \geq 0$, which means that $-\phi_{1}$ is a convex function of $\check{\mathbf{c}}_{1}$. Meanwhile, $-\phi_{1}$ is a symmetric function of $\check{\mathbf{c}}_{1}$. Therefore,
$\phi_{1}$ is a Schur-concave function \cite{Marshalla} of $\check{\mathbf{c}}_{1}$, and then we have $\phi_{1}(\mathbf{1}_{N_{1}},\check{\mathbf{c}}_{2},\boldsymbol{\tilde{\lambda}}_{1},
\boldsymbol{\tilde{\lambda}}_{2}) \geq \phi_{1}(\check{\mathbf{c}}_{1},\check{\mathbf{c}}_{2},\boldsymbol{\tilde{\lambda}}_{1},
\boldsymbol{\tilde{\lambda}}_{2})$ with any  $\check{\mathbf{c}}_{1}$ of descending elements. Similar idea can be applied to show that \eqref{Ihp} is also a Schur-concave function of $\check{\mathbf{c}}_{2}$. Therefore, the optimal power allocation in the  high power case is such that $\check{\mathbf{c}}_{1} = \frac{1}{N_{1}}\mathbf{1}_{N_{1}}$ and $\check{\mathbf{c}}_{2} = \frac{1}{N_{2}}\mathbf{1}_{N_{2}}$.

Also, by applying the same argument,  one can easily prove that \eqref{Ihp} is also a Schur-concave function of $\boldsymbol{\tilde{\lambda}}_{1}$ and $\boldsymbol{\tilde{\lambda}}_{2}$ respectively. Therefore, when $\boldsymbol{\tilde{\lambda}}_{1} = \mathbf{1}_{N_{1}}$ and $\boldsymbol{\tilde{\lambda}}_{2} = \mathbf{1}_{N_{2}}$, \eqref{Ihp} is maximized. In other words, in the high power case, less correlated channel yields a higher secret key rate.

\paragraph{Low Power Case}

For small $P$, we can approximate \eqref{Ip} by its second-order Taylor series expansion at point $P = 0$:
\begin{equation}\label{lp}
\begin{aligned}
I_{2} &= I_{2}|_{P = 0} + \nabla I_{2}|_{P = 0}P
+ \frac{1}{2}\nabla^{2}I_{2}|_{P = 0}P^{2}
+ o(P^{2})
\end{aligned}
\end{equation}
where $\nabla I_{2}$ and $\nabla^{2} I_{2}$ are the first and second order derivatives of \eqref{Ip} with respect to $P$.  It can be easily proved that $\nabla I_{2}|_{P = 0} = 0$ and
\begin{equation}\label{lp1}
\begin{aligned}
&\nabla^{2} I_{2}|_{P=0}\\
& = \frac{2}{\ln 2}\sum_{l=1}^{N_{1}}\sum_{k=1}^{N_{2}}\tilde{\lambda}_{1,l}^{2}\tilde{\lambda}_{2,k}^{2}K^{2}
\check{c}_{1,l}\check{c}_{2,k} \triangleq\phi_{2}(\check{\mathbf{c}}_{1},\check{\mathbf{c}}_{2},
\boldsymbol{\tilde{\lambda}}_{1},\boldsymbol{\tilde{\lambda}}_{2}).
\end{aligned}
\end{equation}
To maximize \eqref{lp}, we just need to maximize the term \eqref{lp1}.
Based on \eqref{lp1} we have $\frac{\partial \phi_{2} }{\partial \check{c}_{1,l}} =K^{2}\tilde{\lambda}_{1,l}^{2}\sum_{j=1}^{N_{2}}\tilde{\lambda}_{2,k}^{2}\check{c}_{2,k} $. Since $\{\tilde{\lambda}_{1,l}\}$ is in descending order,
we know that $\phi_{2}(\check{\mathbf{c}}_{1},\check{\mathbf{c}}_{2},\boldsymbol{\tilde{\lambda}}_{1},\boldsymbol{\tilde{\lambda}}_{2})$ is a Schur-convex function of  $\check{\mathbf{c}}_{1}$ with descending entries, which means it is maximized by putting almost all of the power to $\check{c}_{1,1}$. The reason that ``almost all'' instead of ``all'' is used here is to ensure the positive condition on $\mathbf{c}_a$.   The same conclusion can be drawn about $\check{c}_{2,1}$ for maximizing $\phi_{2}(\check{\mathbf{c}}_{1},\check{\mathbf{c}}_{2},\boldsymbol{\tilde{\lambda}}_{1},
\boldsymbol{\tilde{\lambda}}_{2})$. That is, in the low power case, almost all of the power should be allocated to the strongest stream.

It is also clear that $\phi_{2}(\check{\mathbf{c}}_{1},\check{\mathbf{c}}_{2},\boldsymbol{\tilde{\lambda}}_{1},
\boldsymbol{\tilde{\lambda}}_{2})$ is a Schur-convex function of $\boldsymbol{\tilde{\lambda}}_{1}$ and $\boldsymbol{\tilde{\lambda}}_{2}$ individually. Therefore, in low power region, a higher channel correlation leads to a higher secret key rate.
\subsection{Proof of Theorem \ref{t2}}\label{app_skr}

Refer to Appendix \ref{app_mse}. Assume $\bar{\mathbf{F}} = \sqrt{\alpha_{d}}\mathbf{\bar Q}_m$. With \eqref{q}, the first term of $\boldsymbol{\Gamma}_{i,j}$ in \eqref{gammaij_sc} can be written as
\begin{align}
&\mathbf{S}_{j}\bar{\mathbf{F}}\bar{\mathbf{F}}^{H}\mathbf{S}_{j}^{T}\otimes\mathbf{I}_{N}\notag\\
& =\alpha_{d}(\mathbf{e}_{j}^{T}(MN\mathbf{I}_{M} - N\mathbf{q}_m\mathbf{q}_m^{H})\mathbf{e}_{j})\otimes\mathbf{I}_{N^2}\notag\\
& = \alpha_{d} (M-1)N\mathbf{I}_{N^2}.\label{pair_ij_1}
\end{align}
With \eqref{b_inv}, the second term of $\boldsymbol{\Gamma}_{i,j}$ in \eqref{gammaij_sc} becomes
{\small
\begin{equation}\label{pair_ij_2b}
	\begin{aligned}
	&\big((\mathbf{S}_{j}\bar{\mathbf{F}}\bar{\mathbf{F}}^{H}\bar{\mathbf{S}}_{(i)}^{T})
(\mathbf{I}_{(M-1)N} + \bar{\mathbf{S}}_{(i)}\bar{\mathbf{F}}\bar{\mathbf{F}}^{H}\bar{\mathbf{S}}_{(i)}^{T})^{-1}
(\bar{\mathbf{S}}_{(i)}\bar{\mathbf{F}}\bar{\mathbf{F}}^{H}\mathbf{S}_{j}^{T})\big)\otimes
\mathbf{I}_{N}\\
	& = \alpha_{d}^{2}\bigg(\big((\mathbf{e}_{j}^{T}(MN\mathbf{I}_{M} - N\mathbf{q}_m\mathbf{q}_m^{H})\bar{\mathbf{S}}_{(i)}^{T})\otimes\mathbf{I}_{N}\big)\\
	&\quad \cdot \big(\frac{\big(\mathbf{I}_{M-1} +  \frac{N\alpha_{d}}{1 + N\alpha_{d}  }\mathbf{I}_{M,i}\mathbf{q}_m\mathbf{q}_m^{H}\mathbf{I}_{M,i}^{T}\big)}{(1 + NM\alpha_{d})}\otimes \mathbf{I}_{N}\big)\\
	&\quad \cdot \big((\bar{\mathbf{S}}_{(i)}(MN\mathbf{I}_{M} - N\mathbf{q}_m\mathbf{q}_m^{H})\mathbf{e}_{j})\otimes\mathbf{I}_{N}\big)\bigg)\otimes \mathbf{I}_{N}\\
	& = \frac{\alpha_{d}^{2}(MN\mathbf{e}_{j}^{T} -  Nw_M^{(j-1)m}\mathbf{q}_m^{H})\boldsymbol{\Theta}_{i}(MN\mathbf{e}_{j}- Nw_M^{-(j-1)m}\mathbf{q}_m)}{1 + NM\alpha_{d}}\mathbf{I}_{N^{2}}
	\end{aligned}
\end{equation}}
where $\boldsymbol{\Theta}_{i} \triangleq \mathbf{I}_{M,i}^{T}\mathbf{I}_{M,i} +  \frac{N\alpha_{d}}{1 + N\alpha_{d}  }\mathbf{I}_{M,i}^{T}\mathbf{I}_{M,i}\mathbf{q}_m\mathbf{q}_m^{H}\mathbf{I}_{M,i}^{T}\mathbf{I}_{M,i}$. Note that $\mathbf{I}_{M,i}^{T}\mathbf{I}_{M,i}$ is the identity matrix $\mathbf{I}_{M}$ with its $i$th diagonal element set to zero,  and $\mathbf{I}_{M,i}^{T}\mathbf{I}_{M,i}\mathbf{q}_m$ is $\mathbf{q}_m$ with its $i$th element set to zero. Also  $\mathbf{e}_{j}^{T}\boldsymbol{\Theta}_{i}\mathbf{e}_{j} = 1 + \frac{N\alpha_{d}}{1 + N\alpha_{d}}$, $\mathbf{e}_{j}^{T}\boldsymbol{\Theta}_{i}\mathbf{q}_m = w_M^{(j-1)m}(1 + \frac{N\alpha_{d}}{1 + N\alpha_{d}}(M-1))$, $\mathbf{q}_m^{H}\boldsymbol{\Theta}_{i}\mathbf{e}_{j} = w_M^{-(j-1)m}(1 + \frac{N\alpha_{d}}{1 + N\alpha_{d}}(M-1))$ and $\mathbf{q}_m^{H}\boldsymbol{\Theta}_{i}\mathbf{q}_m = (M-1)(1+\frac{N\alpha_{d}}{1 + N\alpha_{d}}(M-1))$. Then, \eqref{pair_ij_2b} becomes

	\begin{align}
	&  \frac{\alpha_{d}^{2}N^{2}}{1 + NM\alpha_{d}}\big(M^{2}\mathbf{e}_{j}^{T}\boldsymbol{\Theta}_{i}\mathbf{e}_{j} -  Mw_M^{-(j-1)m}\mathbf{e}_{j}^{T}\boldsymbol{\Theta}_{i}\mathbf{q}_m\notag\\
	&\qquad - Mw_M^{(j-1)m}\mathbf{q}_m^{H}\boldsymbol{\Theta}_{i}\mathbf{e}_{j} + \mathbf{q}_m^{H}\boldsymbol{\Theta}_{i}\mathbf{q}_m\big)\mathbf{I}_{N^{2}}\notag\\
	& = \frac{\alpha_{d}^2N^2(\frac{N\alpha_{d}}{1 + N\alpha_{d}} + M^2 -M -1)}{1 + NM\alpha_{d}}\mathbf{I}_{N^{2}}.\label{pair_ij_2}
	\end{align}
Using \eqref{pair_ij_1}, \eqref{pair_ij_2b} and \eqref{pair_ij_2}, $\boldsymbol{\Gamma}_{i,j}$ becomes
\begin{equation}\label{pair_ij_3}
\begin{aligned}
\boldsymbol{\Gamma}_{i,j}= \frac{\alpha_{d} MN - N\alpha_{d}/(1+N\alpha_{d})}{1 +  MN\alpha_{d}}\mathbf{I}_{N^{2}} \triangleq \Gamma\mathbf{I}_{N^{2}}
\end{aligned}
\end{equation}
where $0<\Gamma < 1$ which is invariant to $i,j,m$. Similarly, one can verify that $\boldsymbol{\Gamma}_{T,j,i} = \Gamma\mathbf{I}_{N^{2}}$.
Then we have $(\mathbf{I} - \boldsymbol{\Gamma}_{i,j}\boldsymbol{\Gamma}_{T,j,i})^{-1} = (1 -\Gamma^{2})^{-1}\mathbf{I}_{N^{2}}$.

Using the above results in \eqref{mul_deri}, we have
\begin{equation}\label{mul_deri_3}
\begin{aligned}
&\frac{\partial I(\mathbf{y}_{i};\mathbf{y}_{T,j})}{\partial \bar{\mathbf{F}}}\\
& = \frac{1}{\ln 2\partial \bar{\mathbf{F}}}\left (Tr(\frac{\Gamma}{1 - \Gamma^2}\partial \boldsymbol{\Gamma}_{i,j})+ Tr(\frac{\Gamma}{1 - \Gamma^2}\partial \boldsymbol{\Gamma}_{T,j,i})\right ).
\end{aligned}
\end{equation}
Similar to \eqref{mul_deri_2}, the first term in \eqref{mul_deri_3} (except for a constant factor) can be expressed as
\begin{equation}\label{mul_deri_4}
\begin{aligned}
&\frac{1}{\partial \bar{\mathbf{F}}}Tr\left(\partial \boldsymbol{\Gamma}_{i,j}\right) = 2\big(\boldsymbol{\Gamma}_{i,j}^{(0)} - \boldsymbol{\Gamma}_{i,j}^{(1)}+\boldsymbol{\Gamma}_{i,j}^{(2)}- \boldsymbol{\Gamma}_{i,j}^{(3)}\big)\bar{\mathbf{F}}
\end{aligned}
\end{equation}
where
$\boldsymbol{\Gamma}_{i,j}^{(0)} =  N\mathbf{e}_{j}\mathbf{e}_{j}^{T}\otimes \mathbf{I}_{N}$,
\begin{equation}\label{}
  \boldsymbol{\Gamma}_{i,j}^{(1)} = N\bar{\mathbf{S}}_{(i)}^{T}(\mathbf{I} + \bar{\mathbf{S}}_{(i)}\bar{\mathbf{F}}\bar{\mathbf{F}}^{H}\bar{\mathbf{S}}_{(i)}^{T})^{-1}
  (\bar{\mathbf{S}}_{(i)}\bar{\mathbf{F}}\bar{\mathbf{F}}^{H}\mathbf{S}_{j}^{T})\mathbf{S}_{j},
\end{equation}
\begin{align}\label{}
  \boldsymbol{\Gamma}_{i,j}^{(2)} =& N\bar{\mathbf{S}}_{(i)}^{T}(\mathbf{I} + \bar{\mathbf{S}}_{(i)}\bar{\mathbf{F}}\bar{\mathbf{F}}^{H}\bar{\mathbf{S}}_{(i)}^{T})^{-1}
(\bar{\mathbf{S}}_{(i)}\bar{\mathbf{F}}\bar{\mathbf{F}}^{H}\mathbf{S}_{j}^{T})\notag\\
&\cdot(\mathbf{S}_{j}
\bar{\mathbf{F}}\bar{\mathbf{F}}^{H}\bar{\mathbf{S}}_{(i)}^{T})(\mathbf{I} + \bar{\mathbf{S}}_{(i)}\bar{\mathbf{F}}\bar{\mathbf{F}}^{H}\bar{\mathbf{S}}_{(i)}^{T})^{-1}
\bar{\mathbf{S}}_{(i)}
\end{align}
and $\boldsymbol{\Gamma}_{i,j}^{(3)} = (\boldsymbol{\Gamma}_{i,j}^{(1)})^{T}$. Furthermore,
using $\mathbf{I}_{M,i}^T\mathbf{I}_{M,i}\mathbf{e}_j\mathbf{e}_j^T=\mathbf{e}_j\mathbf{e}_j^T$ for $i\neq j$ and the previous results under $\bar{\mathbf{F}} = \sqrt{\alpha_{d}}\mathbf{\bar Q}_m$, we have
\begin{equation}\label{mul_deri_5}
\begin{aligned}
&\boldsymbol{\Gamma}_{i,j}^{(1)}\\
& = \frac{N\alpha_{d}\big(\boldsymbol{\Theta}_{i}(MN\mathbf{I} - N\mathbf{q}_m\mathbf{q}_m^{H})\mathbf{e}_{j}\mathbf{e}^{T}_{j}\big)\otimes\mathbf{I}_{N}}{1 +NM\alpha_{d}}\\
& = \frac{N\alpha_{d}\big(MN\mathbf{e}_{j}\mathbf{e}^{T}_{j}  -  \frac{N}{1 + N\alpha_{d}  }\mathbf{I}_{M,i}^{T}\mathbf{I}_{M,i}\mathbf{q}_m\mathbf{q}_m^{H}\mathbf{e}_{j}\mathbf{e}^{T}_{j}\big)
\otimes\mathbf{I}_{N}}{1 +NM\alpha_{d}},
\end{aligned}
\end{equation}

\begin{equation}\label{mul_deri_6}
{\small
\begin{aligned}
&\boldsymbol{\Gamma}_{i,j}^{(2)}\\
& = \frac{N\alpha_{d}^{2}\big(\boldsymbol{\Theta}_{i}(MN\mathbf{I} - N\mathbf{q}_m\mathbf{q}_m^{H})\mathbf{e}_{j}\mathbf{e}^{T}_{j}(MN\mathbf{I} - N\mathbf{q}_m\mathbf{q}_m^{H})\boldsymbol{\Theta}_{i}\big)\otimes\mathbf{I}_{N}}{(1+NM\alpha_{d})^{2}}\\
& = \frac{\alpha_{d}^{2}N^3}{(1+NM\alpha_{d})^{2}}\bigg(M^{2}\mathbf{e}_{j}\mathbf{e}^{T}_{j} + \frac{1}{(1 + N\alpha_{d})^{2}}\mathbf{I}_{M,i}^{T}\mathbf{I}_{M,i}\mathbf{q}_m\mathbf{q}_m^{H}\mathbf{I}_{M,i}^{T}
\mathbf{I}_{M,i}\\
&- \frac{M}{1 + N\alpha_{d}}\mathbf{e}_{j}\mathbf{e}^{T}_{j}\mathbf{q}_m\mathbf{q}_m^{H}\mathbf{I}_{M,i}^{T}
\mathbf{I}_{M,i} -\frac{M}{1 + N\alpha_{d}}\mathbf{I}_{M,i}^{T}\mathbf{I}_{M,i}\mathbf{q}_m\mathbf{q}_m^{H}\mathbf{e}_{j}
\mathbf{e}^{T}_{j} \bigg)\\
&\quad\otimes\mathbf{I}_{N}
\end{aligned}}
\end{equation}
where the derivation of \eqref{mul_deri_6} is shown in Appendix \ref{app:d}.

Similarly, one can verify that  $\frac{\partial Tr(\partial \boldsymbol{\Gamma}_{T,j,i})}{\partial \bar{\mathbf{F}}} = 2(\boldsymbol{\Gamma}_{j,i}^{(0)} - \boldsymbol{\Gamma}_{j,i}^{(1)} +\boldsymbol{\Gamma}_{j,i}^{(2)} - (\boldsymbol{\Gamma}_{j,i}^{(1)})^{T})\bar{\mathbf{F}}$.
%

%
Note that
\begin{equation}\label{s1}
\sum_{i=1}^{M-1}\sum_{j=i+1}^{M}\big(\mathbf{e}_{j}\mathbf{e}_{j}^{T} +\mathbf{e}_{i}\mathbf{e}_{i}^{T} \big)\otimes \mathbf{I}_{N} = (M-1)\mathbf{I}_{MN},
\end{equation}
\begin{equation}\label{s2}
\begin{aligned}
&\sum_{i=1}^{M-1}\sum_{j=i+1}^{M}(\mathbf{I}_{M,i}^{T}\mathbf{I}_{M,i}\mathbf{q}_m\mathbf{q}_m^{H}
\mathbf{e}_{j}\mathbf{e}^{T}_{j}+ \mathbf{I}_{M,j}^{T}\mathbf{I}_{M,j}\mathbf{q}_m\mathbf{q}_m^{H}\mathbf{e}_{i}
\mathbf{e}^{T}_{i})\\
& =  (M-2)\mathbf{q}_m\mathbf{q}_m^{H}+ \mathbf{I}_{M},
\end{aligned}
\end{equation}
\begin{align}
&\sum_{i=1}^{M-1}\sum_{j=i+1}^{M}(\mathbf{I}_{M,i}^{T}\mathbf{I}_{M,i}\mathbf{q}_m\mathbf{q}_m^{H}
\mathbf{I}_{M,i}^{T}\mathbf{I}_{M,i} \notag\\
&\quad\quad\quad \quad +\mathbf{I}_{M,j}^{T}\mathbf{I}_{M,j}\mathbf{q}_m\mathbf{q}_m^{H}\mathbf{I}_{M,j}^{T}
\mathbf{I}_{M,j})\\
& =  (M-1)\mathbf{q}_m\mathbf{q}_m^{H}+ 2\mathbf{I}_{M}.\label{s3}
\end{align}
Then, with some further manipulations, we obtain
{\small
	\begin{align}
	\frac{\partial I_{M}}{\partial \bar{\mathbf{F}}} & =\sum_{i=1}^{M-1}\sum_{j=i+1}^M
\frac{\partial I(\mathbf{y}_i;\mathbf{y}_{T,j})}{\partial \bar {\mathbf{F}}} \notag\\
	&  =\frac{2N\Gamma}{(1 - \Gamma^{2})\ln 2} \big(\frac{M-1}{(1+MN\alpha_{d})^2} \notag\\
&\quad\quad\quad\quad + \frac{2N\alpha_{d}(1  + 2N\alpha_{d})}{(1+MN\alpha_{d})^2(1+N\alpha_{d})^2} \big)\bar{\mathbf{F}}. \label{mul_deri_sum2}
	\end{align}}

Then one can verify that the first condition in \eqref{kkt_mul_sk_sc} is satisfied by \eqref{mul_deri_sum2} and  $\mu_{i} = \frac{N\Gamma}{(1 - \Gamma^{2})\ln 2}(\frac{M-1}{(1+MN\alpha_{d})^2} + \frac{2N\alpha_{d}(1 + 2N\alpha_{d})}{(1+MN\alpha_{d})^2(1+N\alpha_{d})^2}) > 0$, and all other conditions in \eqref{kkt_mul_sk_sc} are satisfied by further choosing $\alpha_{d} = \frac{KP}{ N^2(M-1)}$. Therefore, $\bar{\mathbf{F}} = \sqrt{\frac{KP}{ N^2(M-1)}}\mathbf{\bar Q}_m$ is a solution to \eqref{kkt_mul_sk_sc}.

\subsection{Derivation of \eqref{mul_deri_6}}\label{app:d}
From the first equality in \eqref{mul_deri_6}, we have
\begin{align}
&\boldsymbol{\Theta}_{i}\big(M^2\mathbf{e}_{j}\mathbf{e}^{T}_{j} - M\mathbf{q}_{m}\mathbf{q}_{m}^{H}\mathbf{e}_{j}\mathbf{e}^{T}_{j} - M\mathbf{e}_{j}\mathbf{e}^{T}_{j}\mathbf{q}_{m}\mathbf{q}_{m}^{H} + \mathbf{q}_{m}\mathbf{q}_{m}^{H} \big)\boldsymbol{\Theta}_{i}\notag\\
& = M^2\boldsymbol{\Theta}_{i}\mathbf{e}_{j}\mathbf{e}^{T}_{j}\boldsymbol{\Theta}_{i} - M\boldsymbol{\Theta}_{i}\mathbf{e}_{j}\mathbf{e}^{T}_{j}\mathbf{q}_{m}\mathbf{q}_{m}^{H}
\boldsymbol{\Theta}_{i}\notag\\
&\quad -M\boldsymbol{\Theta}_{i}\mathbf{q}_{m}\mathbf{q}_{m}^{H}\mathbf{e}_{j}\mathbf{e}^{T}_{j}
\boldsymbol{\Theta}_{i} + \boldsymbol{\Theta}_{i}\mathbf{q}_{m}\mathbf{q}_{m}^{H}\boldsymbol{\Theta}_{i}.\label{1}
\end{align}

Let $\eta = \frac{N\alpha_{d}}{1 + N\alpha_{d}  }$. Each of the four terms  in \eqref{1} can be simplified as follows:
\begin{align}
&M^2\boldsymbol{\Theta}_{i}\mathbf{e}_{j}\mathbf{e}^{T}_{j}\boldsymbol{\Theta}_{i}\notag\\
& = M^2\bigg(\mathbf{e}_{j}\mathbf{e}^{T}_{j} + \eta\mathbf{e}_{j}\mathbf{e}^{T}_{j}\mathbf{q}_{m}\mathbf{q}_{m}^{H}\mathbf{I}_{M,i}^{T}\mathbf{I}_{M,i}\notag\\
&+\eta\mathbf{I}_{M,i}^{T}\mathbf{I}_{M,i}\mathbf{q}_{m}\mathbf{q}_{m}^{H}\mathbf{e}_{j}
\mathbf{e}^{T}_{j} + \eta^2\mathbf{I}_{M,i}^{T}\mathbf{I}_{M,i}\mathbf{q}_{m}\mathbf{q}_{m}^{H}\mathbf{I}_{M,i}^{T}
\mathbf{I}_{M,i}\bigg),\label{1.1}
\end{align}
\begin{align}
&M\boldsymbol{\Theta}_{i}\mathbf{e}_{j}\mathbf{e}^{T}_{j}\mathbf{q}_{m}\mathbf{q}_{m}^{H}
\boldsymbol{\Theta}_{i}\notag\\
&=M\bigg((\eta(M-1) +1)\mathbf{e}_{j}\mathbf{e}^{T}_{j}\mathbf{q}_{m}\mathbf{q}_{m}^{H}\mathbf{I}_{M,i}^{T}\mathbf{I}_{M,i} \\
&\quad +(\eta^2(M-1)+\eta)\mathbf{I}_{M,i}^{T}\mathbf{I}_{M,i}\mathbf{q}_{m}\mathbf{q}_{m}^{H} \mathbf{I}_{M,i}^{T}\mathbf{I}_{M,i}\bigg),\label{1.2}
\end{align}
\begin{align}
&M\boldsymbol{\Theta}_{i}\mathbf{q}_{m}\mathbf{q}_{m}^{H}\mathbf{e}_{j}\mathbf{e}^{T}_{j}
\boldsymbol{\Theta}_{i}\notag\\
& = M\bigg((\eta(M-1) + )\mathbf{I}_{M,i}^{T}\mathbf{I}_{M,i}\mathbf{q}_{m}\mathbf{q}_{m}^{H}\mathbf{e}_{j}\mathbf{e}^{T}_{j}\notag\\
&\quad + (\eta^2(M-1)+\eta)\mathbf{I}_{M,i}^{T}\mathbf{I}_{M,i}\mathbf{q}_{m}\mathbf{q}_{m}^{H} \mathbf{I}_{M,i}^{T}\mathbf{I}_{M,i}\bigg),\label{1.3}
\end{align}
\begin{align}
&\boldsymbol{\Theta}_{i}\mathbf{q}_{m}\mathbf{q}_{m}^{H}\boldsymbol{\Theta}_{i}
=(\eta(M-1)+1)^2 \mathbf{I}_{M,i}^{T}\mathbf{I}_{M,i}\mathbf{q}_{m}\mathbf{q}_{m}^{H}\mathbf{I}_{M,i}^{T}\mathbf{I}_{M,i}.
\label{1.4}
\end{align}
Applying \eqref{1} - \eqref{1.4}, the second equality of \eqref{mul_deri_6} follows.
\bibliographystyle{IEEEtran}
\bibliography{secret_key_revised}

\end{document}